\documentclass[iop,twocolappendix,numberedappendix,appendixfloats]{emulateapj}

\usepackage{natbib}
\usepackage[bookmarks,breaklinks]{hyperref}
   \hypersetup{
     colorlinks,
     citecolor=blue,
     filecolor=magenta,
     linkcolor=blue,
     urlcolor=cyan
}
\usepackage{multirow}

\shorttitle{Environment and triggering of radio-AGN}
\shortauthors{Karouzos et al.}

\begin{document}


\title{The Infrared Medium-deep Survey II:\\ How to trigger radio-AGN{?} Hints from their environments}


\author{Marios Karouzos$^{1}$, Myungshin Im$^{1}$, Jae-Woo Kim$^{1}$, Seong-Kook Lee$^{1}$, Scott Chapman$^{2}$, Yiseul Jeon$^{1}$, Changsu Choi$^{1}$, Jueun Hong$^{1}$, Minhee Hyun$^{1}$, Hyunsung David Jun$^{1}$, Dohyeong Kim$^{1}$, Yongjung Kim$^{1}$, Ji Hoon Kim$^{1,3}$, Duho Kim$^{1,4}$, Soojong Pak$^{5}$, Won-Kee Park$^{1,6}$, Yoon Chan Taak$^{1}$, Yongmin Yoon$^{1}$, Alastair Edge$^{7}$}
\affil{$^{1}$CEOU - Astronomy Program, Department of Physics \& Astronomy, Seoul National University, Gwanak-gu, Seoul, Republic of Korea}
\affil{$^{2}$Department of Physics and Atmospheric Science, Dalhousie University, Halifax, Nova Scotia, Canada}
\affil{$^{3}$Subaru Telescope, National Astronomical Observatory of Japan, 650 North A'ohoku Place, Hilo, HI 96720, U.S.A.}
\affil{$^{4}$Arizona State University, School of Earth and Space Exploration, PO Box 871404, Tempe, AZ 85287-1404, U.S.A.}
\affil{$^{5}$School of Space Research, Kyung Hee University, Yongin-si, Gyeonggi-do, 446-701, Republic of Korea}
\affil{$^{6}$Korea Astronomy and Space Science Institute, 776 Daedeokdae-ro, Yuseong-gu, Daejeon, Republic of Korea}
\affil{$^{7}$Department of Physics, University of Durham, South Road, Durham, DH1 3LE, UK}
\email{mkarouzos@astro.snu.ac.kr}




\begin{abstract}
Activity at the centers of galaxies, during which the central supermassive black hole is accreting material, is nowadays accepted to be rather ubiquitous and most probably a phase of every galaxy's evolution. It has been suggested that galactic mergers and interactions may be the culprits behind the triggering of nuclear activity. We use near-infrared data from the new Infrared Medium-Deep Survey (IMS) and  the Deep eXtragalactic Survey (DXS) of the VIMOS-SA22 field and radio data at 1.4 GHz from the FIRST survey and a deep VLA survey to study the environments of radio-AGN over an area of $\sim25$ degrees$^2$ and down to a radio flux limit of 0.1 mJy and a $J$-band magnitude of 23 mag AB. Radio-AGN are predominantly found in environments similar to those of control galaxies at similar redshift, $J$-band magnitude, and $(M_{u}-M_{r})$ rest-frame color. However, a sub-population of radio-AGN is found in environments up to 100 times denser than their control sources. We thus preclude merging as the dominant triggering mechanism of radio-AGN. Through the fitting of the broadband spectral energy distribution of radio-AGN in the least and most dense environments, we find that those in the least dense environments show higher radio-loudness, higher star formation efficiencies, and higher accretion rates, typical of the so-called high-excitation radio-AGN. These differences tend to disappear at z$>1$. We interpret our results in terms of a different triggering mechanism for these sources that is driven by mass-loss through winds of young stars created during the observed ongoing star formation.
\end{abstract}



\section{Introduction}
\label{sec:intro}

There is an ongoing debate as to what the possible triggers of activity in the nuclei of galaxy may be. In a fundamental sense, the triggering of active galactic nuclei (AGN) requires the availability of a gas reservoir which feeds the accretion onto the supermassive black hole at the center of a galaxy. Traditionally, the high incidence of merger remnants within samples of powerful quasars (e.g., \citealt{Heckman1986}, \citealt{Canalizo2007}, \citealt{Bennert2008}, \citealt{Karouzos2010}, \citealt{Almeida2011}), the temporal coincidence of cosmic AGN and star formation activity (e.g., \citealt{Hopkins2006c}, \citealt{Richards2006}, \citealt{Aird2010}, \citealt{Kistler2013}), and the phenomenological link between a local population of ultra-luminous infrared galaxies and powerful obscured quasars (e.g., \citealt{Sanders1988}, \citealt{Canalizo2001}) has led to the first-order conclusion that AGN might be triggered through the merging of galaxies (e.g., \citealt{Hernquist1989}, \citealt{Kauffmann2000}, \citealt{Lotz2008}). The gas available during such violent processes should both trigger bursts of intense star formation and consequently a phase of efficient accretion onto the central supermassive black hole (e.g., \citealt{Hopkins2006}), potentially with a time lag between the two phases (e.g., \citealt{Wild2010}).

While this scenario has been successful in explaining how powerful AGN are triggered during galactic mergers (although even this is still heavily debated, e.g., \citealt{Villforth2014}, \citealt{Karouzos2014a}), it soon became evident that such mergers of gas-rich galaxies cannot explain the full population of active galaxies. Especially in the local and intermediate redshift universe, the rate of these major mergers decreases significantly (e.g., \citealt{Lotz2011}). Several studies of the morphologies of moderate luminosity X-ray AGN (i.e., efficient accretors) at low and intermediate redshifts have shown that the morphologies of their host galaxies lack evidence of recent or ongoing mergers (e.g., \citealt{Cisternas2011}, \citealt{Kocevski2011}) but rather show prominent disks. The morphological end product of a potential major merger would be a ``red and dead'' early-type galaxy (e.g., \citealt{Hopkins2008b}). Moreover, a different flavor of active galaxies, AGN exhibiting strong collimated outflows in the radio (radio-AGN; e.g., \citealt{Urry1995}), were understood to be predominantly inefficiently accreting systems (e.g., \citealt{Kauffmann2003}, \citealt{Best2005}, \citealt{Hardcastle2007}), that could be very easily fed with a moderate amount of gas, either cold or hot, potentially originating in the halo of their host galaxy (e.g., \citealt{Hopkins2006b}). As such, there should be a significant population of active galaxies that is unassociated with mergers, their triggering and evolution rather driven by secular processes.

Previous studies on this topic focusing on the morphologies of AGN host galaxies have been plagued by two major drawbacks. Given the potential time lag between the peak of AGN luminosity and the peak of a galactic merger (e.g., \citealt{Sanders1988}), AGN-selected samples should exhibit very faint merger features and tidal distortions (e.g., \citealt{Lotz2008}), if any. The detection of these faint morphological features requires long exposure times and ideal observing conditions. The situation is aggravated by the fact that luminous AGN significantly contaminate, or in some cases even completely overshadow, the light of their host galaxy (e.g., \citealt{Pierce2010}). Therefore very high dynamic range, high resolution imaging is required to disentangle the different emission components. As a result, the number of AGN host galaxies that can be properly observed is constrained both in terms of AGN luminosity and host galaxy stellar mass.

An alternative approach to answer the question of what triggers AGN, over a range of AGN luminosities and independent of host galaxy luminosity, is the study of their environments. The density of the environment within which an AGN is embedded can be used as a proxy not only of past but also of near future mergers. It is expected that if mergers play a significant role in the triggering of AGN, they should be found in denser small-scale\footnote{Here small-scale is defined as $<500$ and is associated with close companions and galaxy group environments. Conversely, large-scale relates to galaxy cluster environments at linear scales of $\sim 1$ Mpc.} environments than other galaxies of similar properties. As in the case of morphology studies, the inferences from the study of AGN environment have largely been dependent on the wavelength in which the AGN were selected. Results range from AGN found in significantly overdense environments (e.g., \citealt{Best2004}, \citealt{Serber2006}, \citealt{Tasse2008}, \citealt{Ellison2011}, \citealt{Bradshaw2011}, \citealt{Galametz2012}, \citealt{Satyapal2014}, \citealt{Pace2014}), to environments that are consistent or even underdense compared to those of non-active galaxies with similar galaxy properties (e.g., \citealt{Miller2003}, \citealt{Kauffmann2004}, \citealt{Tasse2011}). In particular, \citet{Karouzos2014a}, using the first data release of the VIDEO survey (\citealt{Jarvis2013}), showed that the bulk of AGN, irrespective of the wavelength selection, reside in rather unremarkable environments when compared to non-active galaxies of similar stellar mass, out to a redshift of z=3. In the same study, it was shown that among the differently selected AGN, radio-AGN showed the highest degrees of over-densities among radio, X-ray, and mid-IR selected active galaxies. Moreover, radio-AGN were found to preferentially inhabit denser environments at group scales ($<200$ kpc) rather than cluster scales ($\sim 1$ Mpc).

Here we expand upon our previous study (\citealt{Karouzos2014a}), by focusing on a radio-selected sample of AGN, covering a much larger part of the sky than the first data release VIDEO sample ($\sim1$ deg$^{2}$), refining our control sample selection, and expanding the radio luminosity range probed to eight orders of magnitude.  As such, we are in a position to counter the possible cosmic variance effects affecting small-area survey fields. Furthermore, by including the relatively rare very radio-luminous AGN, we test the possible dichotomy of triggering mechanisms between faint and luminous AGN. Finally, we take the study of the environments of radio-AGN a step further by looking at the host galaxy properties of these sources and making a connection between the small scale environments of AGN and the potential feeding mechanisms of their supermassive black holes. 

The paper is organised as follows: in Section \ref{sec:IMS} we describe the data we use in this paper, in particular introducing the Infrared Medium-Deep Survey (IMS). In Section \ref{sec:method} the methodology of this study is described, explaining the calculation of photometric redshifts, the selection of control samples, and the fitting of the broadband spectral energy distribution (SED) undertaken. In Section \ref{sec:env} we present our results concerning the environments of radio-AGN while in Section \ref{sec:SFR} we make the connection between the environments of radio-AGN and their host galaxy properties. Finally, in Sections \ref{sec:discuss} and \ref{sec:conc} we discuss the implications of our results in the context of other studies in the literature and offer our final conclusions, respectively. Throughout the paper, we assume the cosmological parameters $H_{0}=71$ km/s/Mpc, $\Omega_{M}=0.27$, and $\Omega_{\Lambda}=0.73$ (\citealt{Komatsu2011}).

\section{The SA22 field}
\label{sec:IMS}
\subsection{Near-infrared surveys in the SA22 field}
The Infrared Medium-Deep Survey (IMS; Im et al., in prep.) is a recently concluded near-infrared (NIR) imaging survey using the UKIRT Wide Field Camera (WFCAM, \citealt{Casali2007}) on the 3.8m UKIRT telescope in Hawaii, US, the Seoul National University CAMera (SNUCAM, \citealt{Im2010}) on the 1.5m telescope at the Maidanak observatory in Uzbekistan, and the Camera for QUasars in the EArly uNiverse (CQUEAN, \citealt{Park2012}) on the 2.1m Otto-Struve telescope at the McDonald Observatory in Texas, US. The IMS covers $\sim 106$ deg$^{2}$ on the sky in $Y$ and $J$ NIR bands, with a 5$\sigma$ magnitude limit of 23 mag AB. Through the IMS some of the most well-known extragalactic legacy survey fields have been observed in the NIR. These include the XMM Large Scale Structure (LSS), the CFHT Legacy Survey (LS) W2, the Lockman Hole, the Extended Groth Strip, the European Large Area Infrared Survey North 1 and 2, and the SA22 fields.

\begin{figure}[htbp]
\begin{center}
\includegraphics[width=0.48\textwidth,angle=0]{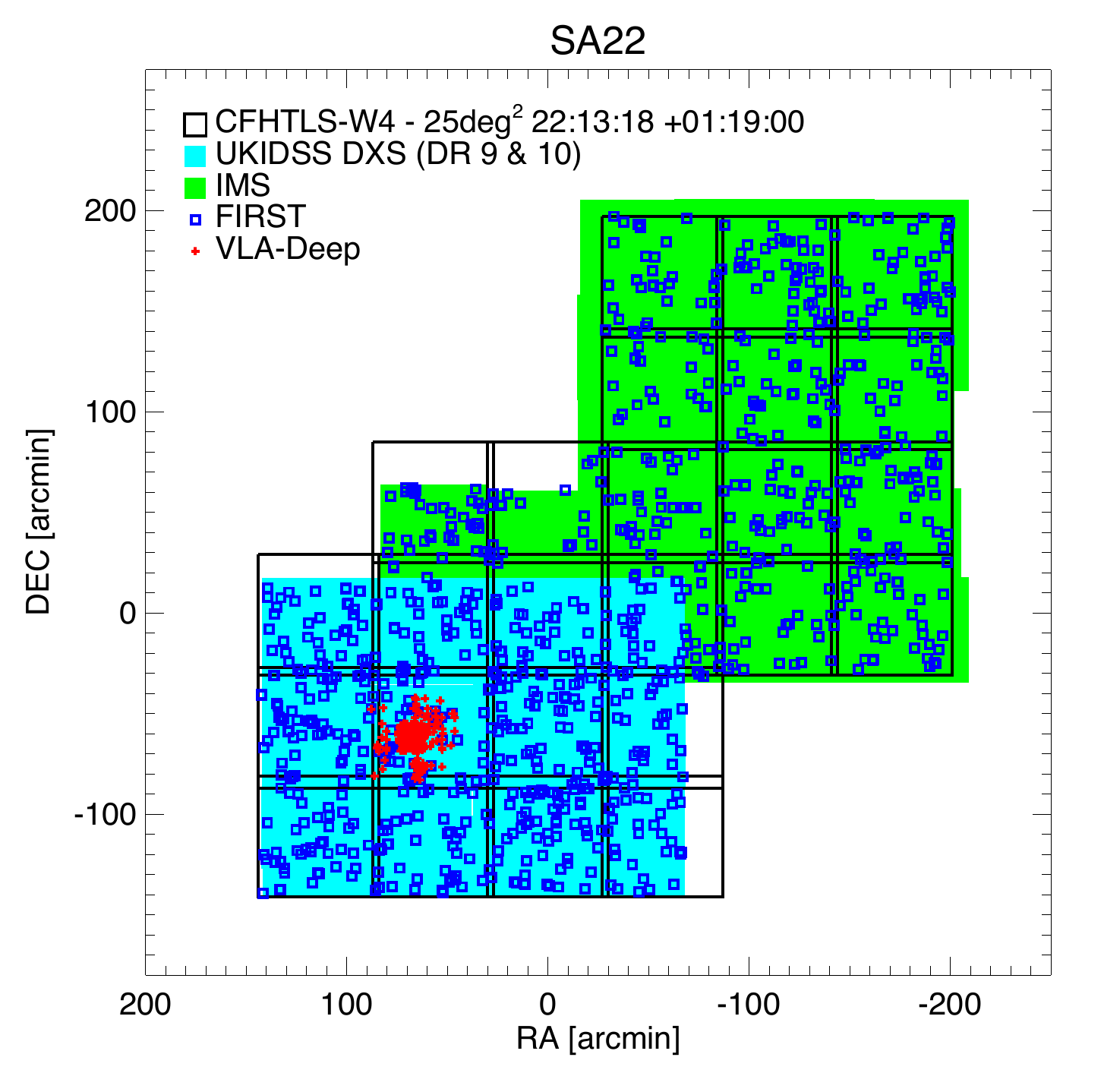}
\caption{Coverage map of the SA22 field  showing the coverage of the IMS and DXS NIR surveys and the CFHT optical survey. The cross-matched radio sources (see Section \ref{sec:cross}) from the VLA-FIRST (blue squares) and VLA-Deep (red circles) radio surveys are also shown. The map is centered at the center of CFHTLS W4 field (22:13:18 +01:19:00).}
\label{fig:cov}
\end{center}
\end{figure}

In particular, here we focus on the SA22 field due to its good wavelength coverage in both the NIR from the IMS ($J$ band, covering a total of 16 deg$^{2}$)) and the optical by the CFHT Legacy Survey (CFHTLS; u, g, r, i, and z). In addition, part of the SA22 field is covered by the Deep eXtragalactic Survey (DXS), part of the UKIRT Infrared Deep Sky Survey (UKIDSS; \citealt{Lawrence2007}), which provides additional $K_s$ band coverage down to a magnitude limit of 22.7 mag AB (5$\sigma$ limit; \citealt{Kim2011}) for a total of 8.75 deg$^{2}$. A coverage map of the SA22 field is shown in Fig. \ref{fig:cov}.

As we are interested in measuring the environment properties of galaxies, the accurate knowledge of distances and hence redshifts is imperative. Given the absence of full spectroscopic coverage of the SA22 field, it is crucial to acquire accurate galaxy colors for the calculation of robust photometric redshifts. Therefore for our analysis we use a point-spread-function (PSF) matched band-merged catalog\footnote{The detailed catalog construction and full catalog will be presented in Kim et al. 2014 (in prep.).}. In short, we use a Gaussian PSF of 1.1 arcsec (equal to the worst seeing among all SA22 observations) to convolve images at all bands. The convolved images are then resampled to match the WFCAM field-of-view and CFHT pixel scale (0.8 deg$^{2}$ and 0.187 arcsec/pix) using the SWARP package (\citealt{Bertin2002}). The original (prior to PSF matching) $J$-band image is then used as the detection image. After that we use the SExtractor software (\citealt{Bertin1996}) in dual-mode to extract sources from the images in the rest of the photometric bands\footnote{An aperture of 2 arcsec is used for aperture magnitude extraction in SExtractor.}. The multiwavelength data available in the SA22 field and their respective limits are shown in Table \ref{tab:sa22}. Our base selection is done in the $J$ band at the IMS survey's limiting magnitude (23 mag AB) over the total area of $\sim25$ deg$^{2}$. 

\begin{table}
\caption{Information about the photometric data available for the SA22 field and in particular for the IR-radio cross-matched sample (VLA-FIRST sample). Column (1) gives the name of the survey and Cols. (2), (3), and (4) the waveband name, central wavelength, and sensitivity (5$\sigma$), respectively. \\$\dagger$ Units for the VLA band are given in units of mJy.\\$\ddagger$ Frequency in units of GHz.}
\begin{center}
\begin{tabular}{|c|c|c|c|}
\hline
Survey							&	Band	&	Wavelength	&	Sensitivity		\\
								&			&	($\mu$m)		&	(AB mag)		\\
\hline
\multirow{5}{*}{CFHTLS}				&	$u*$		&	0.38			&		25.2		\\
								&	$g$		&	0.48			&		25.5		\\
								&	$r$		&	0.62			&		25.0		\\
								&	$i$ 		&	0.75			&		24.8		\\
								&	$z$ 		&	0.88			&		23.9		\\
\hline
IMS								&	$J$		&	1.2			&		23.0		\\
\hline
\multirow{2}{*}{UKIDSS DXS}			&	$J$		&	1.2			&		23.2		\\
								&	$K_s$	&	2.2			&		22.7		\\
\hline
\multirow{2}{*}{WISE-All Sky Survey}	&	$W1$	&	3.4			&		20.0		\\
								&	$W2$	&	4.6			&		19.3		\\
								&	$W3$	&	12.0			&		17.0		\\
								&	$W4$	&	22.0			&		14.7		\\		
\hline
VLA-FIRST						&	$L$		&	1.4$\ddagger$ 	&		1.0$\dagger$	\\
VLA-Deep						&	$L$		&	1.4$\ddagger$ 	&		0.1$\dagger$	\\
\hline
\end{tabular}
\end{center}
\label{tab:sa22}
\end{table}%

\subsection{Radio surveys in the SA22 field}
For the selection of radio sources we use data from the Faint Images of the Radio Sky at Twenty-Centimeters (FIRST, \citealt{Becker1995}), a very wide-field radio survey at 1.4 GHz using the Very Large Array (VLA) radio interferometer. The FIRST survey has a flux density limit of 1 mJy and a resolution of 5 arcsec. We exclude radio sources from the FIRST survey catalog that have a probability, P(S), to be spurious due to side lobes of neighbouring bright sources P(S)$>0.15$ (for the way P(S) is defined and calculated, see \citealt{White1997}). In addition, we use a deeper radio survey of part of the SA22 field using the VLA at two different configurations, A (PI: Chapman) and B (PI: Yun), with a limiting flux density of $\sim0.05-0.12$ mJy beam$^{-1}$ (5$\sigma$ limit), and a resolution of 1.4 and 5 arcsec, respectively. More details on these data is given in Appendix \ref{app:VLA}.

\subsection{Radio-IR Source Cross-matching}
\label{sec:cross}
In order to study the environments and host galaxy properties of radio sources in the SA22 field, we cross-match our base $J$-band catalog with the two radio catalogs for the SA22 field. We use the Poisson probability based method of \citet{Downes1986} (e.g., \citealt{Ivison2007}, \citealt{Hodge2013}). The proximity and magnitude of each individual candidate source are used to calculate the probability that the candidate source is not a background source, P. Around each radio source we calculate the Poisson probability for each near-IR source to be within a circle of radius $r_{c}$. This Poisson probability is defined as:
$$P^{*}=1-e^{-\pi r^{2}N_{m}},$$
where r is the distance of the candidate counterpart from the multi-wavelength source, and $N_{m}$ is the surface number density within radius r and limiting near-IR magnitude m. $r_{c}$ can be defined through the angular resolution of the different instruments (i.e., WFCAM and VLA). Here we assume an $r_{c}=10$ arcsec, which is 2 times the nominal resolution of the FIRST survey\footnote{In practice, the majority of NIR sources ($\sim70\%$) are at distances found below 5 arcsecs from their matched radio source.}. The expected number of events (i.e., near-IR sources) with $P\leqslant P^{*}$ can then be approximated (for a finite search radius $r_{c}$) as 
$$E=P_{c}=\pi r_{c}^{2}N_{T},$$ 
for $P^{*}\geqslant P_{c}$, and 
$$E=P^{*}(1+\ln{P_{c}/P^{*}}),$$ 
for $P^{*}<P_{c}$. $P_{c}$ is a critical Poisson probability, defined by the surface number density, $N_{T}$, at the limiting magnitude of the NIR sample.

Finally, the probability of a chance cross-identification of the source can be calculated as $1-e^{-E}$. The near-IR candidate with the lowest such probability is chosen to be the true counterpart. For the FIRST sample, 1482 sources are matched to an NIR source. Of these, 67 radio sources are matched with 33 NIR sources in double or triple matches (32 cases and one case where the same NIR source was matched with two and three FIRST radio sources, respectively). For the VLA-Deep sample, in total 204 sources are matched to an NIR source. Of these, 16 are part of double cross-matchings with the same NIR source.

\begin{figure}[htbp]
\begin{center}
\includegraphics[width=0.45\textwidth,angle=0]{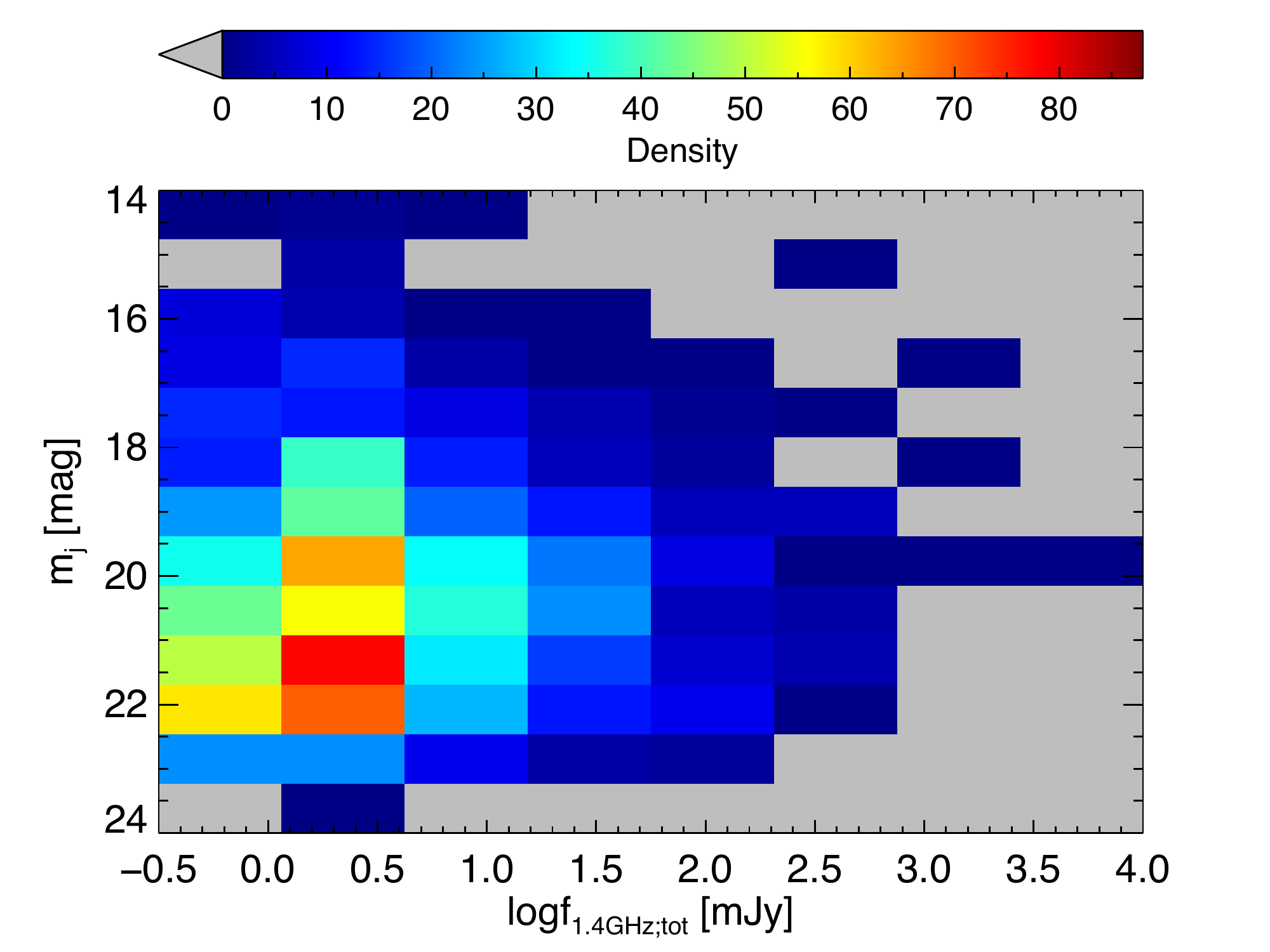}
\includegraphics[width=0.45\textwidth,angle=0,trim=3.5cm 3cm 2.8cm 3cm, clip=true]{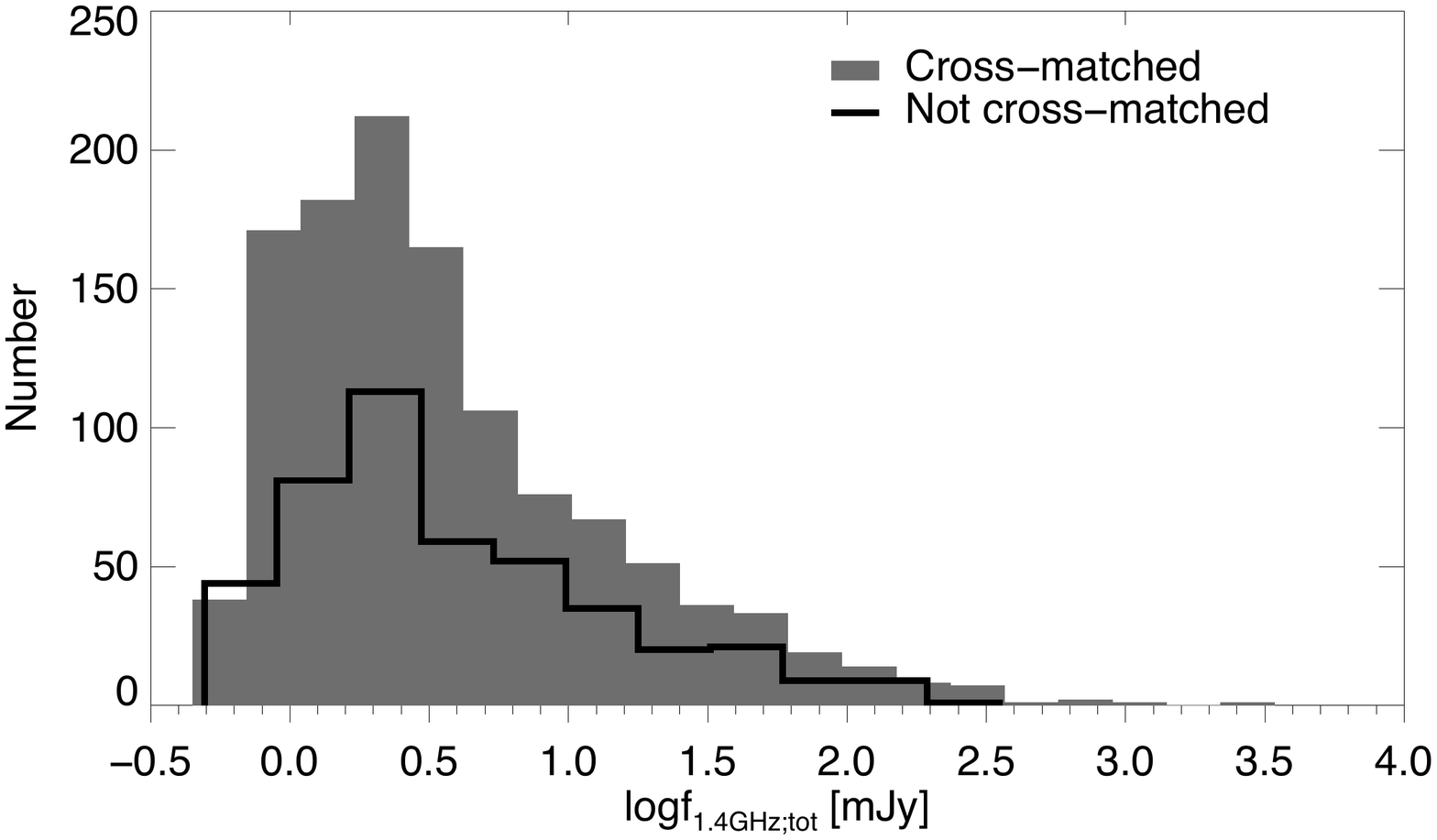}
\caption{Distribution of radio flux densities and $J$-band magnitudes for the cross-matched and non-cross-matched radio samples in the SA22 field. Upper: Density map of the $J$-band magnitude versus flux density at 1.4 GHz (in logarithmic scale). Lower: Flux density distribution at 1.4 GHz {for cross-matched (shaded histogram) and non-matched (open histogram)} radio samples.}
\label{fig:Jradio}
\end{center}
\end{figure}

The occurrence of a double cross-matching of two radio sources with a single NIR source may imply the presence of a double-lobe radio source (Fanaroff-Riley Class-II object, FRII; \citealt{Fanaroff1974}) or a core-jet configuration radio source. We combine the calculated Poisson probabilities and source separation with the classification scheme described in \citet{Best2005} to decide whether a multiple association is a true or a spurious one. 
\begin{enumerate}
\item For cases where one radio source is found very close to the NIR source (r$<5$ arcsec) and has very low probability for chance cross-matching ($<10\%$), while the other(s) is at r$>5$ arcsec and with high probability for chance cross-matching, the latter is considered spurious and the first one is kept. 
\item For cross-matchings where all radio sources have r$>5$ arcsec and probability for chance cross-matching $>10\%$, all of them are considered spurious and removed from the sample.
\item For the rest of the cases, the multiple matching is flagged as a candidate for a multi-component radio source. For these cases we follow the scheme of \citet{Best2005}. In short, if one of the radio sources is at r$<3$ arcsec from the NIR source while the others are not, this is considered a core-jet candidate and a comparison with NVSS (\citealt{Condon1998}) is required. Among our sources we do not have such a case. Alternatively, if the flux-weighted position of the two radio sources is close to the NIR one (here we assume a limit of 1.5 arcsec) and the two radio sources have comparable radio flux densities (here $<1$ mJy) then this is classified as a double-lobe radio source. 
\end{enumerate}
For the FIRST sample, 7 multiple matchings fall within case 1 (one matching considered as true), 8 multiple matchings fall within case 3 (double-lobe radio sources), while the rest fall under case 2 and are discarded as spurious. For the VLA-Deep sample, 5 multiple matchings fall under case 1, with the rest being discarded as spurious as no multiple matchings fulfil the criteria of a double-lobe radio source. Given the small number of actual double-lobe radio sources in our sample and in the absence of multi-frequency information that would confirm or reject this classification, we decide to exclude these 8 double-lobe radio source candidates from any further analysis. In the end, the FIRST-NIR cross-matched sample contains 1424 sources and the VLA-Deep-NIR sample contains 193 sources (the position of these sources within the SA22 field are shown in Fig. \ref{fig:cov}). The radio flux densities and $J$-band magnitudes of the final radio sample are shown in Fig. \ref{fig:Jradio}.

In order to select radio-AGN in these two samples, we employ a radio luminosity limit cut at 1.4 GHz. Following \citet{Condon1992} we can define a limiting non-thermal radio-luminosity above which a source can be classified as an AGN:
\begin{equation}
\left (\frac{L_{N}}{W\cdot Hz^{-1}}\right) \sim 5.3\cdot 10^{21}\left (\frac{\nu}{GHz}\right )^{-\alpha}\left [ \frac{SFR(M\geqslant5M_{\odot})}{M_{\odot}yr^{-1}}\right ],
\end{equation}
where $L_{N}$ is the non-thermal luminosity produced from ongoing star formation with a star-formation rate, \textit{SFR}, at an observing frequency $\nu$ and with a radio spectral index $\alpha=0.8$. We set a threshold of 100 solar masses per year, which gives a critical luminosity of $L_{lim}=1.03\cdot 10^{40}$erg/s. This roughly matches the turn-over power of the local luminosity function of radio sources, above which radio-loud AGN dominate the radio-source population (e.g., \citealt{Best2005a}, \citealt{Mauch2007}).  After K-correcting the 1.4GHz luminosities of the VLA sources {(see Section \ref{sec:redshift} for information on how redshifts were derived) }, objects more luminous than the above limit are assumed to be AGN. Given the absence of multi-frequency radio data, we make the assumption of an average radio spectral index of $\alpha=0.8$ (e.g., \citealt{Condon1992}). There are 777 FIRST sources and 136 VLA-Deep sources classified as radio-AGN according to this criterion. \textbf{Table \ref{tab:samples} gives the names and sizes of the different (sub-)samples used in the rest of the paper, along with the relevant sections where they are first introduced, and a description of their selection criteria.}

\begin{table*}
\caption{A summary of the (sub-)samples used throughout the paper (Column 1), along with the section where they are first introduced (Column 2) and their size (Column 3). For the last two rows, in brackets we give the number of sources identified as with reliable SEDs (see Section \ref{sec:SFR} and Fig. \ref{fig:ssfr_avg}). A summary of the definitions and selection criteria of each sample is shown in the last column, but more details can be found in the respective sections.}
\begin{center}
\begin{tabular}{|c|c|c|c|}
\hline
(Sub-)Sample				&	Section	&	Number	&	Description\\
\hline
FIRST; All					&	3,4		&	1009		&	FIRST/IMS cross-match\\
FIRST; AGN				&	3,4		&	777		&	FIRST/IMS, $L_{\mbox\scriptsize{{1.4GHz}}}\gtrsim10^{40}$ erg s$^{-1}$\\
VLA-Deep; All				&	3,4		&	181		&	VLA-Deep/IMS cross-match\\
VLA-Deep; AGN			&	3,4		&	136		&	VLA-Deep/IMS, $L_{\mbox\scriptsize{{1.4GHz}}}\gtrsim10^{40}$ erg s$^{-1}$\\
DXS; Radio-AGN			&	5		&	689		&	Radio/IMS, $L_{\mbox\scriptsize{{1.4GHz}}}\gtrsim10^{40}$ erg s$^{-1}$, DXS area\\
DXS; Over-Dense			&	5		&	77 (49)	&	Radio/IMS, $L_{\mbox\scriptsize{{1.4GHz}}}\gtrsim10^{40}$ erg s$^{-1}$, DXS area, overdense environments\\
DXS; Under-Dense			&	5		&	31 (24)	&	Radio/IMS, $L_{\mbox\scriptsize{{1.4GHz}}}\gtrsim10^{40}$ erg s$^{-1}$, DXS area, underdense environments\\
\hline
\end{tabular}
\end{center}
\label{tab:samples}
\end{table*}%

\section{Methodology and analysis}
\label{sec:method}

\subsection{Photometric Redshifts}
\label{sec:redshift}
For the study of the environment, information about the distances between sources and from the sources to the observer are needed. The SA22 field has been partly observed by the VIRMOS-VLT Deep Spectroscopic Survey (VVDS; \citealt{LeFevre2005}), down to an $i$-band AB magnitude of 22.5.  In total, there are 6751 spectroscopic redshifts within the SA22 field (we use a cross-matching radius of 1 arcsec to associate spectroscopic sources with NIR sources). In addition to these, we use the publicly available photometric redshift code LePhare (\citealt{Arnouts1999}, \citealt{Ilbert2006}) to calculate photometric redshifts for the rest of the SA22 sources using a total of 7 broadband photometric bands for the full SA22 field and additionally $K_s$ band for the part covered by DXS. 

\begin{figure}[htbp]
\begin{center}
\includegraphics[width=0.4\textwidth,angle=0]{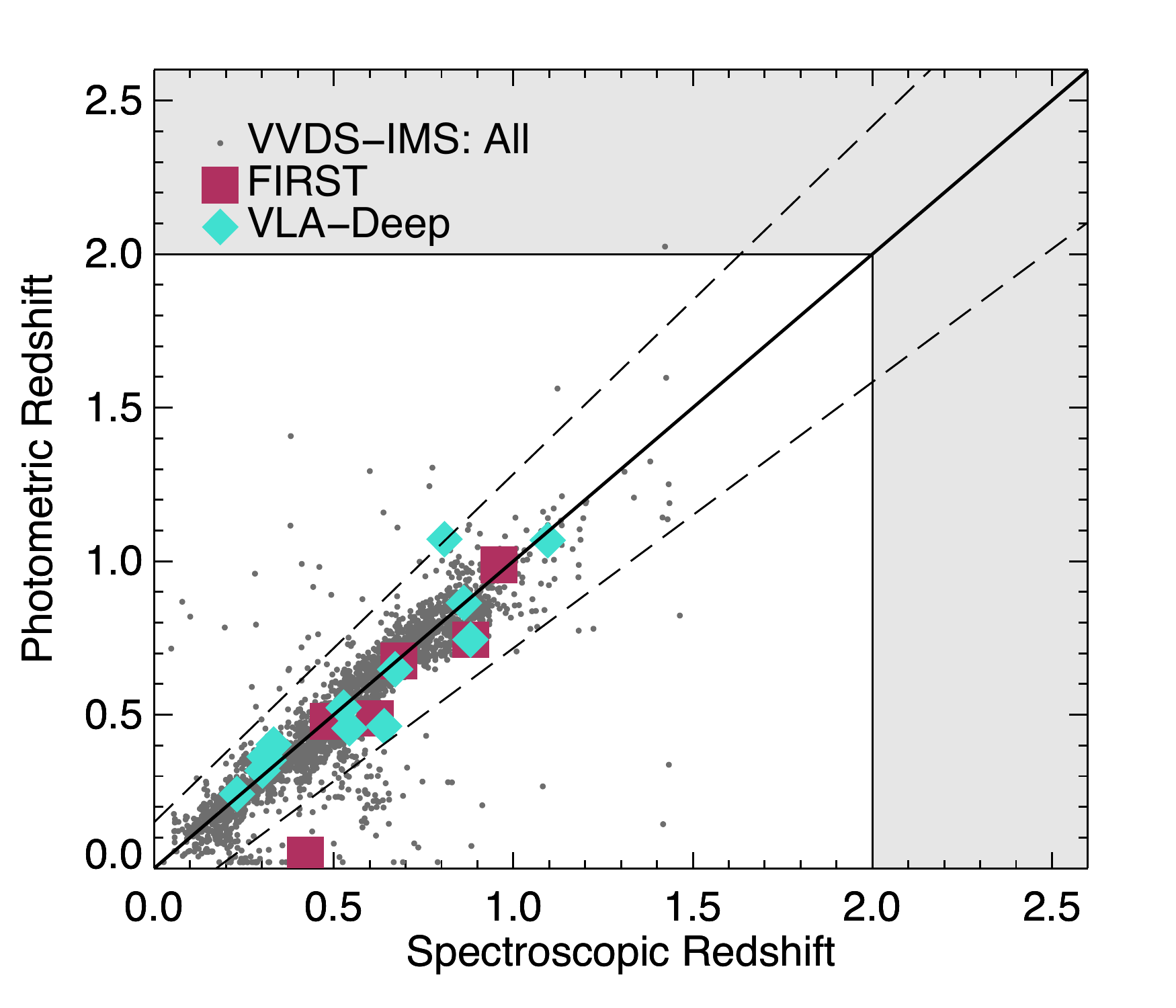}
\caption{Comparison between the spectroscopic redshifts from the VVDS survey and photometric redshifts within the SA22 field. Grey dots show all sources with good quality spectroscopic redshifts (Flag 3 and 4). Large symbols show radio sources from the FIRST and VLA-Deep samples with spectroscopic redshifts (maroon squares and turquoise diamonds, respectively). We show our sample selection at $z<2$ with the vertical and horizontal lines. The diagonal solid line shows the one-to-one relationship and the dashed lines represent a 15\% deviation from the solid line.}
\label{fig:specz}
\end{center}
\end{figure}

For the photometric redshift calculation we use the set of CFHT galaxy SED templates from \citet{Ilbert2006} and the \citet{Polletta2007} and \citet{Salvato2011} AGN templates (these also include composite AGN and star-forming systems). In addition we use stellar template libraries to identify and exclude stars from our sample. We use 3609 reliable spectroscopic redshifts from VVDS to calibrate our photometric redshift calculation, by deriving the possible photometric band offsets which have been shown to exist (e.g., \citealt{Ilbert2006}). The comparison between the spectroscopic and photometric redshifts is shown in Fig. \ref{fig:specz}. 

Using the iterative method of \citet{Ilbert2006} implemented in the LePhare code, we calculate the photometric band offsets. We find these to be mostly below 0.1 mag (only one band, $J$, exhibits an offset of 0.24), with a mean value of 0.06 mag. From the comparison between the spectroscopic and photometric redshifts, we get a normalised median absolute deviation (\citealt{Ilbert2006}) of $\sigma_{NMAD}=0.038$ and an outlier fraction of 4.8\%\footnote{Here we follow the standard definition of outliers as $\delta z/(1+z) >0.15$.}. 

In Fig. \ref{fig:specz} we also show the comparison for the radio sources in the FIRST and VLA-Deep samples with available spectroscopic redshifts. Although we note the small number from which a conclusion can be drawn, all but two of the radio sources show good agreement between their spectroscopic and photometric redshifts, when considering the best-fit photometric redshift from galaxy templates only. This translates nevertheless to an outlier fraction that is double that of the base sample ($\sim10\%$)\footnote{{The two outliers do not appear to stand out in some consistent way in their radio or optical properties from the rest of the radio sample. They both exhibit intermediate optical luminosities ($\sim10^{43}$ erg/s) and, while one appears to be very luminous at 1.4 GHz ($\sim10^{43}$ erg/s), the other one is right on the AGN radio-luminosity limit ($\sim10^{40}$ erg/s).}}. On the other hand, if we consider both galaxy and AGN templates for the determination of the best photometric redshift (in terms of the $\chi^{2}$ result of the fit), the number of outliers increases dramatically resulting in an outlier fraction of $\sim44\%$. Given the relatively few number of radio sources with spectroscopic redshifts, it is impossible to calibrate our AGN template photometric redshift determination any further. As such we decide to consider photometric redshifts derived only from galaxy template fits. This is a reasonable approximation for both star-forming galaxies, found at low radio luminosities, and the majority of radio-AGN, whose optical and NIR SEDs are usually not dominated by an AGN component.

\begin{figure}[htbp]
\begin{center}
\includegraphics[width=0.4\textwidth,angle=0]{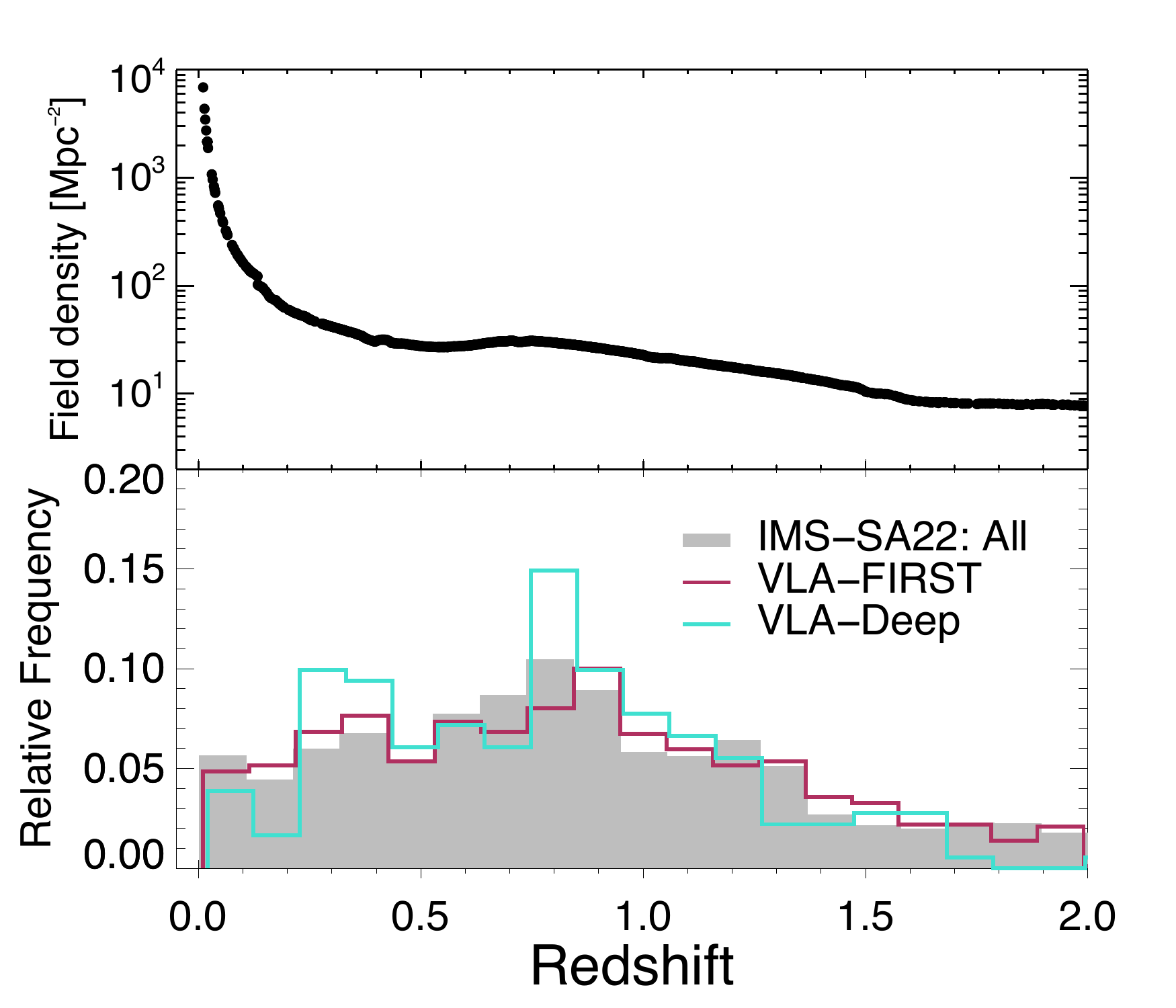}
\caption{Upper: Source number density per unit square area within finite redshift slices of depth $0.2\cdot$(1+z) as a function of redshift, for the total sample. Lower: Redshift distribution for the total SA22 sample (grey-shaded histogram), the VLA-FIRST sample (maroon open histogram), and the VLA-Deep sample (turquoise open histogram).}
\label{fig:histoz}
\end{center}
\end{figure}

In Fig. \ref{fig:histoz} we show the redshift distribution for the total SA22 field and those of the two radio samples. In addition, in Fig. \ref{fig:histoz} the source pseudo-3D density is shown as a function of redshift. The density is calculated within finite slices of redshift with a width of $0.2\cdot$(1+z) and is defined as the total number of sources within a given redshift slice divided by the total area covered by the SA22 survey at that redshift. We see that the field density falls smoothly as a function of redshift, due to the flux-limited natured of our sample. Concerning the redshift distribution of the two radio samples, they do not exhibit obvious differences. The VLA-FIRST sample appears to extend to slightly higher redshifts than the VLA-Deep one. 

\begin{figure}[htbp]
\begin{center}
\includegraphics[width=0.4\textwidth,angle=0]{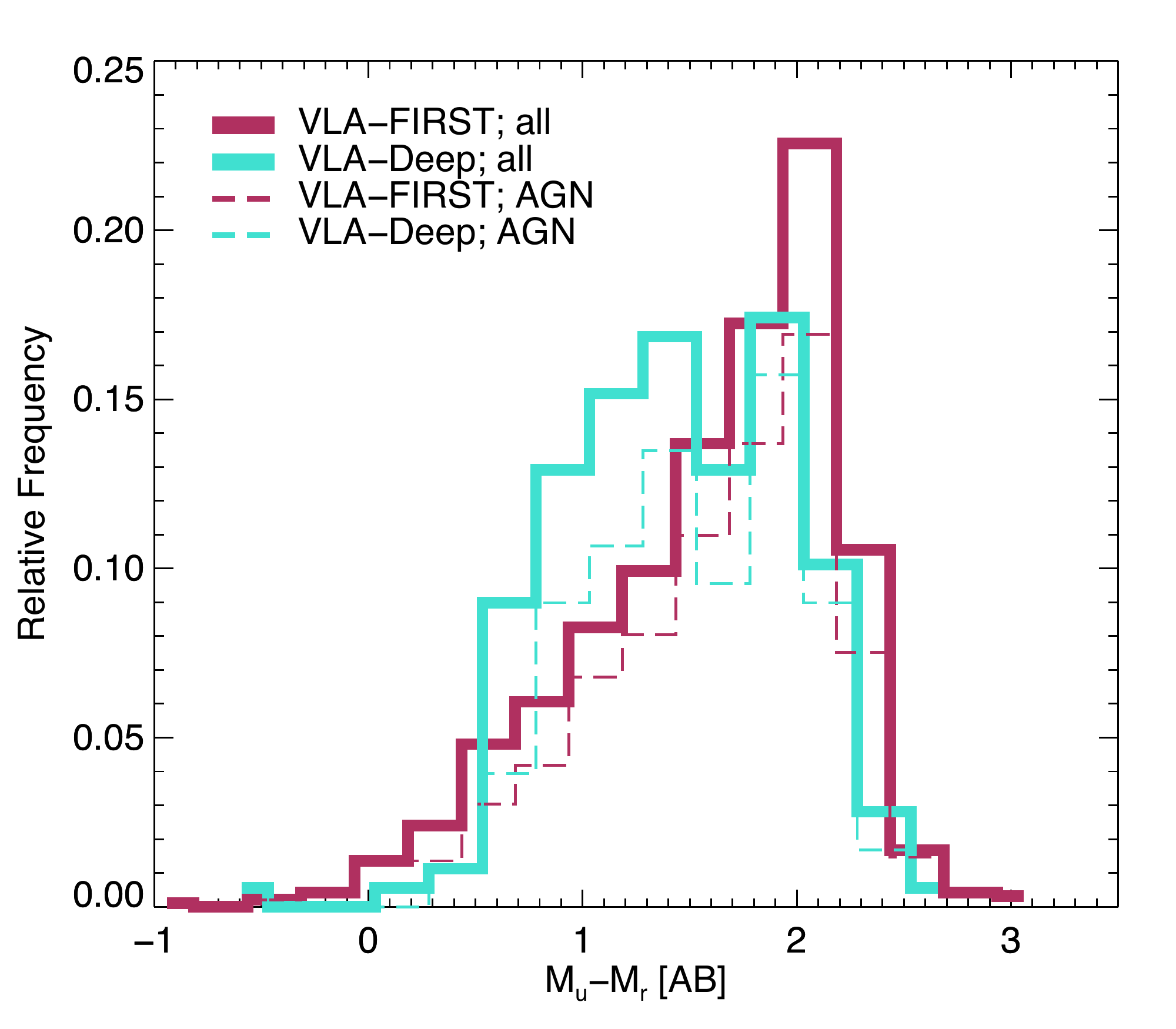}
\caption{Rest-frame absolute magnitude color $(M_{u}-M_{r})$ relative distributions for the FIRST (maroon histograms) and VLA-Deep (turquoise histograms) radio samples. We differentiate between the total samples (solid thick lines) and the AGN sub-samples defined through their 1.4 GHz radio luminosity (dashed thin lines).}
\label{fig:ur}
\end{center}
\end{figure}

Finally, LePhare also provides the estimated absolute magnitudes in the observed filters ($ugrizYJK_{s}$). In particular we use the method detailed in \citet{Ilbert2005}, which should be less template-dependent than other alternatives offered in the LePhare code. As will be explained in Section \ref{sec:control}, we shall use the $(M_{u}-M_{r})$ rest-frame absolute magnitude color to select a control sample for the radio sources. In Fig. \ref{fig:ur} we compare the distributions of $(M_{u}-M_{r})$ rest-frame absolute magnitude color between the FIRST and VLA-Deep samples. As can be seen in Fig. \ref{fig:ur}, the VLA-Deep sources show on average bluer colors. This is expected as deep, small-area surveys with a significant sub-mJy source component tend to be dominated by star-forming galaxies and star-forming/AGN composite systems (e.g., \citealt{Seymour2008a}, \citealt{Padovani2009}).

\subsection{Environment density parameters}
We employ the distance to the \textit{n}th closest neighbor to define the surface density parameter $\Sigma_{n}$, originally introduced by \citet{Dressler1980}. This is defined as:
$$\Sigma_{n}=\frac{n}{\pi d_{n}^{2}}.$$
To identify the closest neighbor and calculate its distance to a radio source, we look within a redshift slice of $\pm0.1(1+z)$. In this way we can calculate a pseudo-3D density. As we are interested in the close environments of radio-AGN, to look for signs of mergers, we concentrate our analysis on the 2nd and 5th closest neighbors\footnote{For our sample, mean projected distance to the second closest neighbor is D$_{2}=127$ kpc and for the fifth closest neighbor it is D$_{5}=217$ kpc. These linear scales are well below typical cluster scales and are therefore more relevant to close companions or group environments.}. Given the finite resolution of our observations and the large difference between the resolution in the NIR and the radio ($\sim1$ arcsec compared to $\sim5$ arcsec, respectively) we do not use the distance to the 1st closest neighbor. Following \citet{Cooper2005}, to minimize the contamination from edge effects, we exclude sources at a distance of 2 arcmins from the field edges. For the calculation of both density parameters and projected distance to the \textit{n}th closest neighbor we consider the total, band-merged, SA22 sample, excluding only sources with stellar-like colours (as identified from the photometric redshift fitting process).

We define the measure of overdensity for our study as the ratio between the density parameter of a radio source (i.e., $\Sigma_i^{radio}$, where i=\{2,5\}) to the mean density parameter of its random and matched control sample (i.e., $\bar\Sigma_i^{cont}$, where i=\{2,5\}). As such, for ${\Sigma_i^{radio}}>{\bar\Sigma_i^{cont}}$ it follows that the overdensity ratio takes values above 1. Similarly, for ${\Sigma_i^{radio}}<{\bar\Sigma_i^{cont}}$ the overdensity ratio has a value below 1. The definition of the overdensity measure in terms of the density ratio between a radio source and its control sources alleviates problems related to the flux-limited nature of our samples. Specifically it addresses the fact that at higher redshifts the selected sample tends to be increasingly dominated by more luminous and hence potentially more massive galaxies.

\subsection{Control samples}
\label{sec:control}
The robust selection of a control sample is fundamental in being able to draw robust conclusions concerning the environment of radio-AGN. The criteria for our matched control sample selection are driven by the needs of our study and the properties of our base sample. \textbf{Therefore we essentially impose a matching of $J$ band luminosity (the reddest band available for the entirety of the SA22 field), as it is, among the available photometry, the band least affected by obscuration.} As we are dealing with radio-AGN, we do not need to worry about a significant contamination in the optical and NIR parts of the SED from the AGN, as the optical and NIR properties of radio-AGN have been found to be largely independent of their radio properties (e.g., \citealt{Best2005}). In a similar manner, we match the $(M_{u}-M_{r})$ rest-frame absolute color, as this is a good measure of star formation in galaxies (essentially covering the 4000 $\AA$ break, e.g., \citealt{Kauffmann2003a}), as well as their morphologies (e.g., \citealt{Strateva2001}). Furthermore, it has been shown that the clustering of galaxies appears to be strongly dependent on their color (e.g., \citealt{Skibba2014}, \citealt{Zehavi2011}, \citealt{Coil2008}), with the general consensus being that redder, and hence lower star formation galaxies, are more strongly clustered than their bluer counterparts.


As such, we select two control samples, one that should reflect the average field density at a given redshift (random control sample) and one that aims to match each AGN with a group of control sources of similar redshift, magnitude, and rest-frame color (matched control sample). In more detail, for the random control sample, for each radio source we select 20-40 random positions within the SA22 field, not necessarily associated with a source. We then compute the environment properties around each of these positions assuming the redshift of the radio source. The average of these is assigned as the random control value of this radio source. Similarly, for the matched control sample, for each radio source we select 20-40 control sources with the SA22 field  that fulfil the following criteria:
\begin{itemize}
\item $|z_{radio}-z_{control}|\leqslant0.1(1+z_{radio})$
\item $|J_{radio;obs}-J_{control;obs}|\leqslant0.2$
\item $|(M_{u}-M_{r})_{radio;rest}-(M_{u}-M_{r})_{control;rest}|\leqslant0.2$
\end{itemize}
We then again calculate the average environment properties for these control sources and assign this value as the matched control value to the radio source. For both samples we pay special attention to edge effects (e.g., \citealt{Cooper2005}) and also avoid selecting random and matched control sources around bright stars and otherwise masked problematic areas of the images. Finally, we also ensure that each of the random positions and matched control sources are not within 60 arcsec of their respective radio source in order to minimise any contamination from local over- or under-densities around the radio source.

\subsection{Broadband spectral energy distributions}
\label{sec:sed}
An important component of this study is the link between the host galaxy properties of radio-AGN to their environments. To that end, we undertake the model-fitting of the broadband spectral energy distributions (SEDs) of the radio sources in our sample. For that we use the full wavelength coverage ranging the optical and the NIR. Given the importance of $K_{s}$ band to constrain the old stellar population component and hence the star formation history of a galaxy, we constrain the SED fitting to galaxies covered by the DXS and thus observed in the $K_{s}$ band. This reduces the available sources to a total of 711 radio sources (530 in the FIRST and all 181 sources from the VLA-Deep survey). {Nevertheless, at the high end of our adopted redshift range, even K$_s$ band fails to probe the NIR stellar bump. To counter this, for sources at redshifts $>1$, we additionally use photometry information from WISE (\citealt{Wright2010}) at observed 3.4 and 4.6 $\mu$m (bands W1 and W2, probing 1.1 and 1.5 $\mu$m rest-frame wavelengths at $z=2$, respectively). We have cross-matched our radio-sample, using the positions of the NIR sources, with the All Sky WISE catalog and requiring a 5$\sigma$ detection in the W1 band. Using a matching radius of 6 arcsec (equal to the PSF size of WISE at 3.4 $\mu$m) we find in total 92 radio sources in the SA22/DXS field with detection in the W1 WISE band.}

The SED-fitting has been done by comparing the observed SEDs of our sample to 
a set of spectral templates from \citet[][hereafter, BC03]{Bruzual2003} 
stellar population synthesis models.
In this work, we use the BC03 model spectra with Padova 1994 evolution track and 
the \citet{Chabrier2003} IMF.
After extracting model galaxy spectra, we apply the Calzetti dust attenuation 
law (\citealt{Calzetti2000}) for internal dust attenuation, and the \citet{Madau1995} law 
to account for the inter-galactic extinction arising by the neutral hydrogen 
in the inter-galactic medium (IGM). 
We allow four metallicity values -- 0.2 $Z_{\odot}$, 0.4 $Z_{\odot}$, 1.0 
$Z_{\odot}$, and 2.5 $Z_{\odot}$ during the fitting procedure.

As for the star-formation history, we assume the parametric form of delayed 
SFHs, which was introduced in \citet{LeeSK2010}, and has been used in previous 
works to analyze observed galaxy SEDs \citep[e.g.,][]{Lee2014,Wiklind2014}. 
The function form of this SFH is   
 
$$\Psi (t,\tau) \varpropto \frac{t}{\tau^{2}} e^{-t / \tau},$$

where $\Psi (t,\tau)$ is the instantaneous SFR.
In this work, we define SFR as the SFR averaged over recent 100 Myr, instead of 
instantaneous SFR, based on the reasoning explained in \citet{Lee2009}.

The parameter $t$ -- which is the time since the onset of the star formation 
-- is allowed to vary from 200 Myr to $t_H$, where $t_H$ is the age of the 
Universe at corresponding redshift of each galaxy. 
$\tau$ is varied from $0.1$ Gyr to 10.0 Gyr with uneven step sizes.

Examples of fits to radio-sources' SEDs are shown in Figs. \ref{fig:sed1} and \ref{fig:sed2}. We use the $\chi^{2}$ output from our SED fitting code to reject fits with $\chi^{2}$ values above 100 {(leading to 54 sources being rejected, $\sim7\%$\footnote{{We checked whether the rejected sources are preferentially found at high radio luminosities, high redshifts, or at more overdense or underdense environments, but this was not the case.}})}. Among sources with acceptable SED fits ($\chi^{2}<100$), $\sim85\%$ have $\chi^{2}<10$.

\section{Environment}
\label{sec:env}
{In this section we investigate the small scale environment of radio sources and more specifically radio-AGN, utilising both the FIRST and VLA-Deep surveys. We will deal with the following samples \textbf{(also see Table \ref{tab:samples})}: 
\begin{itemize}
\item FIRST; All - All cross-matched sources between the FIRST and the IMS surveys,
\item FIRST; AGN - All cross-matched sources between the FIRST and the IMS surveys, above a radio luminosity limit of $\sim10^{40}$ erg/s,
\item VLA-Deep; All - All cross-matched sources between the deep VLA survey and the IMS,
\item VLA-Deep; AGN - All cross-matched sources between the deep VLA survey and the IMS, above a radio luminosity limit of $\sim10^{40}$ erg/s.
\end{itemize}}

\begin{figure*}[htb]
\begin{center}
\includegraphics[width=0.45\textwidth,angle=0]{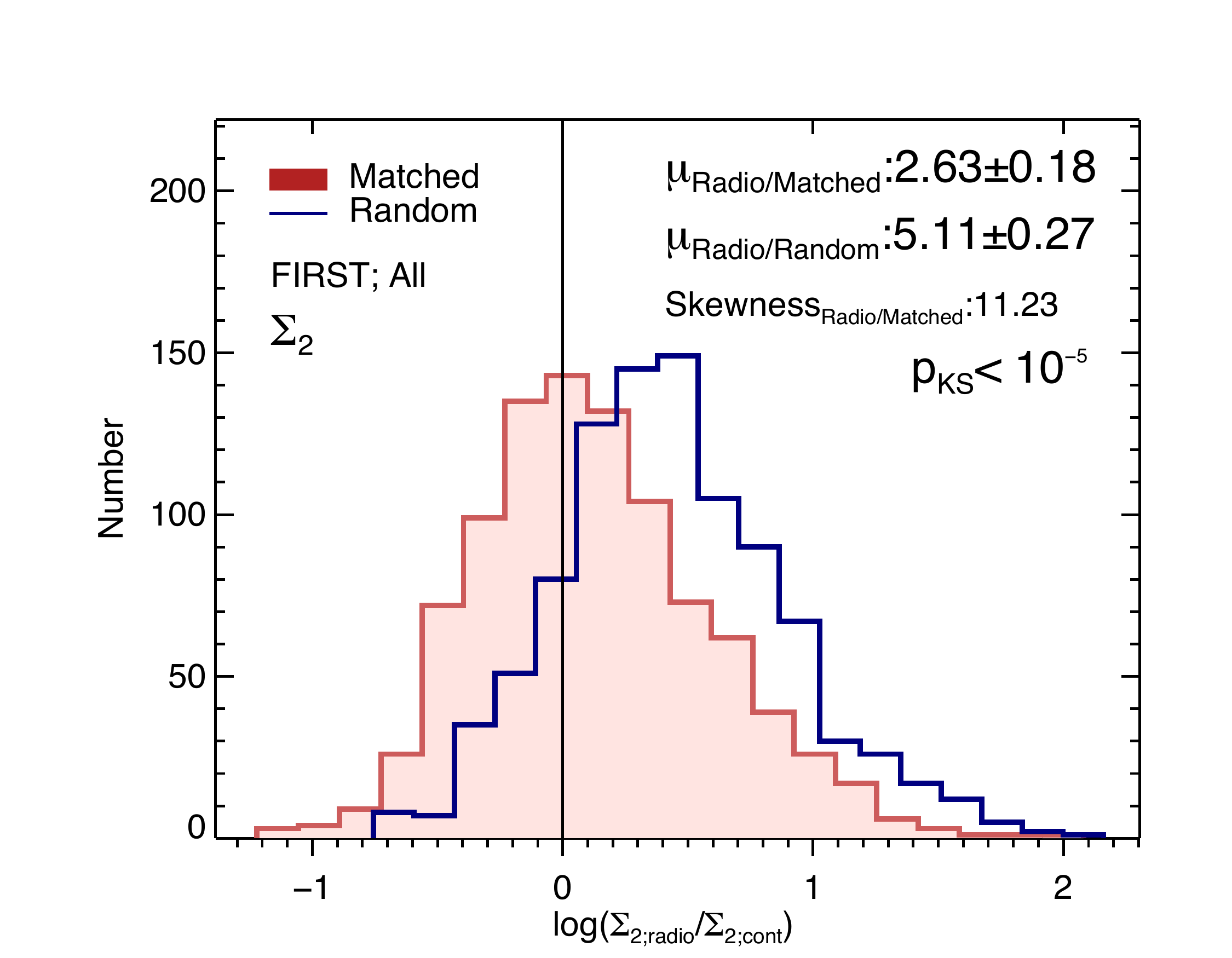}
\includegraphics[width=0.45\textwidth,angle=0]{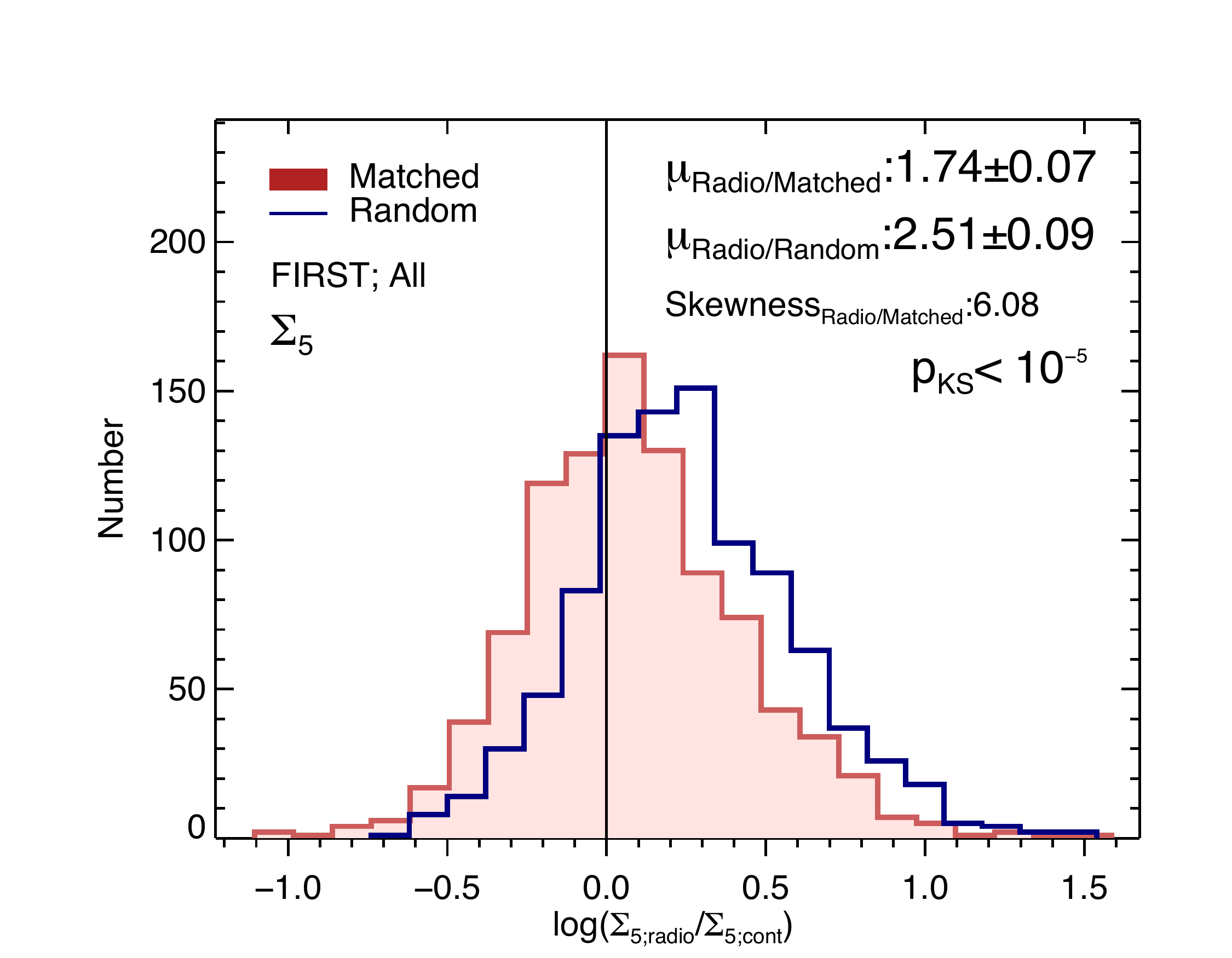}\\
\includegraphics[width=0.45\textwidth,angle=0]{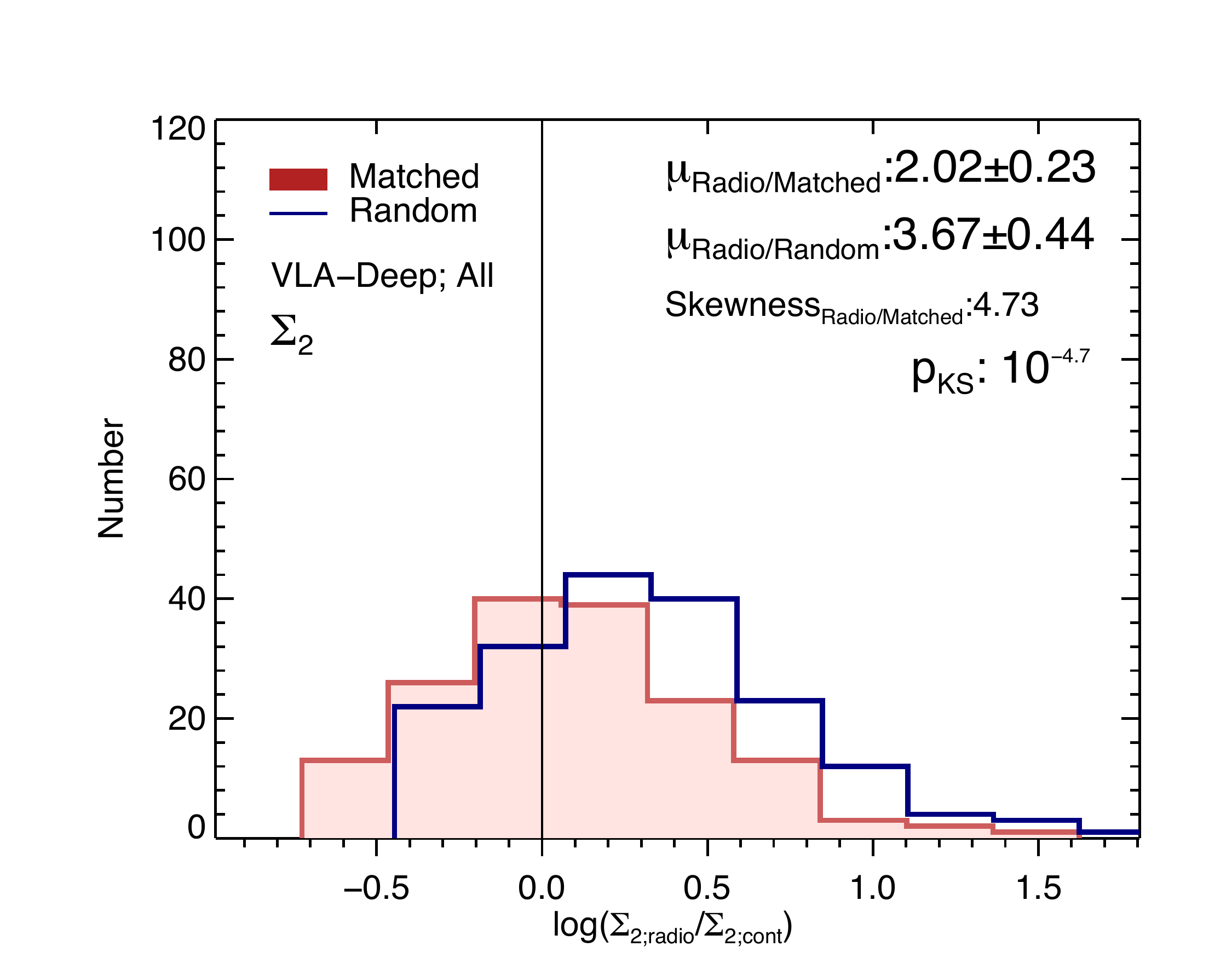}
\includegraphics[width=0.45\textwidth,angle=0]{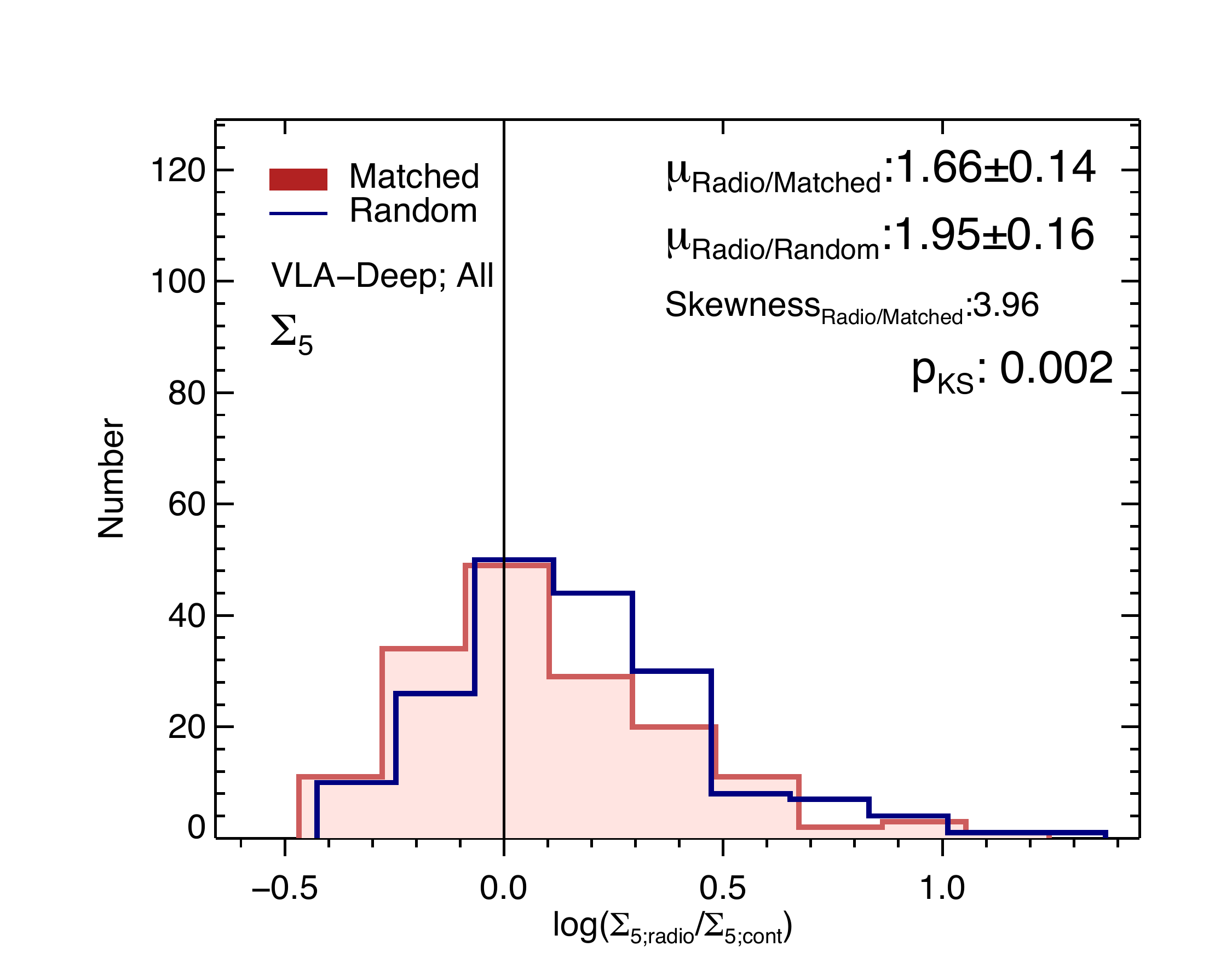}
\caption{Distributions of the overdensity parameters defined as the $\Sigma_{2}$ (left column) and $\Sigma_{5}$ (right column) ratio between each radio source and its control sample, for all sources in the FIRST (top row) and the VLA-Deep (bottom row) samples. The overdensity ratios for the random and the matched control samples are shown with open blue and shaded red histograms, respectively. The average values for each of the distributions, the skewness of the radio-to-matched overdensity ratio distribution, and the probability p that the two distributions are drawn from the same parent sample are also given for each panel. The vertical line denotes an overdensity radio of one, i.e., in which case the radio source resides in an identical environment with its control sample.}
\label{fig:sigma_all}
\end{center}
\end{figure*}

\begin{figure*}[htb]
\begin{center}
\includegraphics[width=0.45\textwidth,angle=0]{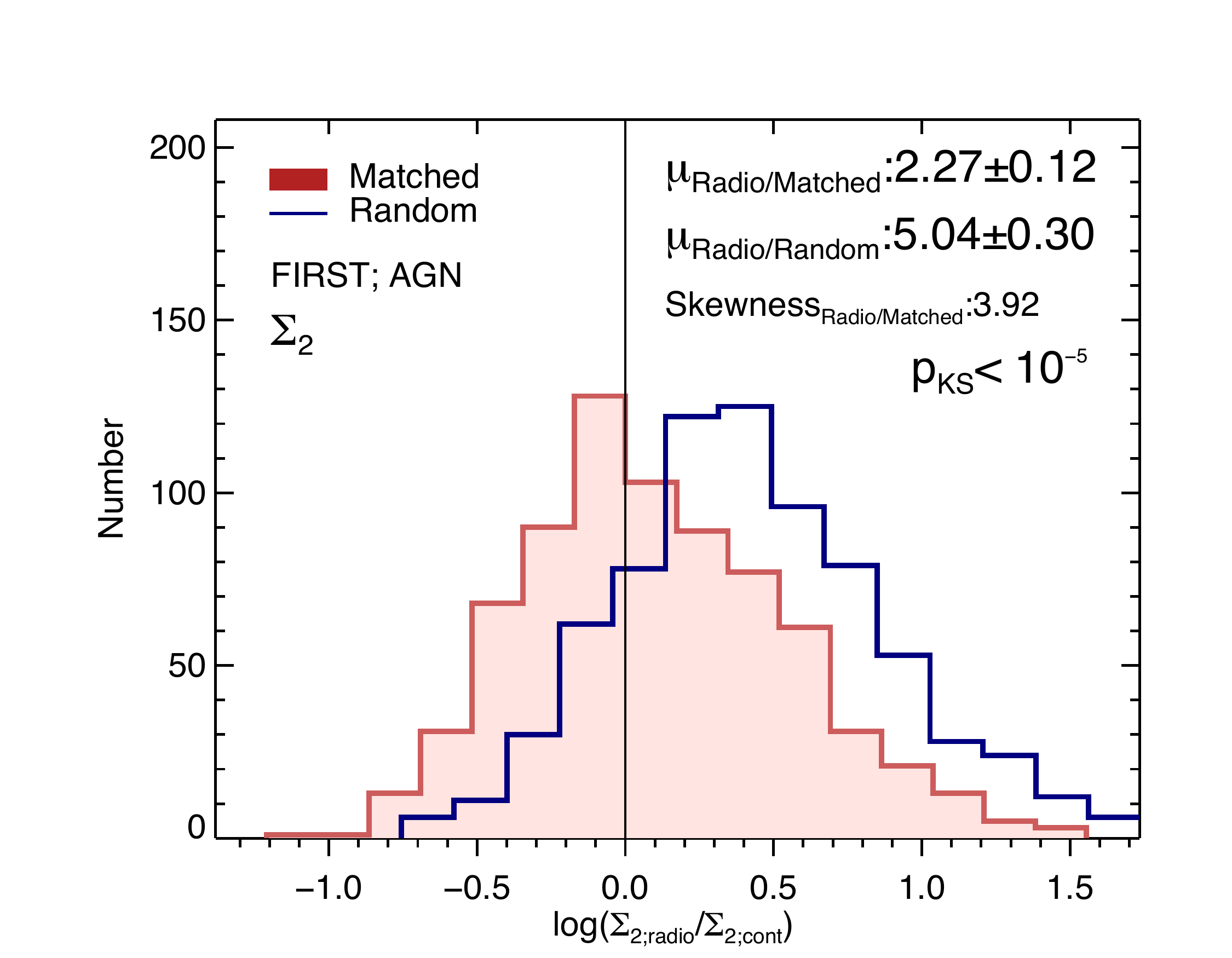}
\includegraphics[width=0.45\textwidth,angle=0]{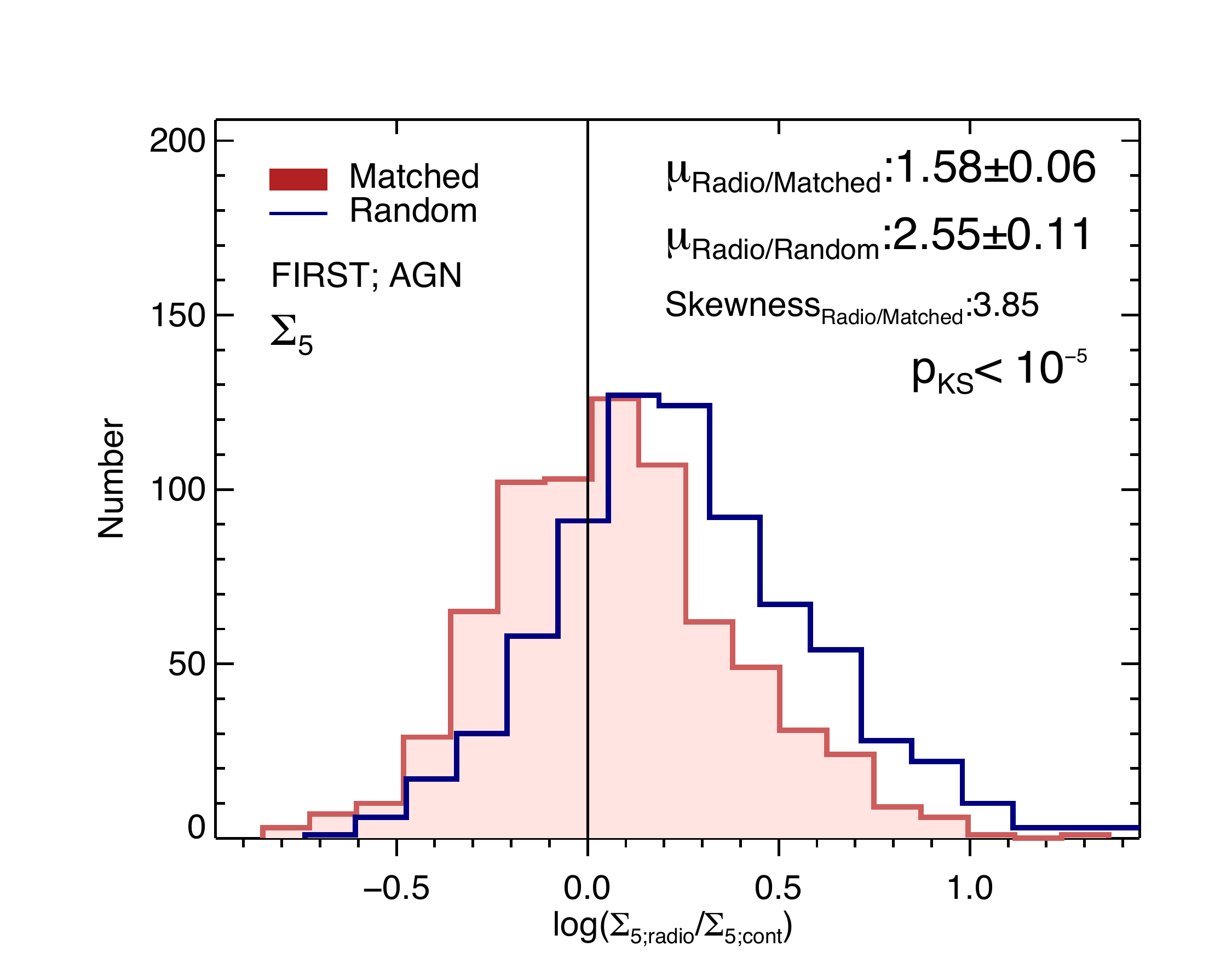}\\
\includegraphics[width=0.45\textwidth,angle=0]{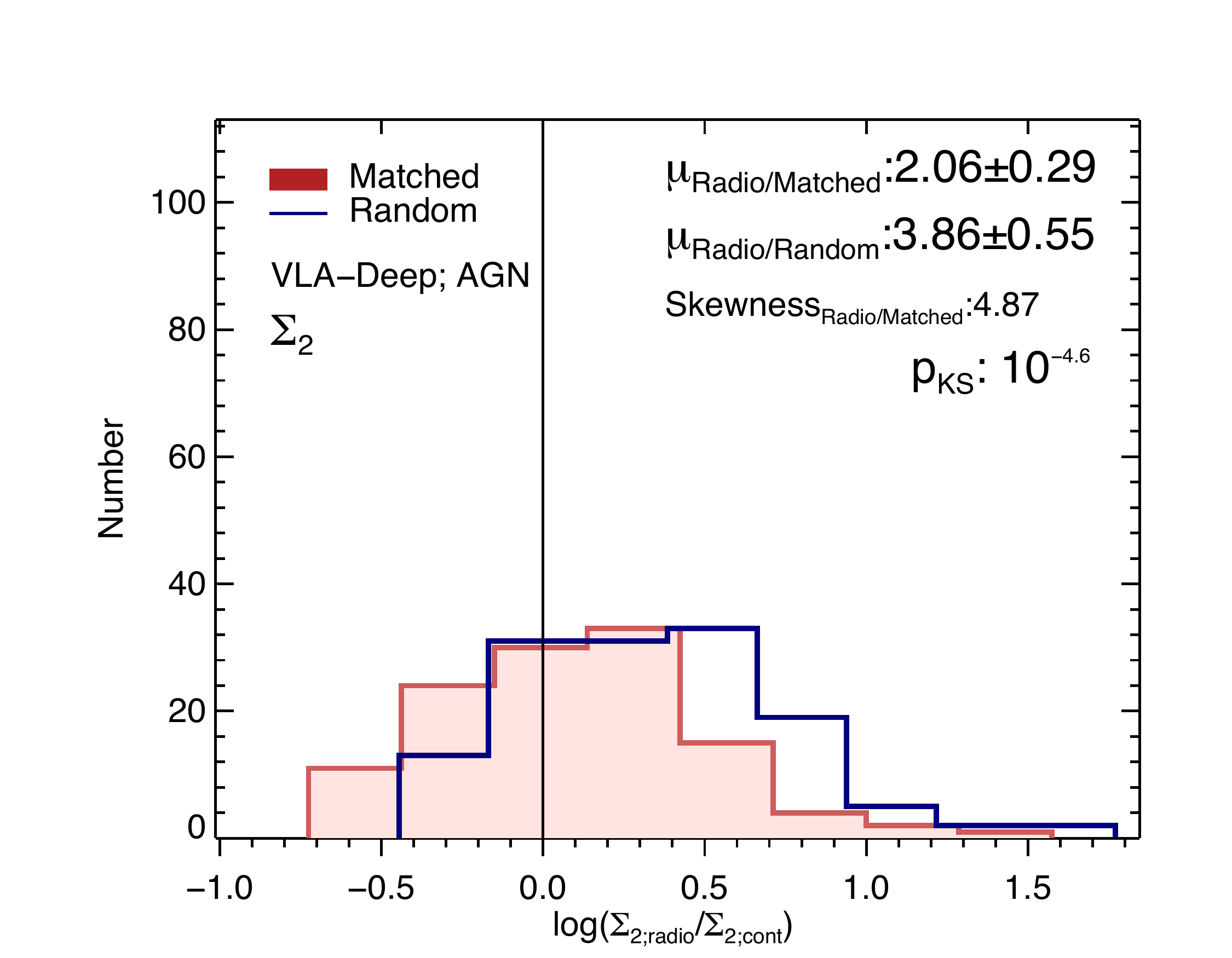}
\includegraphics[width=0.45\textwidth,angle=0]{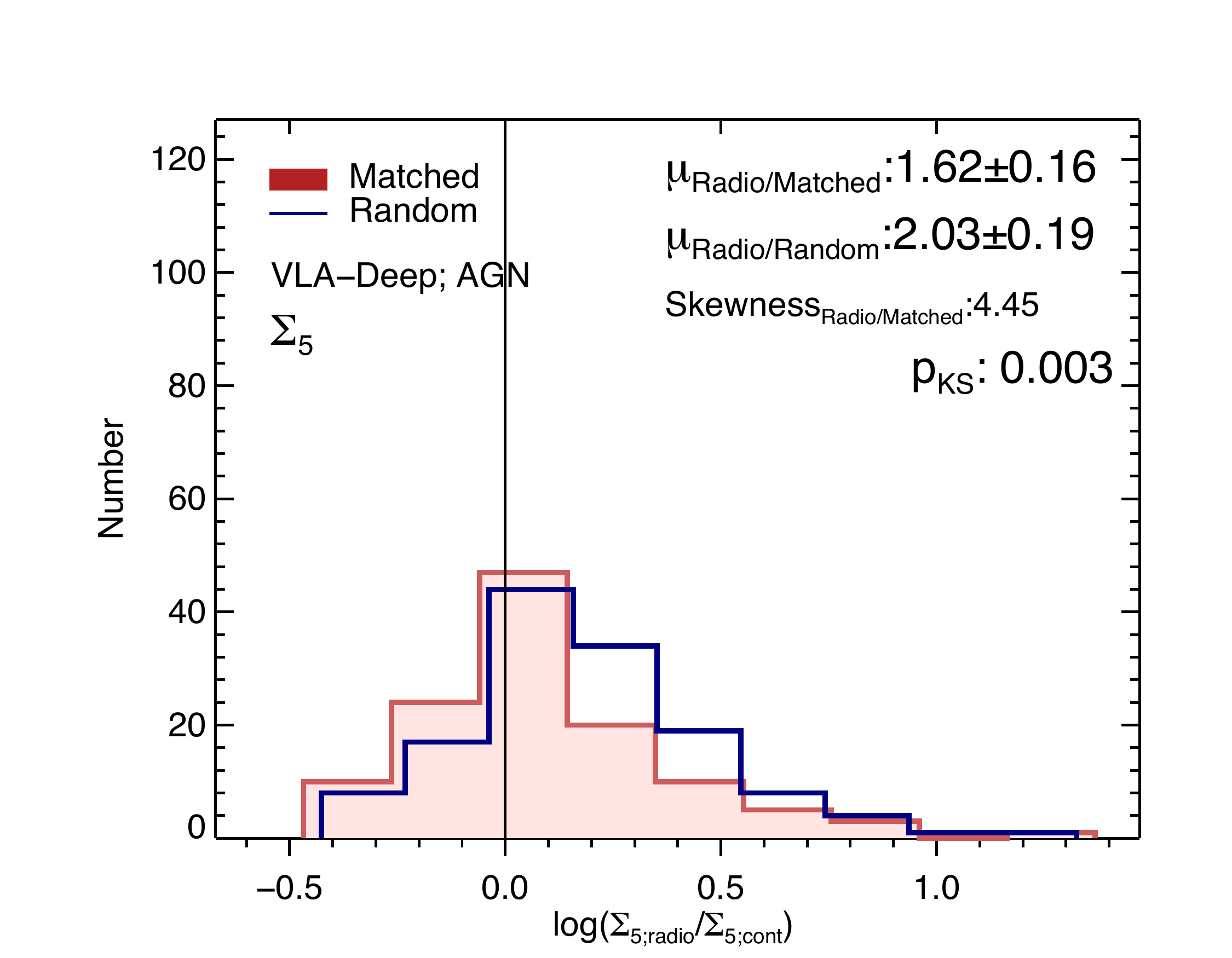}
\caption{Same as in Fig. \ref{fig:sigma_all} but now only radio-AGN, selected through their 1.4 GHz luminosities, are considered.}
\label{fig:sigma_AGN}
\end{center}
\end{figure*}

We start by looking at the $\Sigma_{2}$ and $\Sigma_{5}$ density parameter distributions for the two radio samples. These are shown in Figs. \ref{fig:sigma_all} and \ref{fig:sigma_AGN}. There we plot the distributions of the ratio $\frac{\Sigma_{n;radio}}{\Sigma_{n;control}}$, our chosen measure of overdensity. We consider both the comparison with the matched control sample (red filled histograms) and the random control sample (empty blue histogram). Focusing first on the total samples (both radio-AGN and star-forming galaxies) we see that the distribution of the overdensity ratio peaks at a value of 1, indicating that the bulk of the sources is found in environments very similar to their matched control sources\footnote{We remind here that the matched control sample matches each radio source with 20-40 sources with similar, redshifts, observed $J$-band magnitude, and rest-frame absolute magnitude color $(M_{u}-M_{r})$.}. 

The results are markedly different when we consider the comparison with the random field positions. Here we see that the peak of the distribution, for both samples in the case of $\Sigma_{2}$ and for the FIRST sample for $\Sigma_{5}$, is found at ratios higher than 1, implying that compared to the average field density, radio sources are found predominantly in denser environments. This is not particularly surprising especially since our random field positions are not associated with galaxies and therefore, within the framework of $\Lambda$CDM and a hierarchically clustering Universe (e.g., \citealt{Navarro1997}), not associated with a dark matter halo. It is however obvious that the distributions of both matched and random control sample overdensity ratios are strongly non-Gaussian. They both exhibit strong high overdensity tails that extend to environments for radio sources that are up to a hundred times more dense than those of their control sources. This is reflected in the average overdensity ratio values, also shown for each panel of Fig. \ref{fig:sigma_all}. {The calculated skewness of the distribution is well above 0 for all cases.} A two-sample Kolmogorov-Smirnov (KS) test between the random and matched control sample overdensity ratio distributions rejects the null hypothesis that the two samples are drawn from the same parent distribution with high significance.

If we now focus on the comparison between the two radio samples, the FIRST and VLA-Deep, we find that qualitatively their behaviors in terms of both the distribution shape and the average values are the same. However we note that the degree of overdensity (in terms of highest overdensity ratio and average overdensity ratio) for the FIRST sample appears stronger than for the VLA-Deep. {Moreover, the FIRST overdensity distribution appears more heavily skewed to the right. These differences might }be driven by the fact that we expect the FIRST sample to include more very bright radio-AGN hosted by massive ellipticals (as was also implied by the redder colors for the VLA-FIRST source observed in Fig. \ref{fig:ur}). {This is also corroborated by the radio luminosity distributions of the two radio samples, a larger fraction of the VLA-Deep sources having radio luminosities $<10^{40}$ erg/s.}

Finally, it is interesting to compare how the overdensity ratio distribution changes when considering $\Sigma_{2}$ and $\Sigma_{5}$. One obvious change is that the degree of overdensity appears to dampen as we move to the 5th closest neighbor from the 2nd one (e.g., for the FIRST sample the average values for the matched control sample are $2.63\pm0.18$ and $1.74\pm0.07$ for $\Sigma_{2}$ and $\Sigma_{5}$, respectively). We also see that there is a tendency for the peak of the random control sample overdensity ratio distribution to converge towards a ratio value of 1, with the two distribution peaks coinciding for the case of the $\Sigma_{5}$ for the VLA-Deep sample. This is again qualitatively in agreement with an exponentially declining Navarro-Frenk-White (NFW) dark matter profile (\citealt{Navarro1996}), assuming the galaxy distribution to follow the dark matter distribution, together with the predominantly lower-mass galaxies that should dominate the VLA-Deep sample. Finally we observe that the skewness of the distributions becomes smaller from $\Sigma_{2}$ to $\Sigma_{5}$, indicating fewer sources in extreme overdensities at large scales.

Let us now turn to the overdensity ratio distributions for the radio-AGN sub-sample of the FIRST and VLA-Deep samples. These are shown in Fig. \ref{fig:sigma_AGN} with the same color notation for matched and random control samples as in Fig. \ref{fig:sigma_all}. Most of our observations from Fig. \ref{fig:sigma_all} also hold true when we only consider the radio-AGN in our samples\footnote{We remind that the radio-AGN selection is done by means of a radio-luminosity cut, following the local luminosity function of radio sources.}. There are however some additional points of interest. A first comment concerns the fact that the overdensity ratio for the random control sample, for the VLA-Deep sample, seems to be slightly shifted to higher values (in terms of average values) although still within the uncertainty. This may indicate that through our selection we are preferentially picking up more massive galaxies and therefore the difference in environment between these massive galaxies and empty fields becomes wider. However, when we consider the matched control sample the distributions do not appear to change significantly, with a tendency to actually show slightly lower average values of overdensity ratios and distribution peaks that are shifted towards smaller overdensity ratios (e.g., for the FIRST $\Sigma_{2}$ distribution). {This shift may be explained by the fact that we are excluding strong starburst galaxies, which should have radio luminosities right below the assumed AGN luminosity cut ($\sim10^{40}$ erg/s). These sources are known to be powered by ongoing mergers and are thus preferentially found in dense small-scale environments.} As expected, the two-sample KS test still rejects the null hypothesis.

\begin{figure}[htbp]
\begin{center}
\includegraphics[width=0.48\textwidth,angle=0]{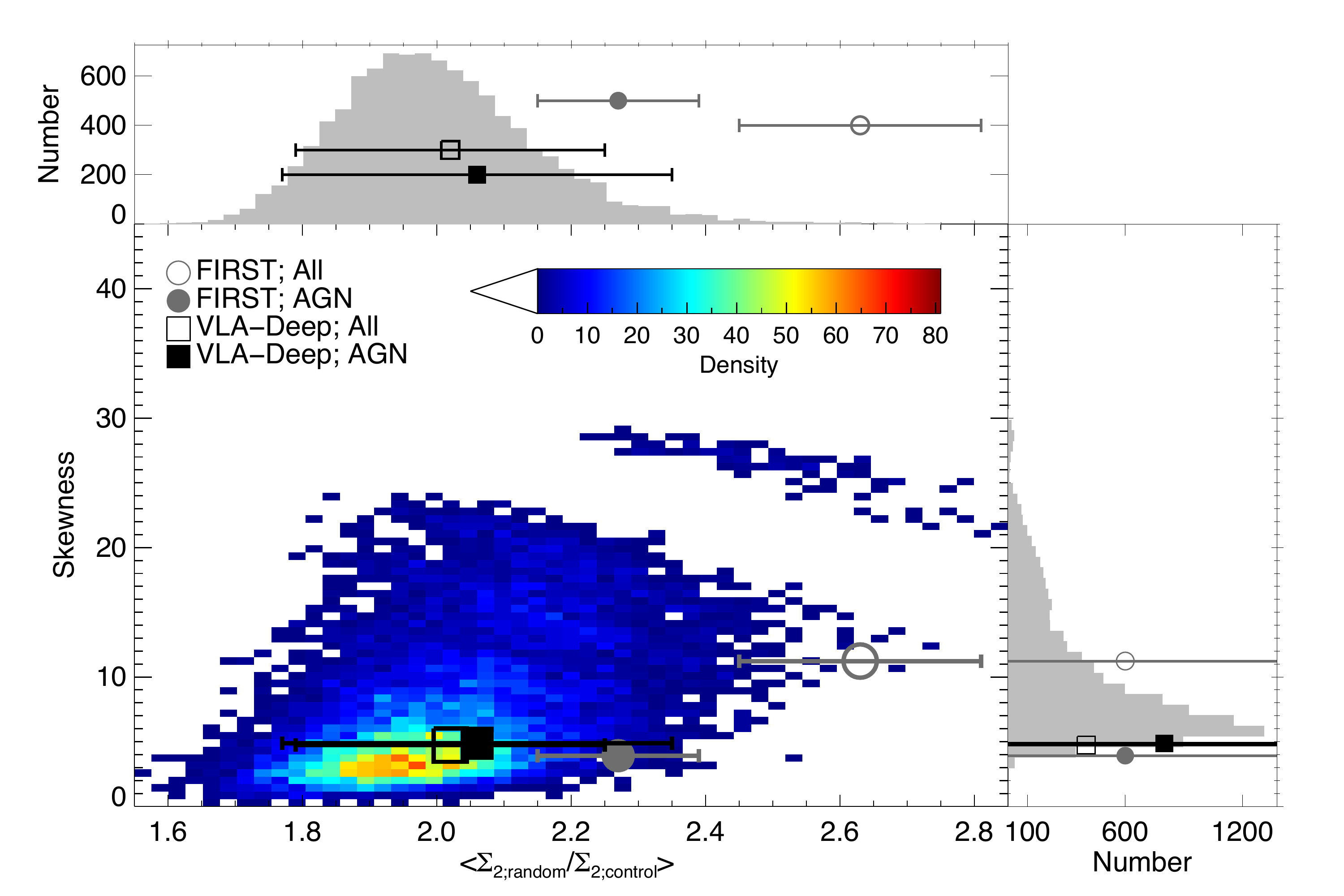}
\caption{Density map of mean overdensity versus skewness values resulting from a Monte Carlo simulations with 10000 runs (central plot). Individual distributions of the average overdensity values and skewness values of the ten thousand samples randomly drawn from the main SA22 sample are also shown (side panels). The corresponding values are shown for the FIRST (grey circles) and VLA-Deep (black squares) samples. These samples are further divided into ``All'' and ``AGN'' sub-samples, shown with open and filled symbols, respectively. Projections of these points on the ``Mean Overdensity'' and ``Skewness'' axis are plotted on the side panels.}
\label{fig:sigma_MC_random}
\end{center}
\end{figure}

We investigate the significance of the high overdensity, non-Gaussian, tail that is observed for all radio-source samples in Figs. \ref{fig:sigma_all} and \ref{fig:sigma_AGN}. In particular, we want to estimate the probability of this high overdensitie tail to be a chance occurrence. To that end, we draw ten thousand random samples from the total IMS-SA22 sample, each comprising of one thousand sources. For each of these random samples we follow the same analysis as described in Section \ref{sec:method}, i.e., for each source we assign 20-40 control sources of similar redshift, J magnitude, and $(M_{u}-M_{r})$ rest-frame absolute magnitude color and then calculate the 2nd and 5th closest neighbor density parameters for the main and matched control samples. We can then calculate the mean value for the overdensity distribution of each of the thousand samples, as well as its skewness that is representative of how heavy tailed the distributions are. The distribution of the mean values and skewness values can tell us the probability, and hence significance, of the average overdensities we derived for the FIRST and VLA-Deep samples. The density map of mean overdensity versus skewness value for our Monte Carlo run is shown, together with the individual distributions of the mean overdensity values and skewness values for the 10000 random samples is shown in Fig. \ref{fig:sigma_MC_random}.

As can be seen in Fig. \ref{fig:sigma_MC_random}, the distribution peaks at a mean overdensity value of $\sim1.9$ and a skewness value of $\sim6$. This implies that, independent of the sample selection, we can expect an average overdensity $>1$ with a heavy-tailed distribution. Furthermore, we observe that there is a broad correlation between the mean overdensity and how heavy-tailed the resulting distribution is. Let us now compare these numbers to the mean overdensities calculated for the FIRST and VLA-Deep samples. For the FIRST sources (both full sample and AGN sub-sample) their mean value is at overdensities well above the peak of the mean overdensity distribution derived. If we take skewness also in account, we see that both the full and AGN FIRST samples are found on the edge of the combined distribution. We calculate the combined probability that a random sample has a mean overdensity value and skewness value equal or larger than that of the FIRST sample,
$$P_{FIRST;all}[\frac{\Sigma_{2;radio}}{\Sigma_{2;matched}}\cap Skewness]=0.1\%$$
and
$$P_{FIRST;AGN}[\frac{\Sigma_{2;radio}}{\Sigma_{2;matched}}\cap Skewness]=4.4\%$$
for the full sample and the AGN sub-sample respectively. On the other hand the VLA-Deep sample shows a wide range (due to the small number of sources and hence the large standard error of its mean overdensity value), covering the peak of the mean overdensity distribution of Fig. \ref{fig:sigma_MC_random}. The calculated skewness values for the VLA-Deep samples are close to the peak of the skewness distribution. The combined probability to get such mean values randomly is calculated to be 40.5\% and 30.3\%, for the full sample and AGN sub-sample, respectively. We can conclude that while both radio samples appear to inhabit overdense environments compared to their control sources, only the FIRST sample appears to do so in a statistically significant manner.  

\subsection{Environment and AGN luminosity}
\begin{figure}[htb]
\begin{center}
\includegraphics[width=0.48\textwidth,angle=0]{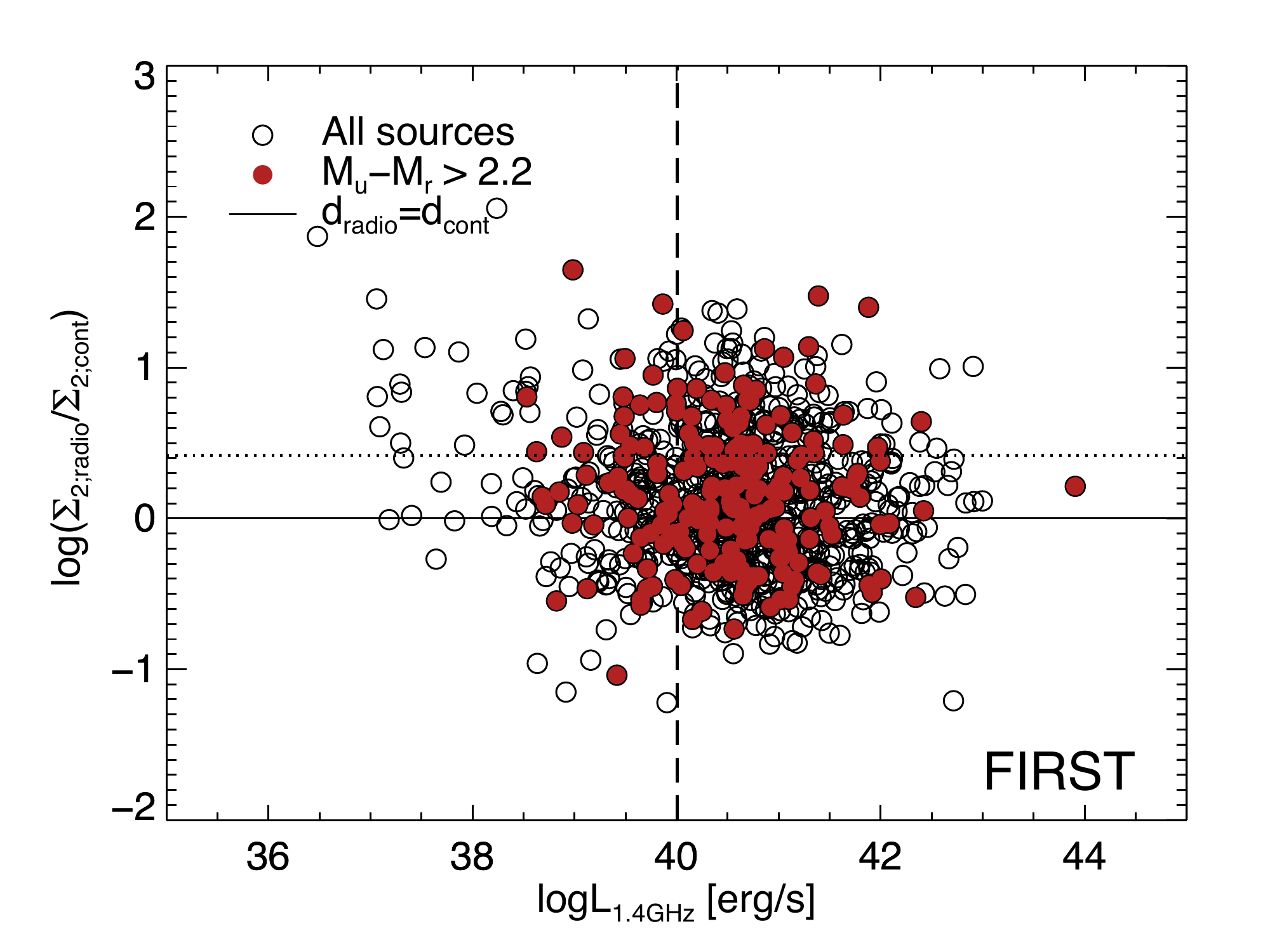}\\
\includegraphics[width=0.48\textwidth,angle=0]{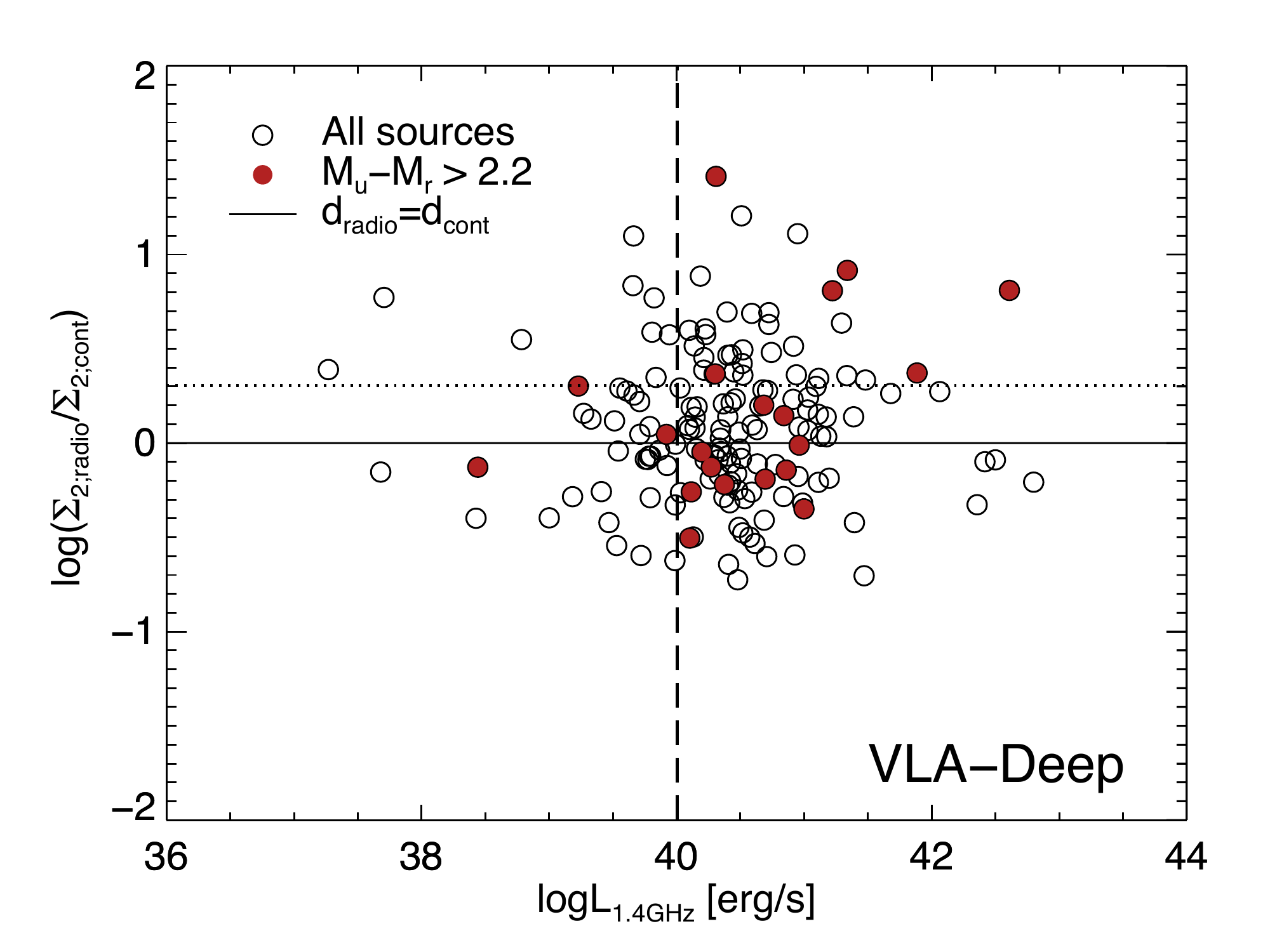}
\caption{Overdensity ratio parameter, between radio sources and their matched control sources, as a function of radio luminosity at 1.4 GHz for the FIRST (top) and VLA-Deep (bottom) radio sources. The horizontal lines denote a ratio value of one (solid line) and the mean overdensity value for the whole sample (dotted line). We also separate sources according to their rest-frame absolute magnitude color $(M_{u}-M_{r})$, showing with filled red circles sources with red colors ($>$2.2).}
\label{fig:sigma_lum}
\end{center}
\end{figure}

As was discussed in Section \ref{sec:intro}, galaxy evolution models have been put forth that support the triggering of luminous quasars by gas-rich major mergers (e.g., \citealt{Hopkins2006}). In addition, evidence have been provided for the most luminous radio-AGN to be also highly associated with ongoing or recent mergers (e.g., \citealt{Karouzos2010}, \citealt{Almeida2011}). Here we test a possible link between the radio luminosity of our radio samples with the overdensity ratios, whose distributions we studied above. This comparison is shown in Fig. \ref{fig:sigma_lum} for both the FIRST (top) and the VLA-Deep (bottom) samples. In addition, we in these plots we separate between all sources and those that are particularly red (rest-frame absolute magnitude color $(M_{u}-M_{r})>2.2$). {\citet{Strateva2001} showed that local SDSS galaxies with colors $u-r>2.2$ are mainly early-type, elliptical galaxies with relatively old stellar populations. These should then be the typical ``red and dead'' galaxies usually associated with bright radio-AGN in the local Universe.}

We see that both samples do not exhibit any appreciable trend with radio luminosity. We see again that on average they are higher than the ratio-equals-one value (average value shown with the dotted line). However, it seems that on first order the color of these sources does not seem to affect their place on this plot, i.e., red radio-sources are not preferentially found in more dense environments. We find no evidence for the most high radio-luminosity sources ($>10^{42}$ erg/s) to be in overdense environments either.

Conversely, we note that the low radio-luminosity sources ($<10^{38}$ erg/s) appear to be preferentially found in overdense environments. As star formation should dominate that radio luminosity regime, this confirms our previous statement that powerful star-forming galaxies are driving the shift of the overdensity ratio distribution peak from Fig. \ref{fig:sigma_all} to Fig. \ref{fig:sigma_AGN}.

\section {Host galaxies of radio-AGN}
\label{sec:SFR}
{In this section we focus on the properties of the radio-AGN host galaxies, derived through SED fitting. For the following we combine the FIRST; AGN and VLA-Deep; AGN samples (i.e., sources with radio luminosities $>10^{40}$ erg/s). Moreover, as described in Section \ref{sec:sed}, we limit ourselves to the DXS area within the SA22 field, in order to benefit from the uniform $K_s$-band coverage \textbf{(see Table \ref{tab:samples})}.}

\begin{figure}[htbp]
\begin{center}
\includegraphics[width=0.4\textwidth,angle=0]{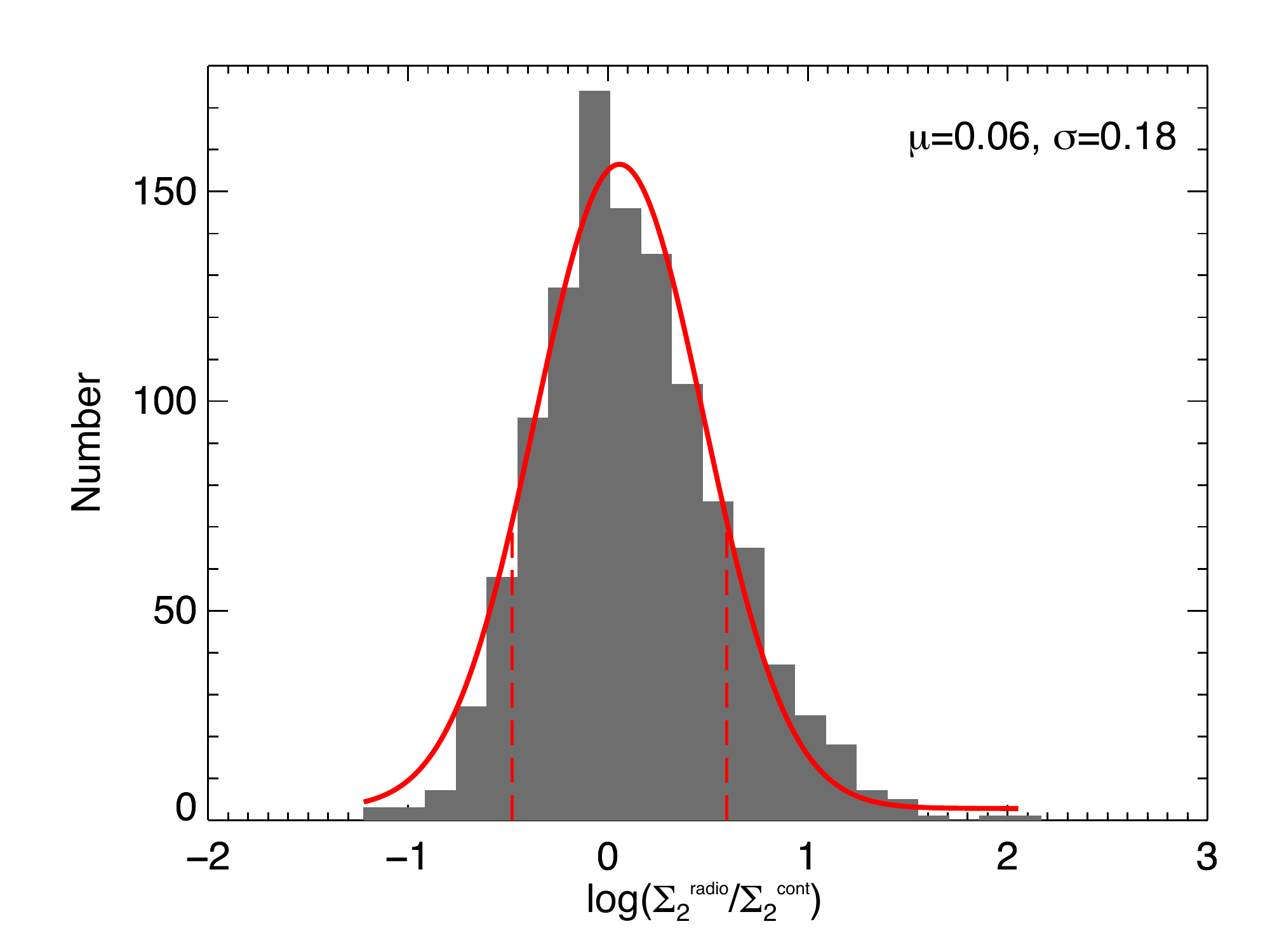}
\caption{Distribution of the overdensity ratio based on the $\Sigma_{2}$ density parameter for the full combined FIRST and VLA-Deep radio samples. The distribution is fitted with a standard Gaussian profile (shown with the red curve). The mean value and standard error, in logarithmic scale, of the Gaussian are also given. The most underdense and most overdense samples are defined according to the 3$\times$SE margins shown with the dashed lines on the plot.}
\label{fig:sigma_comb}
\end{center}
\end{figure}

We have shown that while on average most radio-AGN appear to inhabit environments similar to galaxies of similar properties, there does exist a significant sub-population of radio-AGN that inhabit environments up to a hundred times more dense than their control sources. We can define two sub-samples of radio-AGN that inhabit the most overdense and least overdense (i.e., most underdense) environments compared to their control sources. This separation is done based on the $\Sigma_{2}$ density parameter ratio as we are interested in the close neighbors of radio-AGN. In Fig. \ref{fig:sigma_comb} we show the $\Sigma_{2}$ ratio for the combined FIRST and VLA-Deep samples, which we have fitted with a standard Gaussian profile. Using the derived SE of the Gaussian distribution, we define the most overdense and most underdense sub-samples as those sources with overdensity ratios above and below the 3$\times$SE margin with respect to the mean value of the fitted Gaussian. 

\begin{figure}[htbp]
\begin{center}
\includegraphics[width=0.4\textwidth,angle=0]{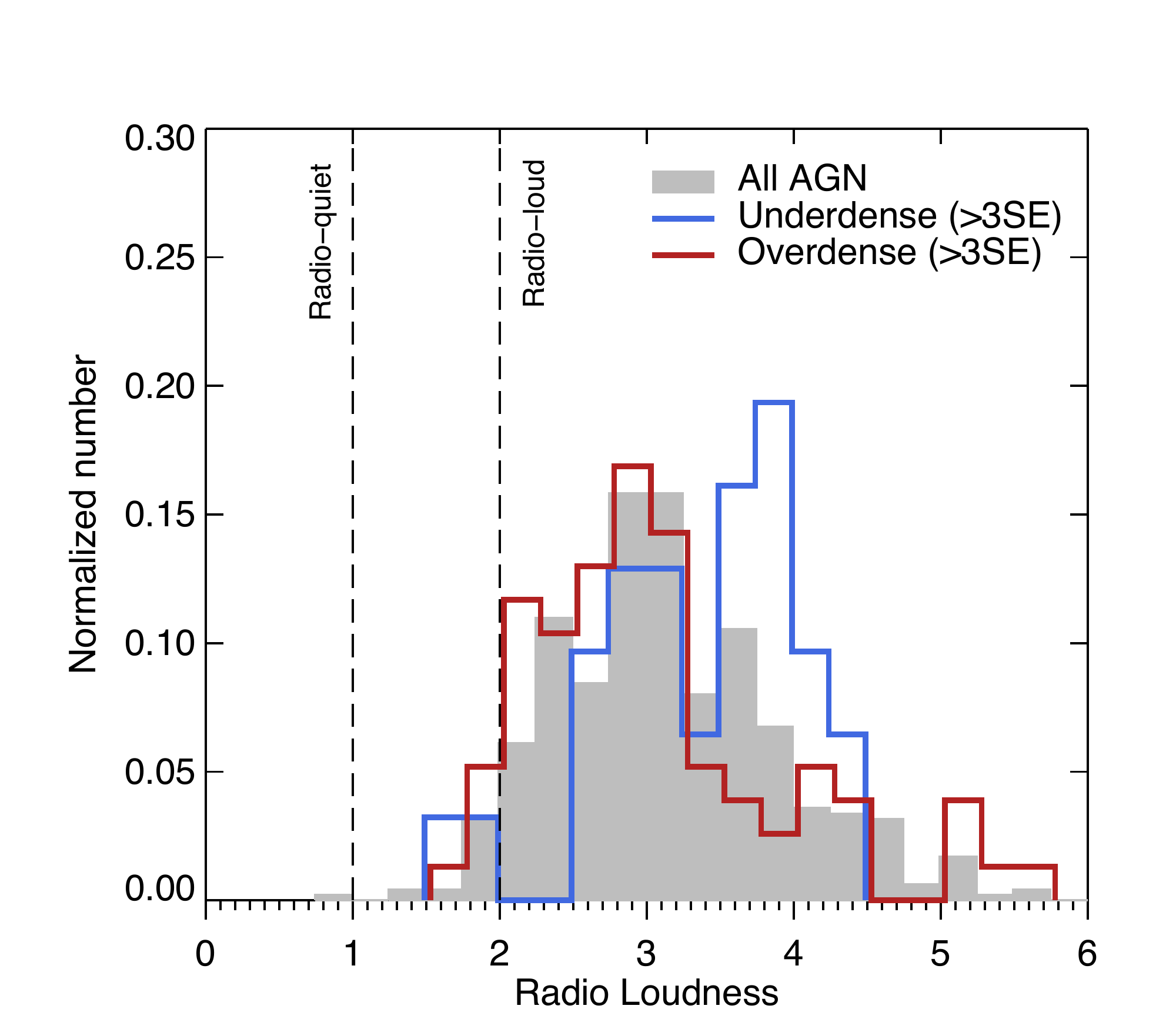}
\caption{Normalised distribution of the radio-loudness of radio-AGN found in the most overdense (red open histogram) and most underdense (blue open histogram) environments, compared to their control sources. We also plot the distribution for the total luminosity-selected radio-AGN sample (grey filled histogram). The limits for radio-loud and radio-quiet sources are shown with dashed lines.}
\label{fig:RL}
\end{center}
\end{figure}

By defining two samples in terms of the two extremes of the overdensity ratio distribution in Fig. \ref{fig:sigma_comb} and also imposing a radio luminosity limit to select only those sources qualifying as radio-AGN, we can compare the properties of radio-AGN in significantly different environments. For brevity, in the following we shall call these two samples as OD and UD, for radio-AGN in the most overdense and most underdense environments, respectively. We first look at the radio-loudness distributions of the two sub-samples. Radio loudness is a measure of the power of the radio-jet in an AGN and its dominance over the overall energy output of the nucleus (\citealt{Kellermann1989}). Prominent radio-jets classify an AGN as radio-loud, while radio-quiet sources are thought to lack or have a very weak jet component. Here we use the definition of radio-loudness from \citet{Ivezic2002},
$$R_{i}=log\left(\frac{f(1.4GHz)}{f(7480\AA)}\right),$$ 
where $\lambda=7480\AA$ is the central wavelength of the $i$-band in the optical. For both radio and optical fluxes, we use the observed values. Under that definition, we can classify sources with $R_{i}>2$ as radio-loud, while for $R_{i}<1$ a source is classified as radio-quiet. The $R_{i}$ distributions of the OD and UD samples are shown in Fig. \ref{fig:RL}.

As expected, most of our luminosity-selected AGN fall within the radio-loud AGN, with only a few found in the intermediate regime between radio-quiet and radio-loud limits. Interestingly, we observe that the peak of the $R_{i}$ distribution of the OD sample is at a value of $\sim3$. The UD sample on the other hand shows on average higher values of $R_{i}$, with its distribution peaking at $\sim4$. A two-sample KS test rejects the null hypothesis at a probability value of p=0.02. Conversely, a similar comparison between the radio-luminosity distributions does not reveal any significant differences, with a two-sample KS test failing to reject the null hypothesis. In other words, radio-AGN in the UD sub-sample, although more radio-loud, are not more luminous in the radio compared to their OD counterparts.

\begin{figure}[htbp]
\begin{center}
\includegraphics[width=0.4\textwidth,angle=0]{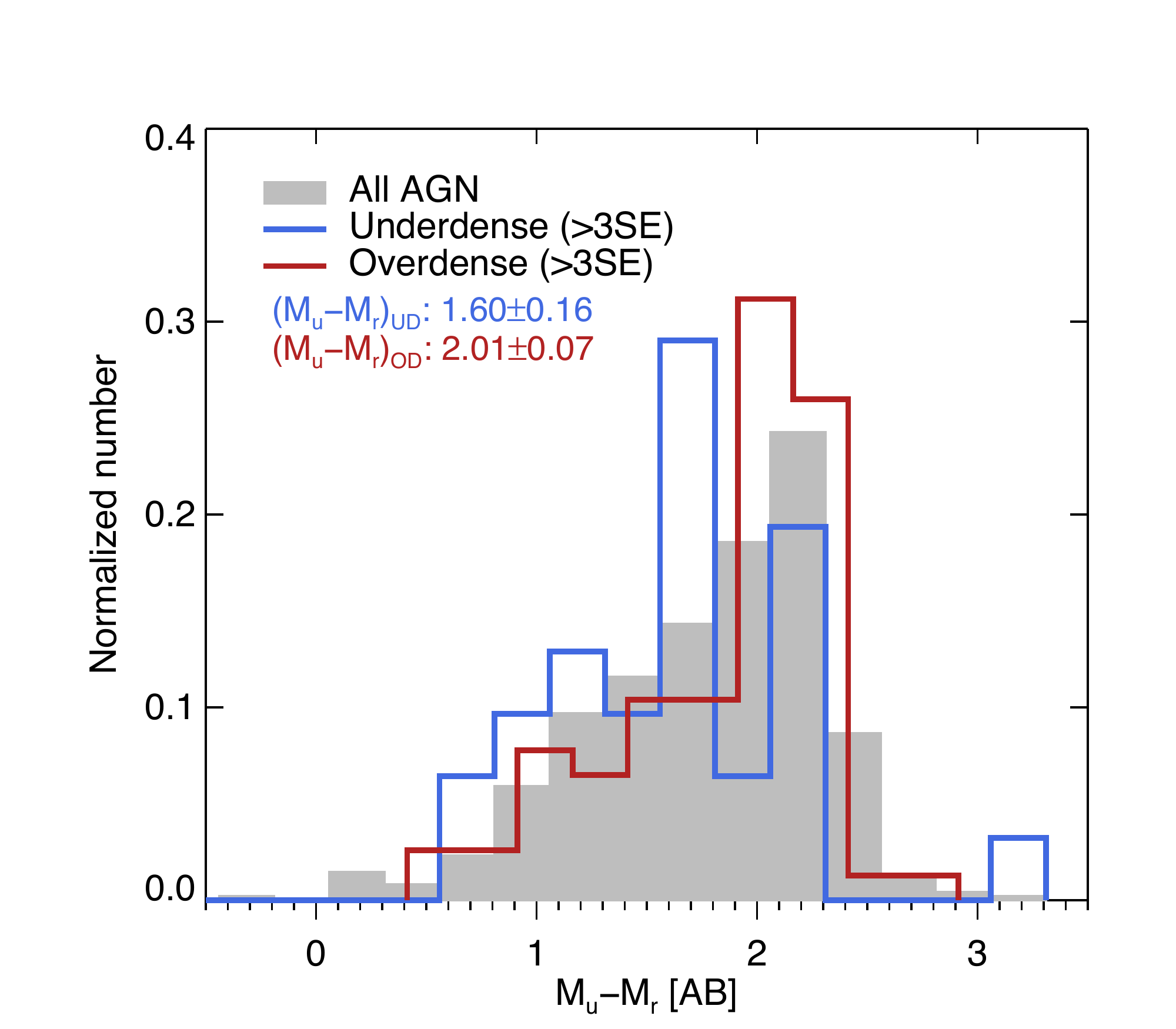}
\caption{Absolute AB magnitude color M$_u$-M$_r$ for the UD (blue) and OD (red) sub-samples of radio-AGN. The median values of the two distributions are also shown on the plot. For comparison we also plot the normalised distribution of the total luminosity-selected radio-AGN sample (grey filled histogram).}
\label{fig:ur_comb}
\end{center}
\end{figure}

We now turn our attention to the host galaxy properties of the two sub-samples. In Fig. \ref{fig:ur_comb} we plot the rest-frame absolute magnitude color (M$_{u}$-M$_{r}$) for the UD and OD sub-samples. As discussed previously, the $(M_{u}-M_{r})$ color should cover the 4000 $\AA$ break and therefore reflect the age of the dominant stellar populations in these galaxies as well as trace potential ongoing star formation. We observe that for the OD sample the majority of sources are found at a color of $\sim2$, while sources in the UD sample have a wider spread, reaching the bluest colors. A comparison between their median colors reflects this difference. This is further corroborated by a value p=0.001 derived from a two-sample KS test. While our sample covers a significantly wider redshift range, the differentiation between the two sub-samples in terms of their rest-frame $(M_{u}-M_{r})$ color, together with the findings by \citet{Strateva2001}, imply a difference in host galaxy morphologies and star-formation histories.

\begin{figure}[htbp]
\begin{center}
\includegraphics[width=0.45\textwidth,angle=0]{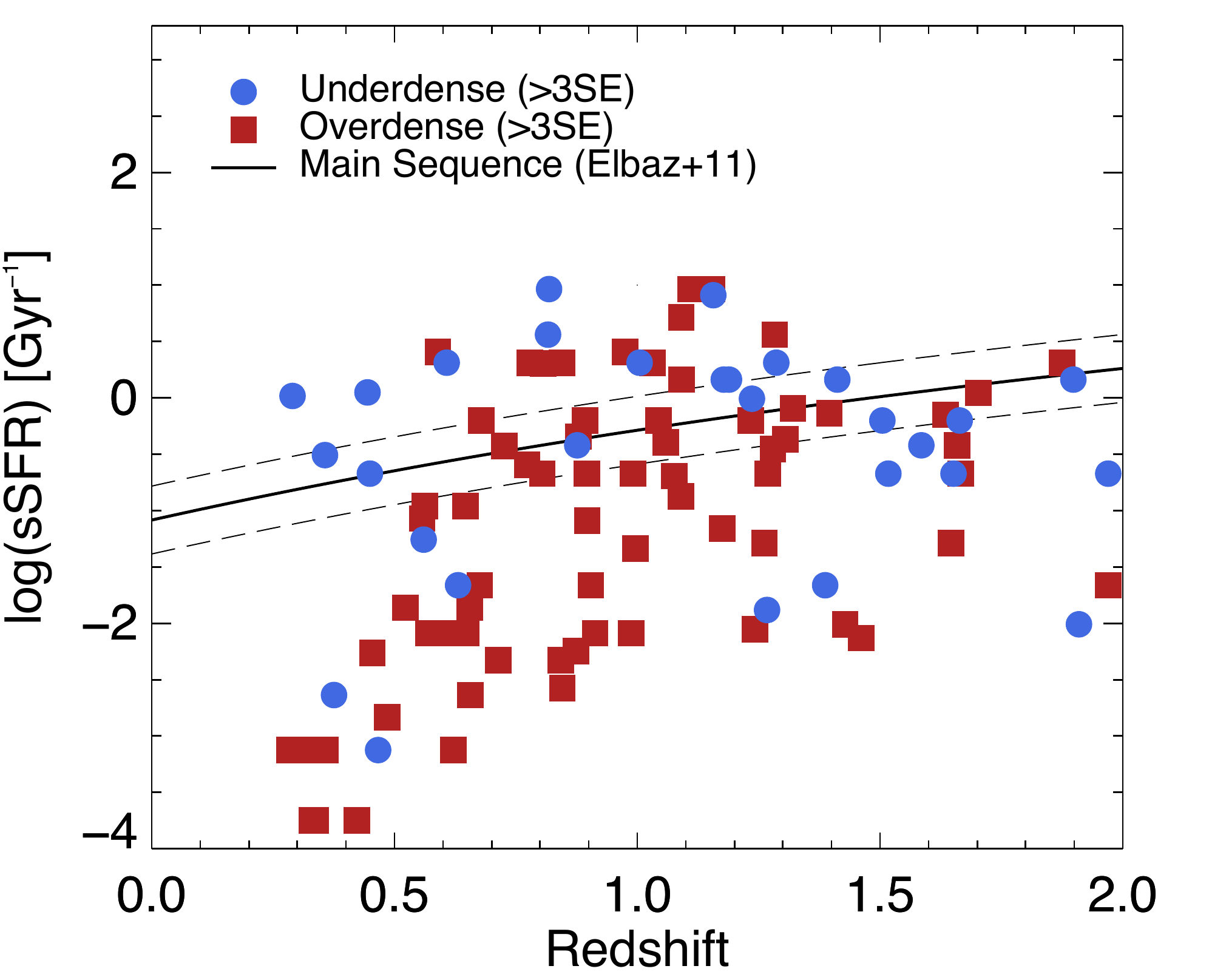}
\caption{Specific star-formation rate as a function of redshift for the two sub-samples of radio-AGN, OD (red) and UD (blue). The average sSFRs values of the matched control sample are also shown (with open blue and red circles for the UD and OD samples, respectively). The solid black line shows the calculated ``Main Sequence'' of star formation, as reported in \citet{Elbaz2011}, while the dashed black lines show the 3$\sigma$ margins of that relation.}
\label{fig:ssfr}
\end{center}
\end{figure}

We can use the results of our SED fitting to calculate the star formation efficiency of the radio sources in the OD and UD sub-samples. The star-formation rate (SFR) per unit of stellar mass, the specific SFR or sSFR, can be used as a measure of the star formation efficiency. We plot the sSFR of the two sub-samples as a function of their redshift in Fig. \ref{fig:ssfr}. In particular, we want to compare these to the known ``Main Sequence'' (e.g., \citealt{Rodighiero2010b}, \citealt{Elbaz2011}) of star formation, on which normally star-forming galaxies lie at a given redshift. This is shown with the solid line in Fig. \ref{fig:ssfr}, with the dashed lines showing the 3$\sigma$ margins of the ``Main Sequence''. Sources above the upper dashed line in the same plot are considered to be vigorously star-forming galaxies (starbursts), while sources below the lower dashed line are usually red ellipticals, with zero or minimal ongoing star formation (both dashed lines and source classification taken from \citealt{Elbaz2011}). As we can see the radio AGN are found both above and below the ``Main Sequence''. We observe however that radio-AGN in the UD sub-sample are preferentially found around or above the ``Main Sequence'', while the bulk of the OD radio-AGN are found below it.

\begin{figure*}[htbp]
\begin{center}
\includegraphics[width=0.32\textwidth,angle=0]{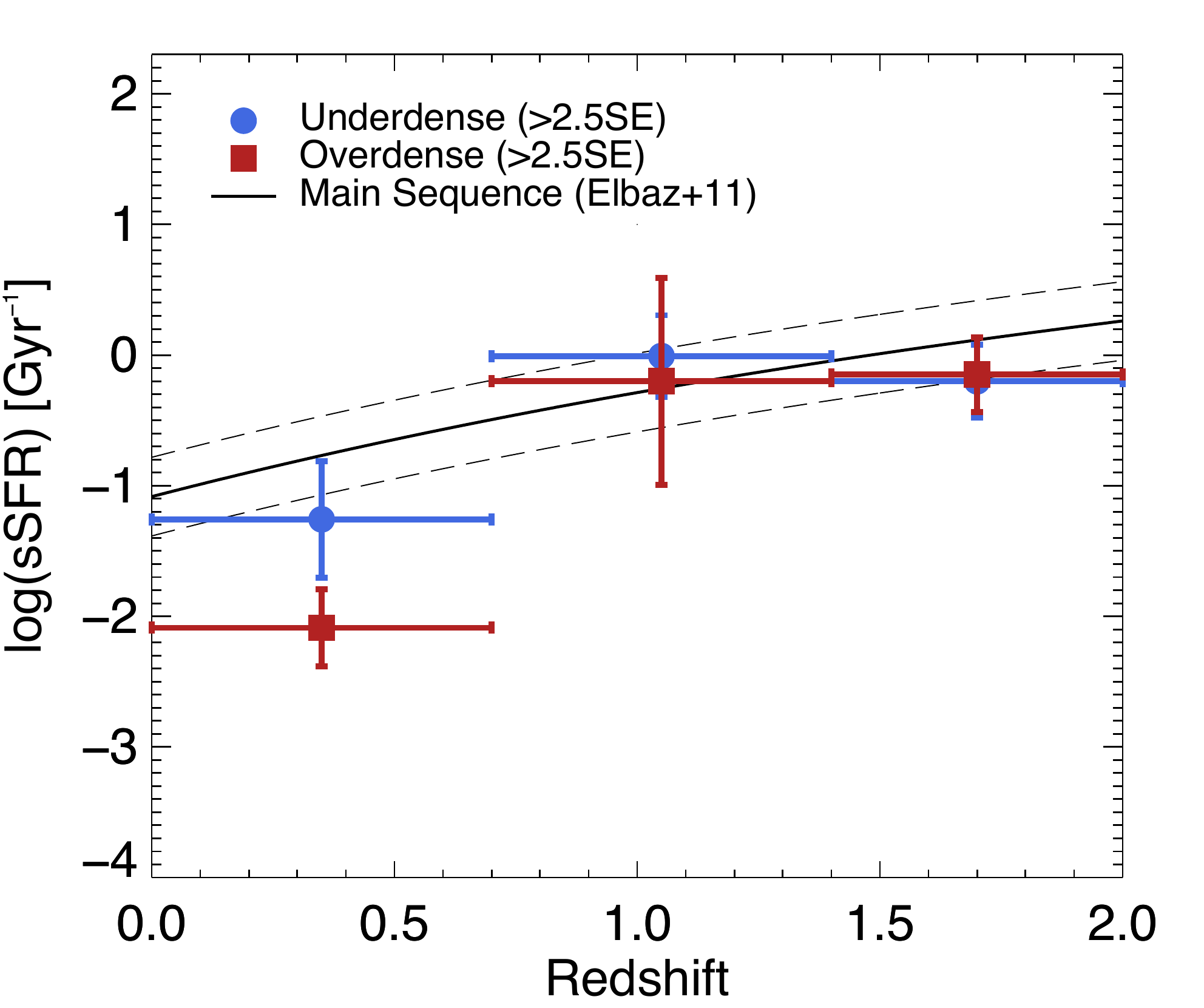}
\includegraphics[width=0.32\textwidth,angle=0]{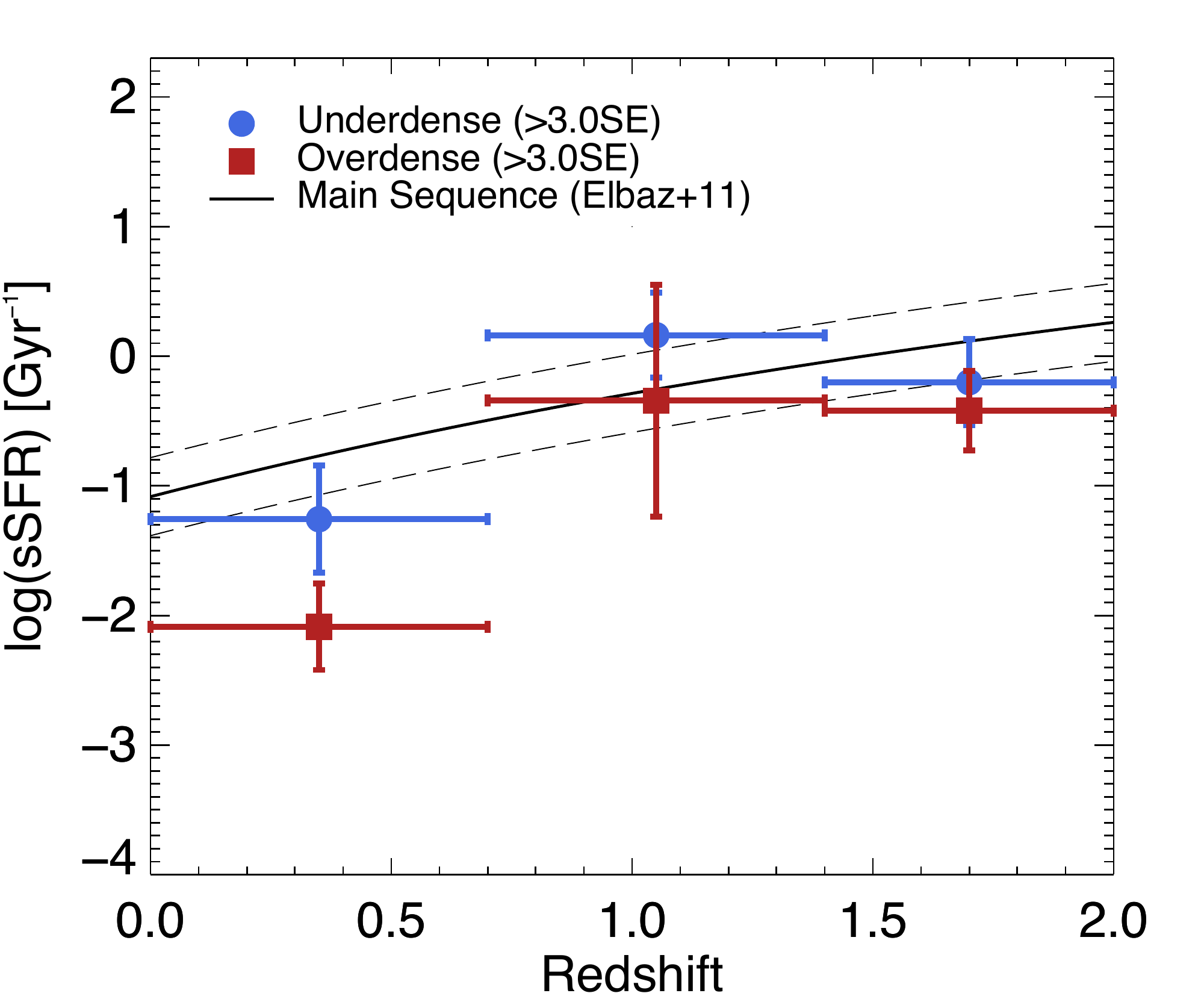}
\includegraphics[width=0.32\textwidth,angle=0]{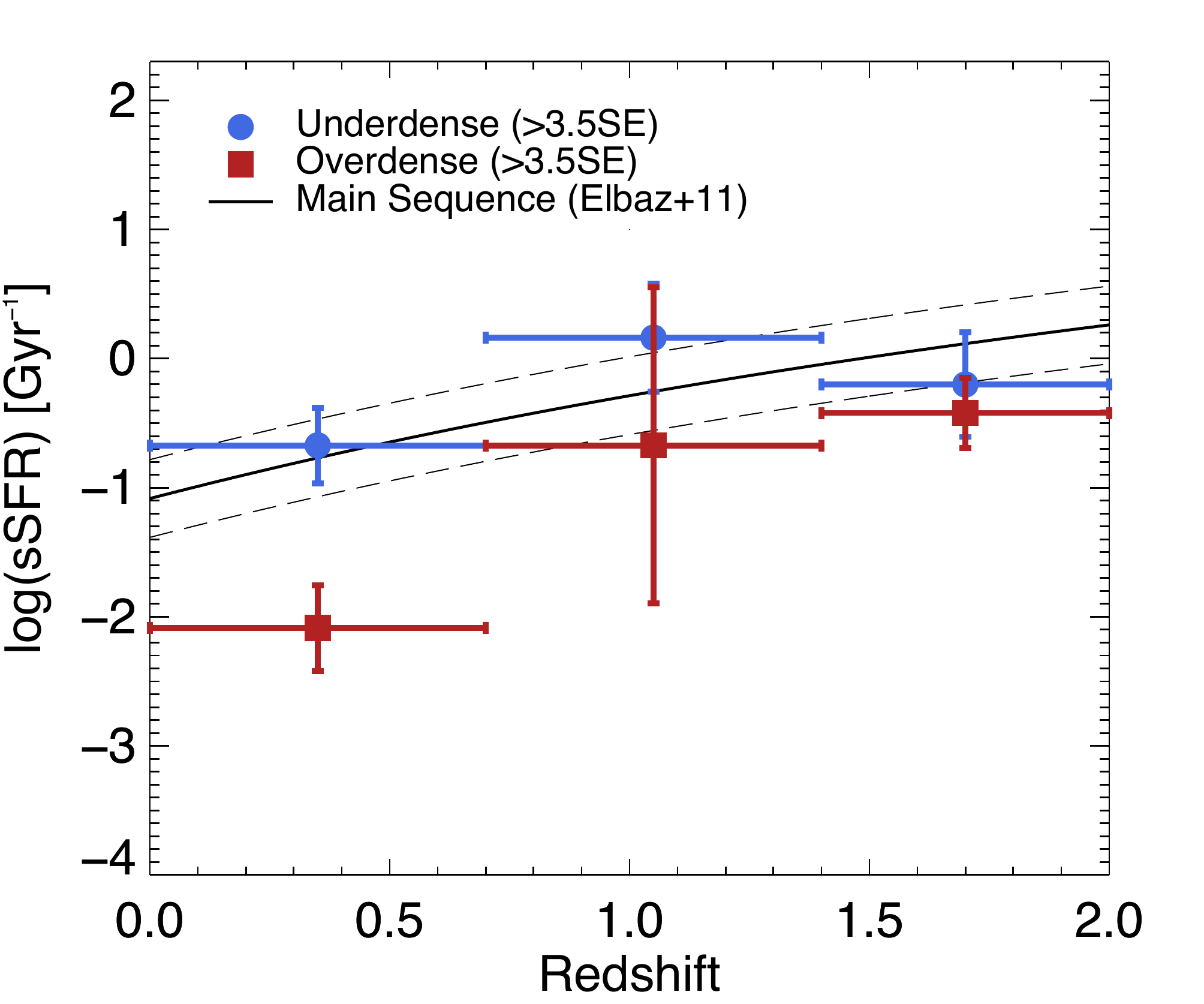}
\caption{As in Fig. \ref{fig:ssfr}, the specific star-formation rate is plotted as a function of redshift for the two sub-samples of radio-AGN, OD (red) and UD (blue), but in redshift bins. The solid black line shows the calculated ``Main Sequence'' of star formation, as reported in \citet{Elbaz2011}, while the dashed black lines show the 3$\sigma$ margins of that relation. From left to right, the selection limit for overdense and underdense sources changes from 2$\times$ (left), to 3$\times$ (middle), to 3.5$\times$ the SE (right). For all panels, the radio-AGN found in the most underdense environments show on average higher sSFR than those in the most overdense environments. This is particularly significant at redshifts $<1$.} 
\label{fig:ssfr_avg}
\end{center}
\end{figure*}

We investigate this further in Fig. \ref{fig:ssfr_avg}, where instead of individual sources, median values over redshift bins are plotted. We take an {additional} number of measures to ensure that bad SED fits and contamination from a strong optical AGN component do not affect our results. For the average values, we constrain ourselves to sources with:
\begin{itemize}
\item detections in at least 5 of our total 7 photometric bands {(7 out of 9, in the cases where WISE photometry is used)}, 
\item low $\chi^{2}$ values for their photometric redshift fits ({$\chi^{2}<1000$}\footnote{{We have looked at the fraction of outliers and $\sigma_{NMAD}$ as a function of the $\chi^{2}$ cut-off. However, both of them show a very weak, if any, dependence on the $\chi^{2}$.}}), 
\item photometric redshift with $\chi^{2}$ from the galaxy templates fit lower than the $\chi^{2}$ from the AGN templates fit, and
\item low $\chi^{2}$ values for their broadband SED fitting {(as already explained in Section \ref{sec:sed})}. 
\end{itemize}
{The above additional restrictions, while further reducing the number of sources from which we draw conclusions, err on the side of caution. We want to ensure} trustworthy SED fitting results, both in the sense of accurate photometric redshifts and minimising any AGN contamination that would affect our estimation of the sSFR of a source. We can now directly compare between the two sub-samples of radio-AGN. We can see that there is a significant difference between the two sub-samples, in terms of their sSFR, below redshift $\sim1$. At higher redshifts, while  the same trend for UD radio-AGN to show higher sSFR compared to their OD counterparts persists, it is not as significant. We can see how the definition of the UD and OD samples affects our results. We shift our selection limit of the two sub-samples from 2.5$\times$, to 3$\times$, and 3.5$\times$ the SE (shown in the left, middle, and right panel of Fig. \ref{fig:ssfr_avg}, respectively). It is clearly seen that the gap in sSFR between the two sub-samples grows wider as the contrast in the overdensity ratio between the two sub-samples increases.

As we are doing our selection based on the environment overdensity of these radio-AGN, the effect seen in Figs. \ref{fig:ssfr} and \ref{fig:ssfr_avg} may just reflect the difference in the large scale environment of these sources. A systematic difference between the star-formation properties of galaxies in the field and in cluster environments is known to exist (e.g., \citealt{Blanton2009}, \citealt{Peng2010}, \citealt{Alberts2014}). Therefore it is conceivable that the same radio-AGN that show the higher small-scale overdensities are also embedded in larger-scale overdensities and hence exhibit quenched star formation due to their large-scale environment. 

\begin{figure}[htbp]
\begin{center}
\includegraphics[width=0.4\textwidth,angle=0]{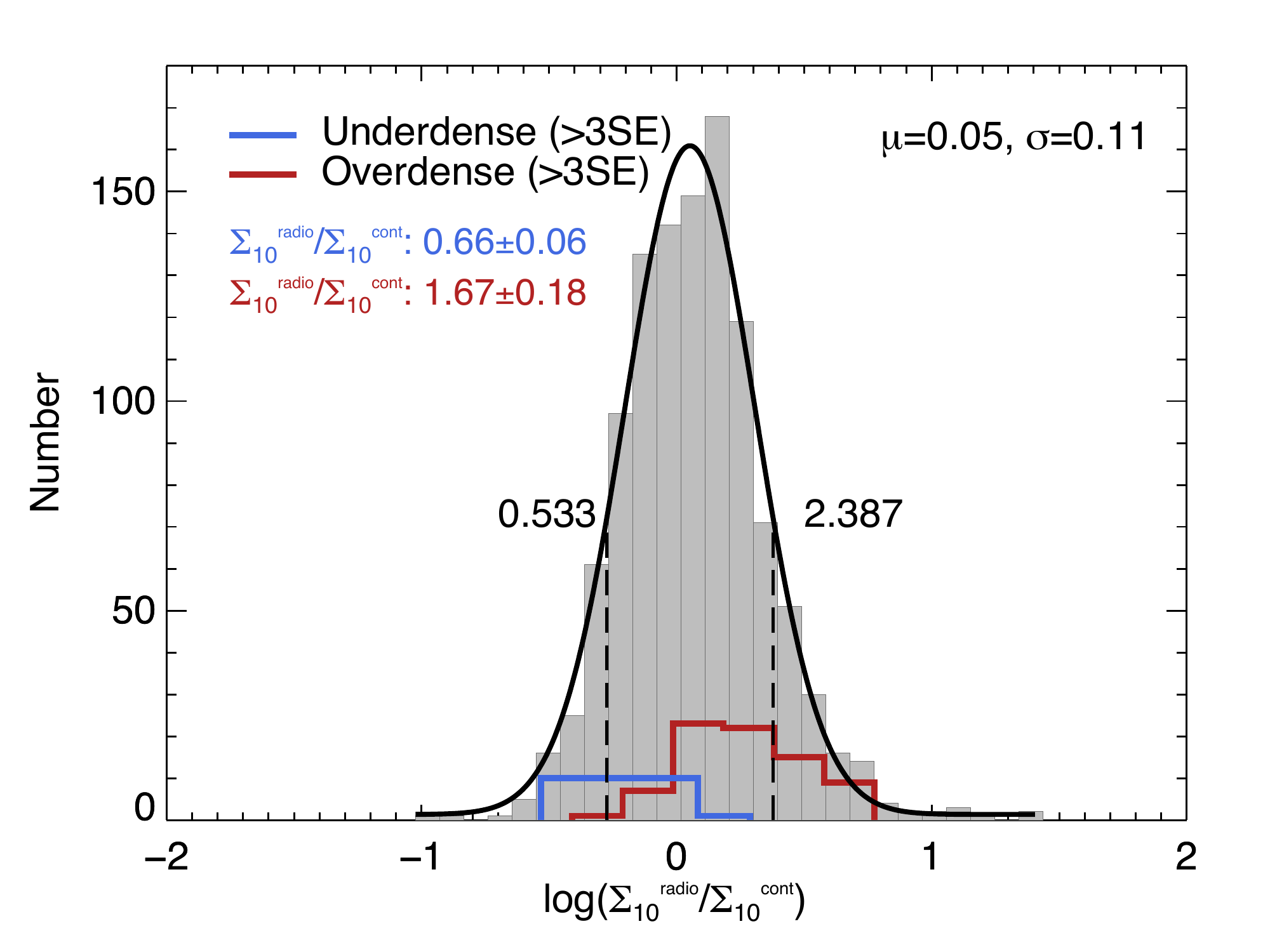}
\caption{Distribution of the overdensity ratio based on the $\Sigma_{10}$ density parameter for the full combined FIRST and VLA-Deep radio samples (filled grey histogram). The distribution is fitted with a standard Gaussian profile (shown with the black curve). The mean value and standard error, in logarithmic scale, of the Gaussian are also given. We also mark the 3$\times$SE margins for Gaussian profile (dashed lines). The distribution of the overdensity ratio of the UD and OD sub-samples\footnote{We note that the UD and OD sub-samples are defined in terms of their $\Sigma_2$ overdensity ratio.} of radio-AGN are also shown (in blue and red open histograms, respectively), together with their median overdensity ratio values.}
\label{fig:sigma10}
\end{center}
\end{figure}

We test this in Fig. \ref{fig:sigma10}, where the overdensity ratio {(defined as the ratio of the density around a radio source and the average density around its matched control sources)}  based on the 10th closest neighbor is plotted for the whole radio sample. Unlike for the 2nd closest neighbor, the 10th closest neighbor overdensity ratio should reflect the large-scale environment properties of these sources. We follow the same exercise as before to define the 3$\times$SE margin of the distribution. As a next step, we overplot the $\Sigma_{10}$ overdensity ratio distributions for the UD and OD radio-AGN sub-samples (in blue and red open histograms). It becomes immediately obvious that the bulk of these radio-AGN are not found in the most overdense large-scale environments. In fact, the median overdensity ratio values of the two subsamples are both within the 3$\times$SE margins of the full sample distribution.

\section{Discussion}
\label{sec:discuss}

Let us summarize briefly the main findings of our study:
\begin{itemize}
\item The bulk of the radio sources, both for the FIRST and VLA-Deep samples, appear to inhabit environments that are very similar to their matched control sources (Fig. \ref{fig:sigma_all}). Conversely, when matched to random field positions they exhibit strong overdensities.
\item There is a component of radio-sources that is found in the most dense environments, exhibiting densities up to 200 times that of their control sources. The same holds true if we consider just the luminosity-selected radio-AGN (Fig. \ref{fig:sigma_AGN}).
\item The average overdensity ratio value for the FIRST sample is significantly above 1 with a very low probability for this to be a chance result ($2.62\pm0.18$ with a chance probability of 0.1\% and $2.27\pm0.12$ with a chance probability of 5.3\% for all radio-sources and just radio-AGN, respectively). While for the VLA-Deep sample the average overdensity ratio is also above 1, it is not significantly so (Fig. \ref{fig:sigma_MC_random}).
\item We find no correlation between the radio-luminosity of a source and its small-scale environment. This holds true for both all sources and just radio-AGN, up to the highest radio luminosities probed here ($10^{44}$ erg/s; Fig. \ref{fig:sigma_lum}).
\item Radio-AGN in the most underdense (UD) environments exhibit, on average, higher radio-loudness and bluer $(M_{u}-M_{r})$ colors than then most overdense (OD) radio-AGN (Figs. \ref{fig:RL} and \ref{fig:ur_comb}, respectively).
\item Radio-AGN in the UD sub-sample show, on average, higher sSFRs compared to radio-AGN in the OD sub-sample, at or above the ``Main Sequence'' of star formation. This difference is more prominent at redshifts $<1$, although it persists even out to redshift $\sim2$, albeit at lower significance (Fig. \ref{fig:ssfr_avg}).
\item The UD and OD sub-samples of radio-AGN do not correspond to the most underdense and overdense large-scale environments, as probed by the 10th closest neighbor overdensity ratio (Fig. \ref{fig:sigma10}). Their sSFR differences cannot thus be attributed to large-scale environment quenching.
\end{itemize}

\subsection{Possible shortcomings of the analysis}
Before we continue with the comparison with previous studies and the discussion of the above summarised results, we wish to investigate possible biases and systematical effects that may influence our results. These can be broadly divided in terms of our selection and in terms of our analysis. We will discuss each of the two separately.
\subsubsection{Selection effects}
Our main selection is done in the radio, 1.4 GHz, and the near-infrared, $J$ band. Radio selection, especially at lower frequencies, tends to pick up either star-forming galaxies (due to their relatively steep spectrum) or radio-quiet AGN at low radio luminosities, starburst galaxies and weak radio-AGN at intermediate radio luminosities, and AGN with powerful radio jets at the highest radio luminosities. Here we have combined two different radio surveys, one shallow but wide and one deep that however focuses on a small area on the sky. As such we can efficiently probe down to at least 0.1 mJy (the 5$\sigma$ limit for the B-configuration VLA-Deep observations), effectively including many star-forming and composite (star formation + AGN) galaxies in our sample (e.g., \citealt{Seymour2008a}). 

We assume a radio luminosity cut in order to select radio-AGN and study their environment. At radio luminosities $>10^{40}$ erg/s the sky is dominated by radio-AGN rather than star-forming galaxies (e.g., \citealt{Best2005a}, \citealt{Mauch2007}). We expect a contamination from star-forming galaxies in our luminosity-selected radio-AGN sample of $<10\%$. While our radio luminosity limit for AGN selection is based on the local radio luminosity function, it is expected that, due to the evolution of the luminosity function of both AGN and star-forming galaxies with redshift, this limit should shift towards higher values at earlier epochs. However, due to the redshift-luminosity relation imposed by the flux-limited nature of our sample, we do not expect a significant effect in our selection.

Due to the flux limit of our radio observations, we are by definition missing fainter radio sources. As it has been shown (e.g., \citealt{Seymour2008a}, \citealt{Padovani2009}) at flux densities below 1 mJy, the sky is increasingly dominated by star-forming galaxies and potentially radio-quiet AGN. As such, we do not expect to be missing a significant component of radio-AGN from our study, especially out to our relatively modest assumed redshift limit.

We combine the radio selection with a magnitude-limited sample in the near-infrared from the IMS survey. As such, our secondary selection is in $J$ band, at $\sim1.2$ $\mu$m. {This is the longest wavelength available over the full area of the IMS survey in the SA22 field and hence the least prone to obscuration. Furthermore, for radio sources at low and moderate redshifts ($<0.5$), $J$ band probes the light in the near-infrared}, which should be dominated by emission from old stellar populations. We know that radio-AGN are usually found in massive galaxies (e.g., \citealt{Best2005}) and therefore we do not expect our results to be severely affected by the exclusion of low-mass galaxies. Another factor we need to consider is contamination of the $J$-band light from powerful AGN emission. This is particularly relevant for the most luminous AGN, quasars, which are usually point-like and their whole spectral energy distribution is dominated by their nuclear emission. The maximum optical luminosity of our sources (in $i$ band) is below $10^{46}$ erg/s and therefore well below the bright quasar luminosity regime ($>10^{47}$ erg/s).  Only 19 sources in our sample exhibit optical luminosities $>10^{45}$ erg/s, the nominal limit above which quasars dominate over their host galaxy emission (e.g., \citealt{Shen2011}). Therefore there should not be significant contamination (mainly in terms of the control sample selection) from AGN emission in the optical and near-infrared.

\subsubsection{Methodology caveats}
There are two main points that can introduce uncertainties and systematics to our results, (1) the estimation of photometric redshifts and (2) the fitting of the broadband spectral energy distributions and the derivation of stellar masses and star formation rates for galaxies. As we showed in Section \ref{sec:method}, we have taken a number of steps usually employed in studies of photometric redshifts to minimise the effects of photometric uncertainties and also calibrate photometric redshifts to an accurate set of spectroscopic redshifts. We have further showed that for the majority of radio sources with spectroscopic redshifts, the estimated photometric redshifts agree well within $15\%$. We calculated the fraction of miscalculated redshifts for our radio sources (in terms of outlier fraction) to be $\sim10\%$. This translates to a roughly 10\% uncertainty of the calculated fraction of AGN in overdense or underdense environments. Given that there is no preferential bias in the miscalculated photometric redshifts (i.e., we see in Fig. \ref{fig:specz} of the two radio-source outliers one exhibits higher and one lower photometric redshift compared to their true redshifts), we do not expect a systematic effect on our results. Moreover, Fig. \ref{fig:specz} is based on VVDS data and therefore all sources included have an optical magnitude of at least 22.5 AB. As most of our radio sources are actually fainter than this limit and it is known that at lower optical magnitudes the potential contamination from a powerful nucleus is smaller, we expect the outlier fraction calculated from Fig. \ref{fig:specz} to be a conservative upper limit. In reality, we expect lower optical luminosity radio sources to exhibit more accurate photometric redshifts.

We showed that the uncertainty of our photometric redshifts for the main sample is 0.038. This means that the width of the redshift slices used for the calculation of the environment density parameters, $\pm0.1(1+z)$, is $>2.5\sigma$ of the photometric redshift. Finally, as our overdensity parameters are expressed in terms of a density ratio between each radio-source and its control sources, any potential systematic effects induced by the photometric redshifts will cancel out.

The second point we need to consider is possible problems with the results of our spectral energy distribution fitting. Assuming accurate redshifts and relatively low contamination from the nuclear emission (both points covered previously), the next pitfall that need be considered is possible degeneracies between the different parameters of the fit and the accuracy of the properties derived, i.e., how well can the stellar mass and star formation rate be constrained. As was discussed previously, the near-infrared emission is necessary to estimate the stellar mass of a galaxy. Given our redshift range (z$<2$) and the fact that we require full near-infrared coverage to perform the SED fitting (i.e., up to 2.2 $\mu$m for z$<1$ {and up to 4.6 $\mu$m for z$>1$}), we are confident in our estimation of the stellar masses (e.g., \citealt{Shapley2005}, \citealt{Lee2009}). 

The calculation of absolute star formation rates is not as straightforward. Due to the dusty nature of the birthing sites of stars, a large fraction of young stellar emission is absorbed and re-radiated in the mid- and far-infrared regime. As such, we are at a disadvantage with our current dataset missing any data points above 2.2 $\mu$m (4.6 $\mu$m, respectively). \citet{Karouzos2014b} showed that in absence of far-infrared data, star formation rates in radio AGN can be underestimated by at least an order of magnitude. However, it is important to note here that star formation rates are used in our study to compare between the different sub-samples of UD and OD radio sources and therefore this difficulty in calculating absolute SFRs should not affect as severely. Furthermore, given our selection, we do not expect significantly different dust properties between the UD and OD sub-samples. This implies that the missing component of ``hidden'' star formation should not be preferentially larger for one of the two sub-samples.

\subsection{Comparison with Other Studies}
\subsubsection{Environment of radio-AGN}
A direct comparison can be drawn between this study and our previous study on the environment of radio-AGN (\citealt{Karouzos2014a}). We have improved upon that work in two key aspects: we have done a more vigorous control sample selection (introducing the color matching) and we have expanded the radio luminosity range probed by at least 2 orders of magnitude (maximum luminosity probed here is $\sim10^{44}$ erg/s compared to the $\sim10^{42}$ erg/s previously). In addition, we have expanded upon previous works to look at the host galaxy properties, differentiating between those radio-AGN in the most overdense and most underdense small-scale environments.

Our key results in terms of the environments properties of radio-AGN match very well with those from \citet{Karouzos2014a}, essentially showcasing the relatively ordinary environments that the bulk of radio-AGN reside in. The average overdensity ratios found here match remarkably well with those for the similar luminosity-selected radio-AGN sample of \citet{Karouzos2014a}. The only exception to that is the overdensity ratio for the $\Sigma_{2}$ density parameter, where the VIDEO results  give a significantly higher value ($5.4\pm1.3$ compared to $2.27\pm0.12$ for this study). This difference might arise due to the different (and more vigorous) matching criteria used here for the selection of the control sources. A better matching in terms of the host galaxy properties of the radio-AGN to their control sample, under the assumption that radio-AGN on average do not inhabit different environments that their non-active counterparts, would lead to a decreased mean overdensity ratio. 

A mean overdensity of radio-AGN above one agrees also well with previous studies of the environment of radio-AGN by \citet{Best2004}, \citet{Tasse2008}, \citealt{Donoso2010}, \citet{Bradshaw2011}, \citet{Lietzen2011}, \citealt{vanVelzen2012}, \citet{Almeida2013}, \citet{Worpel2013}, and \citet{Pace2014}. We should note however that each of these studies has followed a different way of both selecting their radio sample and defining their control sources. As a result, both the host galaxy stellar mass and radio-AGN luminosity ranges probed are different for each one of these studies. Moreover, they all use different measures of environment overdensity. Therefore a direct comparison is difficult. However, there appears to be a consensus that, especially at small scales ($<1$ Mpc), radio-AGN are on average embedded in more dense environments than galaxies of similar mass, color, and morphology.

Despite our expanded radio luminosity range, the absence of any appreciable trend between the AGN radio luminosity and its environment overdensity persists. This may be somewhat puzzling and in contrast to prevalent models of powerful AGN triggering through mergers. It is in line with the results of \citet{Karouzos2014a} and also in good agreement with the study of \citet{Villforth2014}. \citet{Almeida2011} however showed that up to $\sim90\%$ of their sample of powerful radio-AGN (2 Jy sample) show morphological distortions, signs of recent mergers, much more than their control sample. Given that the radio brightness of these galaxies and the fact that the majority of them show optical emission lines, may indicate a different flavor of AGN than the ones we are dealing with here. 

\subsubsection{Environment and host galaxies}
The host galaxies of radio-AGN have been studied intensively, both in terms of their morphologies and the stellar populations that they host (e.g., \citealt{Best2005}). It is now believed that there is a fundamental division within the general population of radio-AGN, driven mainly by the accretion rate onto their central supermassive black holes, classifying them into high- and low-excitation radio-AGN (e.g., \citealt{Hardcastle2007}). In turn, this classification reflects the different modes of gas feeding towards the nucleus of these galaxies. The latter type of radio-AGN are believed to be fed through hot gas, have low accretion rates, and weak or absent optical emission lines. The former type on the other hand exhibit strong optical emission lines, implying efficient accretion. It has been found that this dichotomy in feeding is also accompanied by a dichotomy in host galaxy properties, with high-excitation radio-AGN being characterised by relatively young stellar populations and ongoing star formation. This is in contrast to their low-excitation counterparts, which show old stellar populations (e.g., \citealt{Herbert2010}, \citealt{Best2012}, \citealt{Hardcastle2013}).

A connection between the above and the environments of radio-AGN has been attempted in previous studies. \citet{Tasse2008} argued that low-excitation radio-AGN live in overdensities and are being fed by the ample inter-cluster gas available in such environments. On the other hand, low-mass, high-excitation radio-AGN are found in large-scale underdensities hinting towards close pair interactions driving the triggering of these AGN. This is somewhat corroborated by the study of \citet{Almeida2013} that find galaxies  to show smaller degree of spatial clustering around strong-lined radio-AGN (high-excitation) compared to their weak-lined (low-excitation) counterparts. In our study we make a link between the environment of radio-AGN and star formation in their host galaxies. Our finding that radio-AGN in the densest small-scale environments show the lowest sSFR and relatively moderate-power jets is in agreement with the findings by \citet{Almeida2013}, as these sources should be the equivalent of weak-line radio-AGN. As expected, these sources have low sSFR and are expected to be mainly fed by the accretion of hot gas from their halo.

On the other hand, radio-AGN with high sSFR and high-power jets (as exhibited by their radio loudness distribution) may be triggered by mergers, as is implied by \citet{Almeida2013} for their sample of radio-AGN with recent star formation activity (based on results by \citealt{Dicken2012}) and further corroborated by the results of high spatial clustering around these galaxies in their study. Our results however may point towards a different direction. We showed that these sources exhibit high sSFR, on par with or above the ``Main Sequence'', but at the same time are found in the most underdense small-scale environments within our sample. This would discard close pair interactions and mergers as the driver of nuclear activity in these sources. Moreover, we showed that on average, these most underdense sources at small scales, live in unremarkable large-scale environments. This therefore also rejects the possibility that cooling flows (e.g., \citealt{Fabian1994}) within a cluster environment are responsible for the ongoing star formation and triggering of the radio-AGN. This population of UD radio-AGN may be related to the low optical luminosity radio-loud quasars found to show a star formation excess, compared to their radio-quiet counterparts, by \citet{Kalfountzou2014}. The authors find that radio-loud quasars at optical luminosities $<10^{45.5}$ erg/s show an excess of star formation (as reflected by the far-infrared properties), on average lower dust temperatures, and potentially higher dust masses.

\subsection{How to trigger a radio-AGN?}
We have shown evidence that the environments of radio-AGN are not particularly different from those of non-active galaxies with similar properties (both in terms of their masses and star formation rates). Nevertheless we find a significant sub-population of radio-AGN which do appear to inhabit very dense environments at small scales. For these galaxies, close pair interactions and galaxy mergers might be important for the triggering of the active nucleus. How can we then differentiate between the different processes responsible for triggering these radio-AGN?

The star formation properties of radio-AGN in the most and least dense environments might be able to provide the necessary clues. As it became apparent from our comparison with previous studies in the literature, while radio-AGN are predominantly believed to reside in somewhat overdense environments, what this means in terms of their triggering is heavily debated. Below we address two different aspects of this problem before attempting to draw a single consistent picture:
\begin{itemize}
\item \textit{Jet power:} We do not have a clear picture of what drives the power of radio-AGN jets. A dense gas screen with which the jet can interact can lead to increased radio emission. Such would be the case of radio-AGN residing in rich clusters with dense ICM. Alternatively, the spin of the jet-producing supermassive black hole may also play a role in how powerful a jet is. Our understanding of spin processes is very poor, although semi-analytic models imply that galaxies with rich merging histories (as is the case for example for central cluster galaxies) may host more rapidly spinning supermassive black holes (e.g., \citealt{Fanidakis2011}).
\item \textit{AGN power:} We know that the availability and the temperature of gas in galaxies decides the power of a triggered AGN. Efficient accretion of cold gas, either through a cooling flow (in a cluster where radio-AGN feedback has not kicked in), a gas-rich merger (in group environments), or from mass-loss of young stars and along cold filaments (in any environment), can trigger powerful AGN with Eddington accretion ratios close to 1. On the other hand, inefficient hot gas accretion, either from the inter-cluster medium (in cluster environments) or from the galaxy's own halo (in ``normal'' environments), leads to low-luminosity AGN with low ($<0.1$) accretion ratios.
\end{itemize}

We propose that the existence of a population of radio-AGN in underdense environments compared to their control sources with high sSFRs and high radio-loudness offers compelling evidence for triggering through stellar feedback (e.g., \citealt{Norman1988}). As we explained above, their underdense environments discard mergers and galactic interactions as the dominant effect for triggering these systems. Moreover, their ``normal'' large-scale environments also preclude an important role of hot or cold gas accretion from a cluster environment. Cold gas accretion along filaments, which can extend to large scales (e.g., \citealt{Keres2005}), has been argued to provide galaxies with a significant part of their cold gas reservoir. However, as \citealt{Keres2005} has shown, this mode of gas accretion is particularly relevant for galaxies at mass $M_{stel}<10^{10.3}M_\odot$. Both the UD and OD sub-samples have mean stellar masses above that limit, with most of sources of both samples at $M_{stel}>10^{10}M_\odot$. We thus do not expect accretion along cold filaments to be a viable feeding mechanism for the majority of these radio-AGN. On the other hand, the presence of ongoing vigorous star formation offers a gas reservoir which can be tapped into to trigger an AGN. In particular, it has been shown that mass-loaded winds from relatively low-mass stars in their AGB phase can dominate the mass outflow during intense episodes of star formation (e.g., \citealt{Winters2003}, \citealt{Wild2010}) and can lead to the triggering of an AGN, albeit with a certain time lag (e.g., \citealt{Davies2007},\citealt{Wild2010}). The last missing piece of the puzzle may come from the difference of the radio-loudness distribution of these sources. Systems that are experiencing intense star formation that may be feeding the central AGN are expected to have high degrees of obscuration (e.g., \citealt{Wild2007}). This would lead to a decreased optical emission and therefore a shift of their radio-loudness towards higher values compared to the relatively unobscured radio-AGN counterparts residing in dense environments. {Additionally, radio luminosities of the UD radio-AGN may be contaminated by their ongoing SF, which should also contribute to the total radio emission measured by the large VLA beam, over the whole galaxy.  }

\begin{figure}[htbp]
\begin{center}
\includegraphics[width=0.4\textwidth,angle=0]{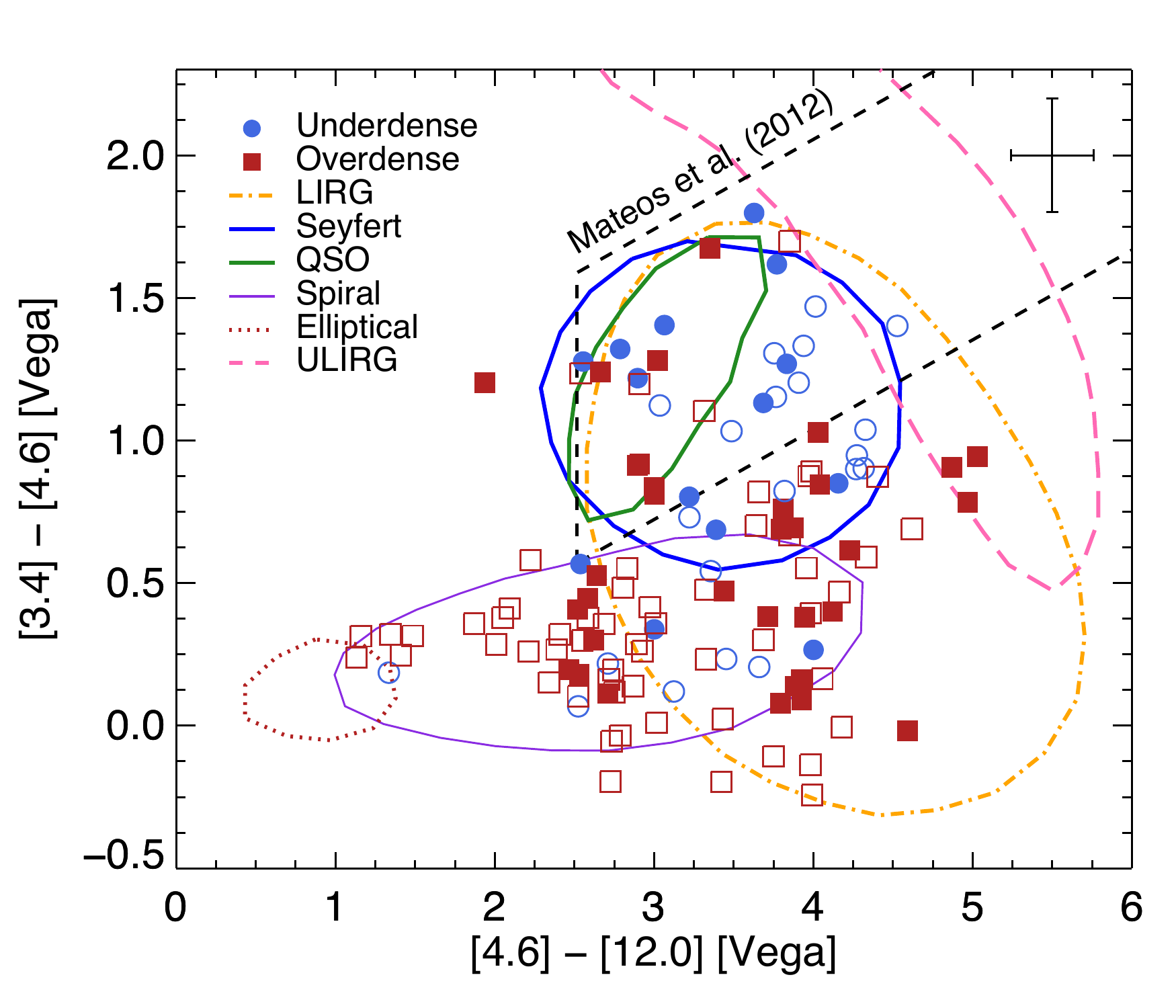}
\caption{WISE color-color plot for the 3.4, 4.6, and 12 $\mu$m bands. We show sources in the UD (blue circles) and OD (red squares) sub-samples of radio-AGN. We separate sources that are detected in the longest wavelength band (12 $\mu$m; filled symbols) from those with only upper limits (open symbols). The loci of different types of astronomical objects are taken from \citet{Wright2010} (solid lines). The dashed line shows the AGN selection locus from \citet{Mateos2012}. The typical uncertainty of the measurements is shown in the upper right corner of the plot.}
\label{fig:WISE_col}
\end{center}
\end{figure}

It is important to note here that Fig. \ref{fig:ssfr_avg} shows not only a difference in sSFR between the UD and OD sub-samples but also implies a redshift evolution of this apparent subdivision among radio-AGN. Above redshift of z$\sim1$ the two sub-samples appear to be consistent with each other (within the measured uncertainties). Moreover, the OD sample shows higher sSFR compared to $z<1$, consistent with normally star-forming galaxies at similar redshifts. This reflects the expected evolution of the dominant triggering mechanism with redshift. Moreover, it agrees well with the general trend for any environmental dependence of star formation in galaxies to wash out at redshifts $>1$ (e.g., \citealt{Scoville2013}).

\begin{figure*}[htbp]
\begin{center}
\includegraphics[width=0.4\textwidth,angle=0]{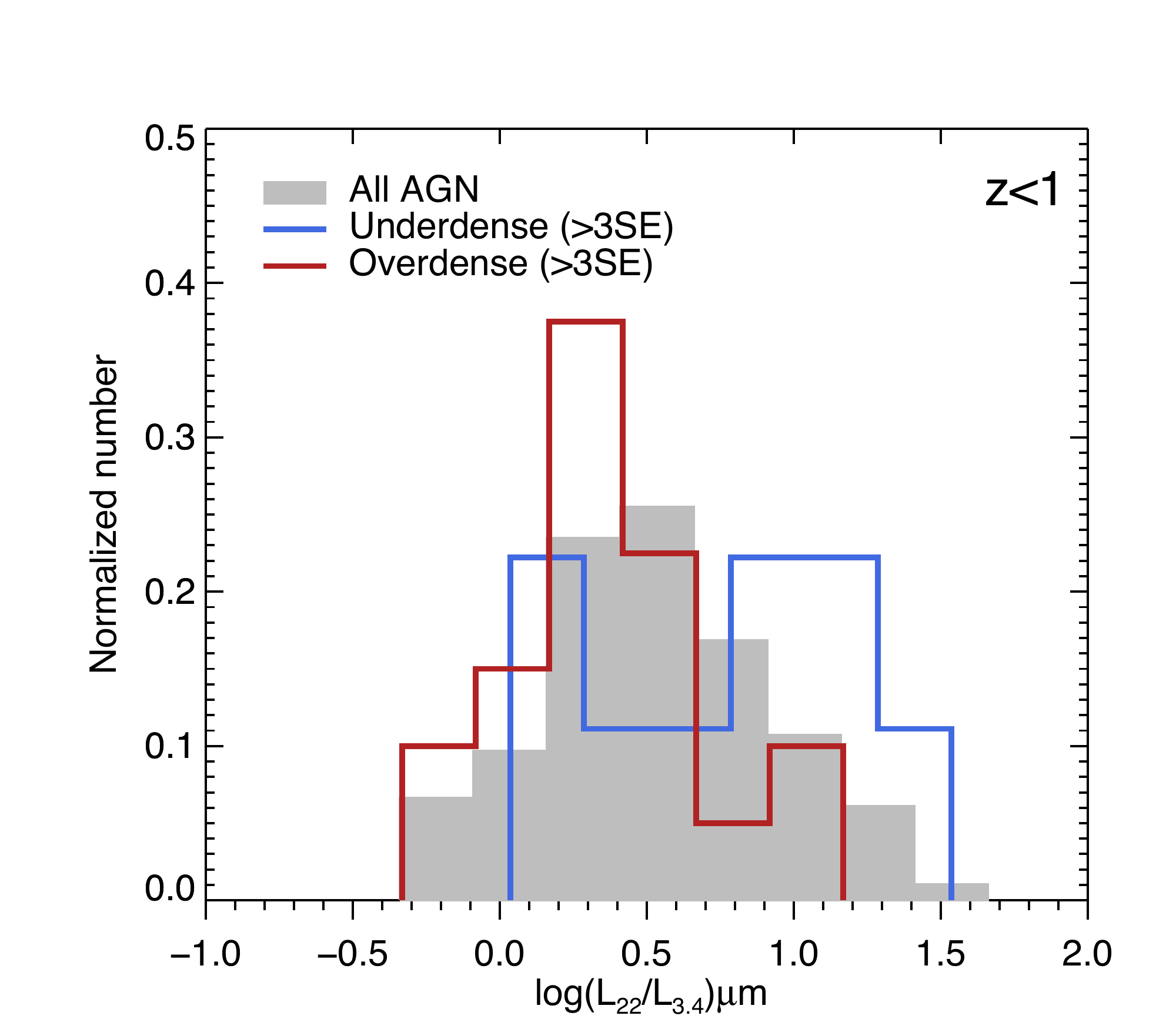}
\includegraphics[width=0.4\textwidth,angle=0]{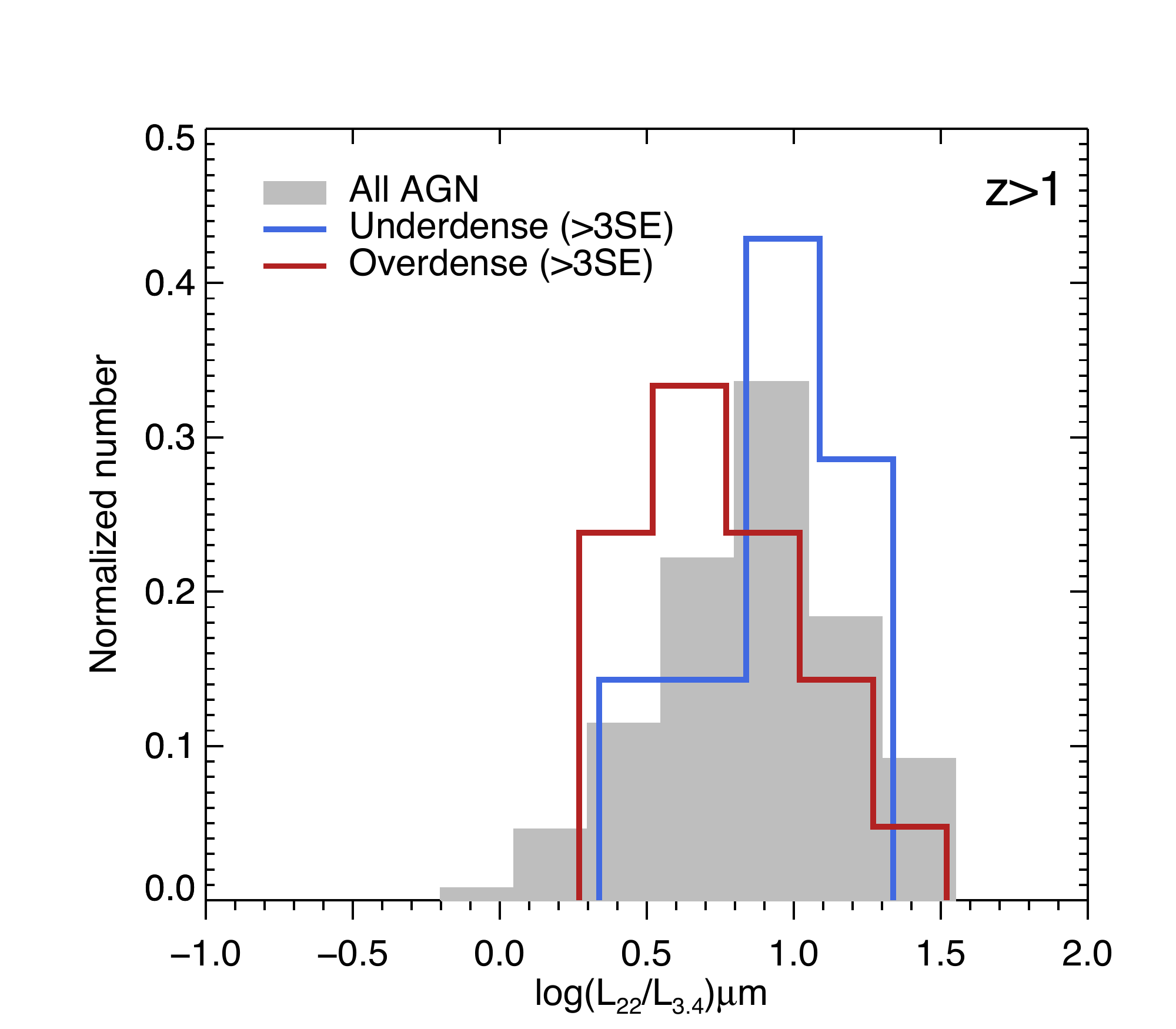}
\caption{Distribution of the luminosity ratio, in logarithmic scale, between the 22 $\mu$m and 3.4 $\mu$m bands of WISE ($W4$ and $W1$, respectively). The distributions of the total radio-AGN sample (grey shaded histogram), and the UD and OD sub-samples (blue and red open histograms, respectively) are shown. Sources at redshifts $<1$ and $>1$ are shown separately (left and right panel).}
\label{fig:wise}
\end{center}
\end{figure*}

We are now ready to outline a scheme of how radio-AGN can be triggered. In the local universe it seems that the availability of gas is the controlling factor that drives differences between different flavors of radio-AGN. Radio-AGN embedded within dense environments, either small- or large-scale, are fed through accretion of hot gas either from their own halo or from the intra-cluster medium, respectively. These are the low-excitation radio-AGN mentioned before and they exhibit low sSFRs, well below the ``Main Sequence'' of star formation. On the other hand, we have radio-AGN found in the most underdense environments (UD sub-sample), whose activity should be unrelated to their environments. The high sSFRs exhibited by these sources imply that the ongoing star formation might provide the necessary fodder for the central supermassive black hole. These are the termed high-excitation AGN.

The situation at higher redshifts appears to be different, as potentially gas-rich mergers and more readily available cold gas within galaxies shift accretion in radio-AGN towards a more efficient mode, accompanied by ongoing star formation. We therefore observe a much less significant difference not only of the environment but also the star formation properties of these sources.

\subsection{An infrared point of view}
Mid-IR colors can be used to differentiate between different flavors of AGN. \citet{Gurkan2014} showed that the WISE luminosity at 22 $\mu$m and luminosity ratio between 22 and 3.4 $\mu$m (WISE bands $W4$ and $W1$, respectively) can be used as a proxy for Eddington accretion ratio and can thus be used to differentiate between low and high accretion rate radio-AGN.  To get the maximum number of radio sources with WISE information, we performed a match between our total radio sample and the All Sky WISE catalog (containing detections above 5$\sigma$) and the All Sky WISE Reject table (containing detections below 5$\sigma$). The cross-matching was done in the same manner as described in Section \ref{sec:sed}. We find in total 759 radio sources in the SA22 field with detection in at least one WISE band at the 3$\sigma$ sensitivity limit. In addition, we also find 32 more matches with lower significance detections. Assuming the latter to be upper limits, we use a total of 791 SA22 radio sources to calculate the luminosity ratio between 22 and 3.4 $\mu$m.

In Fig. \ref{fig:WISE_col} we show the ($4.6-12$) $\mu$m and ($3.4-4.6$) $\mu$m color-color plot for the UD and OD sub-samples of radio-AGN. We see that the two sub-samples occupy distinct regions of the WISE color parameter space. Radio-AGN in the UD sub-sample are mainly contained within the locus of Seyfert galaxies, with relatively red ($3.4-4.6$) $\mu$m colors of $\sim0.8$. On the other hand, OD radio-AGN are mostly found inside the spiral and luminous infrared galaxies loci, with only a few consistent with AGN colors. The AGN selection by \citet{Mateos2012} mostly picks up radio-AGN in the UD sub-sample but largely misses radio-AGN in the most overdense environments.

The distinction between the UD and OD radio-AGN in our sample becomes even clearer in Fig. \ref{fig:wise}. In this plot we show the distribution of the 22$\mu$m to 3.4 $\mu$m luminosity ratio distribution for the two sub-samples below (left panel) and above (right panel) a redshift of $z=1$. For a redshift $z<1$, where a significant difference in sSFR is observed between the UD and OD sources, we also see that their WISE luminosity ratio distributions are significantly different. While OD sources show a distribution peaked at ratio values of $\sim2$, the UD sources show a double-peaked distribution, with the first peak at around a ratio of $\sim1$ and a second broad peak at ratio values of $\sim10$, while extending to much higher ratios than their OD counterparts. A two-sample KS test gives a p=0.03.

If we now turn at the same distribution but for sources at redshifts $z>1$ (right panel of Fig. \ref{fig:wise}), the differences between OD and UD sources become smaller, with the peak of the UD distribution remaining at a ratio value of $\sim10$ but the peak of the OD distribution shifting to a ratio value of $\sim5$. Furthermore, both distributions are missing the low-value tails and both extending to the highest luminosity ratios values within their parent sample. A two-sample KS test fails to reject the null hypothesis at high significance. We note here that, while the OD sources show lower luminosity ratio values, they still have ratios which are high compared to the low-excitation radio-AGN from \citet{Gurkan2014}. Figure \ref{fig:WISE_col} implies that the mid-IR emission of the OD radio-AGN may be dominated by their star formation component, rather than the AGN. Therefore, the red histograms of Fig. \ref{fig:wise} may be actually overestimating the Eddington ratios of these sources. This can reconcile our OD radio-AGN with the low-excitation radio-AGN of \citet{Gurkan2014}.

Nevertheless, we can conclude that sources in the most underdense environments and with high sSFRs, also show on average higher accretion rates, typical of efficient accretion of cold gas. This is in agreement with the scenario put forth above and matches with the far-IR properties of low optical luminosity radio-loud quasars of \citet{Kalfountzou2014}.

\section{Conclusions}
\label{sec:conc}
We have used data from the new IMS near-infrared survey of the VIMOS-SA22 field together with data from the DXS survey within the same field to study the environment and host galaxy properties of radio sources over a wide dynamic range of radio luminosities and environments. In particular we have focused our investigation on the small-scale environment of these sources, showing that the majority of radio-sources is found in environments consistent to the environments of sources with similar $J$-band magnitude, redshift, and rest-frame absolute magnitude $(M_{u}-M_{r})$ color. Nevertheless, we concluded that there is a significant sub-population of radio-AGN that is found in very overdense environments at small scales, up to 100 of times denser than their control sources. We showed that this component is significant and is not due to any statistical effects of our analysis. 

In addition to the above, we have also investigated the probability that there is a link between the radio luminosity of radio-AGN and their environment. Despite a wider radio luminosity range than previous studies, we do not find any appreciable trend between more radio-luminous AGN and more overdense environments. 

We then turned our attention to the host galaxy properties of the most overdense and the most underdense radio-AGN in our sample. By looking at the star formation efficiency of these sources and by utilising our environment results and those of similar studies in the literature we have put forth a scenario where, in the local universe, a significant component of radio-AGN should be triggered by the feeding of their central supermassive black holes through the mass ejection, in the form of stellar winds, from ongoing star formation in their host galaxies. On the other hand, we expect that the rest of the AGN should be triggered through accretion of gas from their galactic environment and inter-cluster medium, albeit with a lower accretion efficiency. This scenario agrees well with previous stipulations about the different origins and phenomenology of low- and high-excitation radio-AGN. 

Conversely, the picture seems to change at redshifts above $z\sim1$. At higher redshifts we do not observe any significant differences between either the sSFR or the accretion efficiency of radio-AGN in the most overdense and most underdense environments. This leads us to believe that at higher redshifts the availability of cold gas, either in the form of untapped gas reservoirs in galaxies or through galactic interactions and mergers, plays an increasingly important role in the triggering of radio-AGN, independent of their environment.

Follow-up spectroscopic observations of those radio-AGN in the most underdense environments can elucidate the true nature and properties of their nuclear activity, in terms of high- or low-excitation, and can also constrain the star formation in their host galaxies. We plan to pursue a continuation of this study employing a dataset with better wavelength coverage in the mid- and far-infrared regime, in order to better constrain the star formation efficiency through detailed modelling of their spectral energy distributions. Complimentary integral field spectroscopy observations of nearby radio-AGN in the most underdense environments can provide crucial insight about the actual processes fuelling these black holes by looking at the density, velocity, and velocity dispersion of the circumnuclear gas in these galaxies. 

\acknowledgments{We thank the anonymous referee whose comments have significantly improved this manuscript. This work was supported by the National Research Foundation of Korea (NRF) grant, No. 2008-0060544, funded by the Korea government (MSIP). D.K. acknowledges the fellowship support from the grant NRF-2014-Fostering Core Leaders of Future Program, No. 2014-009728 funded by the Korean government. This work is based in part on data obtained as part of the UKIRT Infrared Deep Sky Survey. The United Kingdom Infrared Telescope is operated by the Joint Astronomy Centre on behalf of the Science and Technology Facilities Council of the U.K. This research has made use of NASA's Astrophysics Data System Bibliographic Services.}
\clearpage

\appendix
\section{Deep VLA survey in SA22 field}
\label{app:VLA}
For this work we use radio observations of the VIMOS-SA22 field with the VLA radio interferometer at 1.4 GHz (\citealt{Chapman2004a,Chapman2004b}). As these observations and catalogs have not been explicitly presented before, we give here a brief description of the data acquisition, reduction, and source extraction.

The process of obtaining and reducing deep, high-resolution,
wide-field 1.4-GHz images is complicated by bandwidth smearing,
necessitating the use of spectral-line, pseudo-continuum correlator
modes at the National Radio Astronomy Observatory's\footnote{NRAO is
operated by Associated Universities Inc., under a cooperative
agreement with the National Science Foundation.} (NRAO) VLA, by
interference (man-made and solar), and by the presence of dozens of
bright (often structurally complex) sources in the primary beam.

For the field under consideration here, SA22, the problems encountered during data reduction were due to the fact that the field is
 crowded with bright sources (the
central 100\,arcmin$^2$ field contains a $\sim$\,200-mJy radio galaxy as well as several structurally complex FR\,I/II
sources). The field also has relatively poor nearby phase/amplitude
calibrators, the best of which is resolved on some
baselines. Fortunately, the presence of bright sources allowed self
calibration of the data, correcting the poor initial phase/amplitude
calibration. 

Data were taken every 5\,s in 3.25-MHz channels, 28 in
total, centred at 1.4\,GHz, recording left-circular and right-circular
polarisations. 3C\,84 and 3C\,286 were used for flux calibration. The
phase/amplitude calibrators, 1625+415 and 1035+564, were observed
every hour. 45\,hr of integration was
obtained in A configuration (maximum
baseline, 27\,km), and 20\,hr in B configuration
(maximum baseline, 9\,km)\footnote{Date of observations were '2003-07-06' and '1998-10-03' for A and B configurations, respectively.}.

After standard spectral-line calibration and editing of the data and
their associated weights, using {\sc aips}, the wide-field imaging
task, {\sc imagr}, was used to map the central $10\times 10$ arcmin$^{2}$ field in A-config (double the field size for B-config). 
These maps, made
with {\sc robust = 0} weighting of the visibilities, were used to
position {\sc clean} boxes around the sources, and {\sc imagr} was
re-run with 10,000 interations of the {\sc clean} algorithm (\citealt{Hogbom1974}, \citealt{Clark1980}). The {\sc clean} components thus produced were used
as a model for self calibration (in phase only) using {\sc calib} with
a relatively long integration time ($\sim$\,1--2\,min) and a low
signal-to-noise threshold (3--4\,$\sigma$).  Mapping was then
repeated, after checks on the {\sc clean} boxes. The new {\sc clean}
components were subtracted from the visibilities and the data were
clipped to remove spikes, then added back to the {\sc clean}
components.  The {\sc imagr/calib} loop was then repeated a further
four times (though without further clipping), steadily decreasing the
integration time and increasing the signal-to-noise threshold, the
final pass of {\sc calib} including both amplitude and phase (with the
mean gain modulus of the applied calibration set at unity).  This
iterative method resulted in the loss of less than 5 per cent of the
data. The A- and B-configuration data were dealt with separately and were not 
co-added since the complicated noise response was very different in the two configurations.
The resulting maps  have average noise levels of
8.8\,$\mu$Jy\,beam$^{-1}$, with 1.4 arcsec resolution, and 
22.4\,$\mu$Jy\,beam$^{-1}$, with 5.0 arcsec resolution. 

The area around the brightest objects, which cause the manifestation of sidelobes that could be detected as spurious sources, has been masked and sources within this area are disregarded for the following. In total we had to mask 3 vertical stripe-shaped regions (1 for A-configuration map and 2 for B-configuration), losing a total of 16.56 arcsec$^2$ and 522 arcsec$^2$ for the two configurations, respectively. Due to nature of overlap of the two configuration and the location of the masked regions, there is only a handful of sources which are common between the final catalogs of the two configuration. For these cases, we retain the detected sources in the A-configuration due to its superior sensitivity and resolution.

\section{SED fitting}
\begin{figure*}[htbp]
\begin{center}
\includegraphics[width=0.33\textwidth,angle=0]{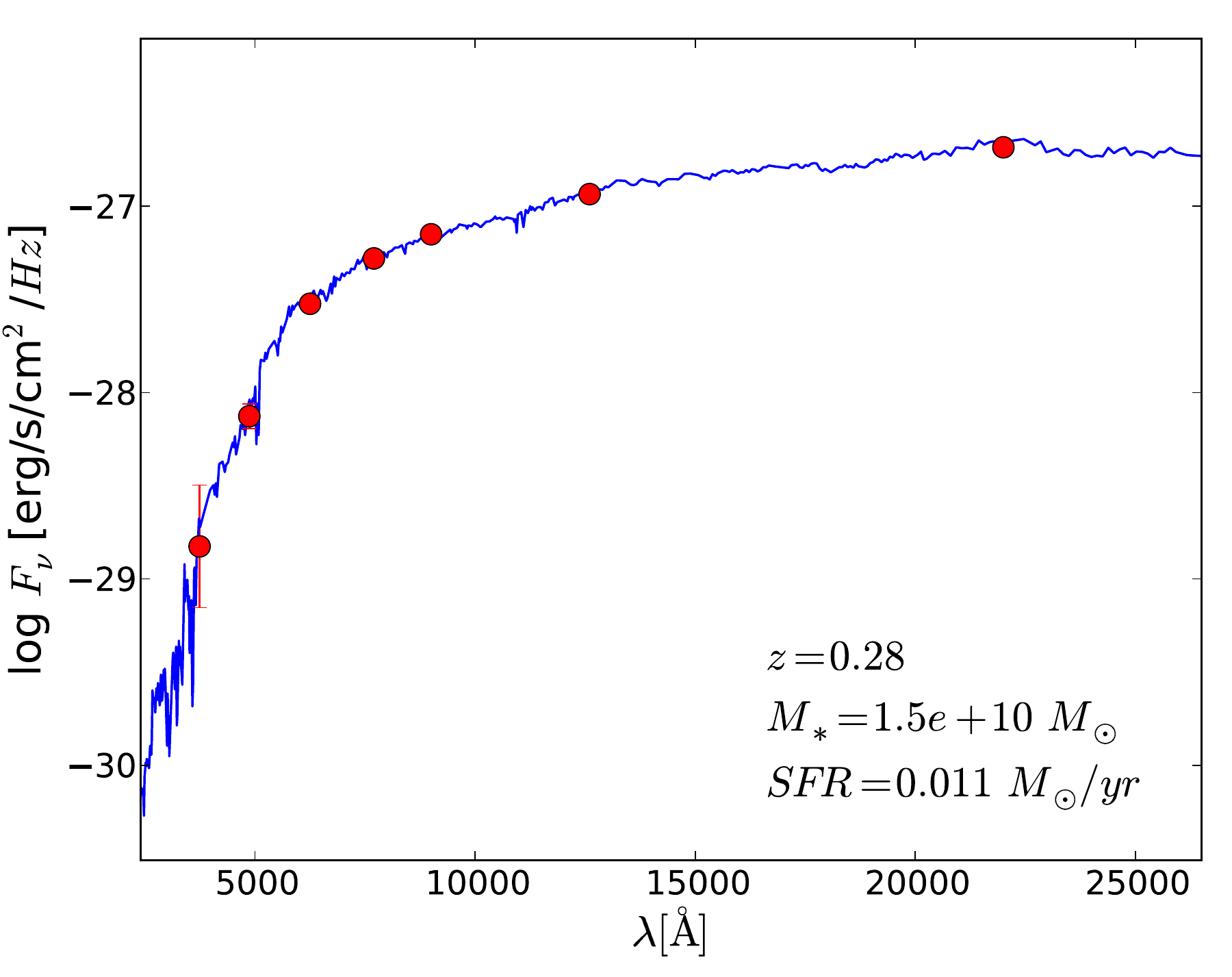}			
\includegraphics[width=0.33\textwidth,angle=0]{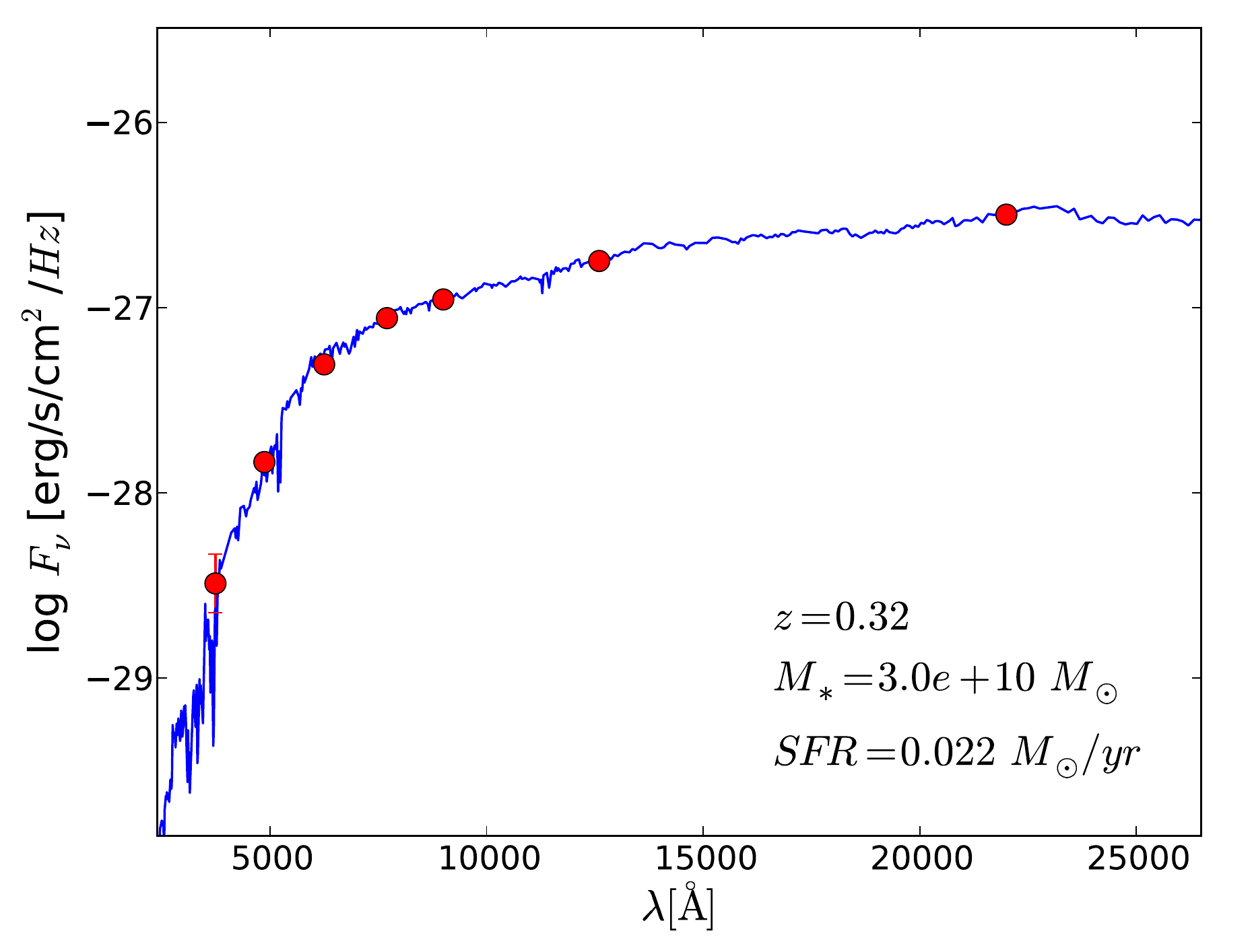}			
\includegraphics[width=0.33\textwidth,angle=0]{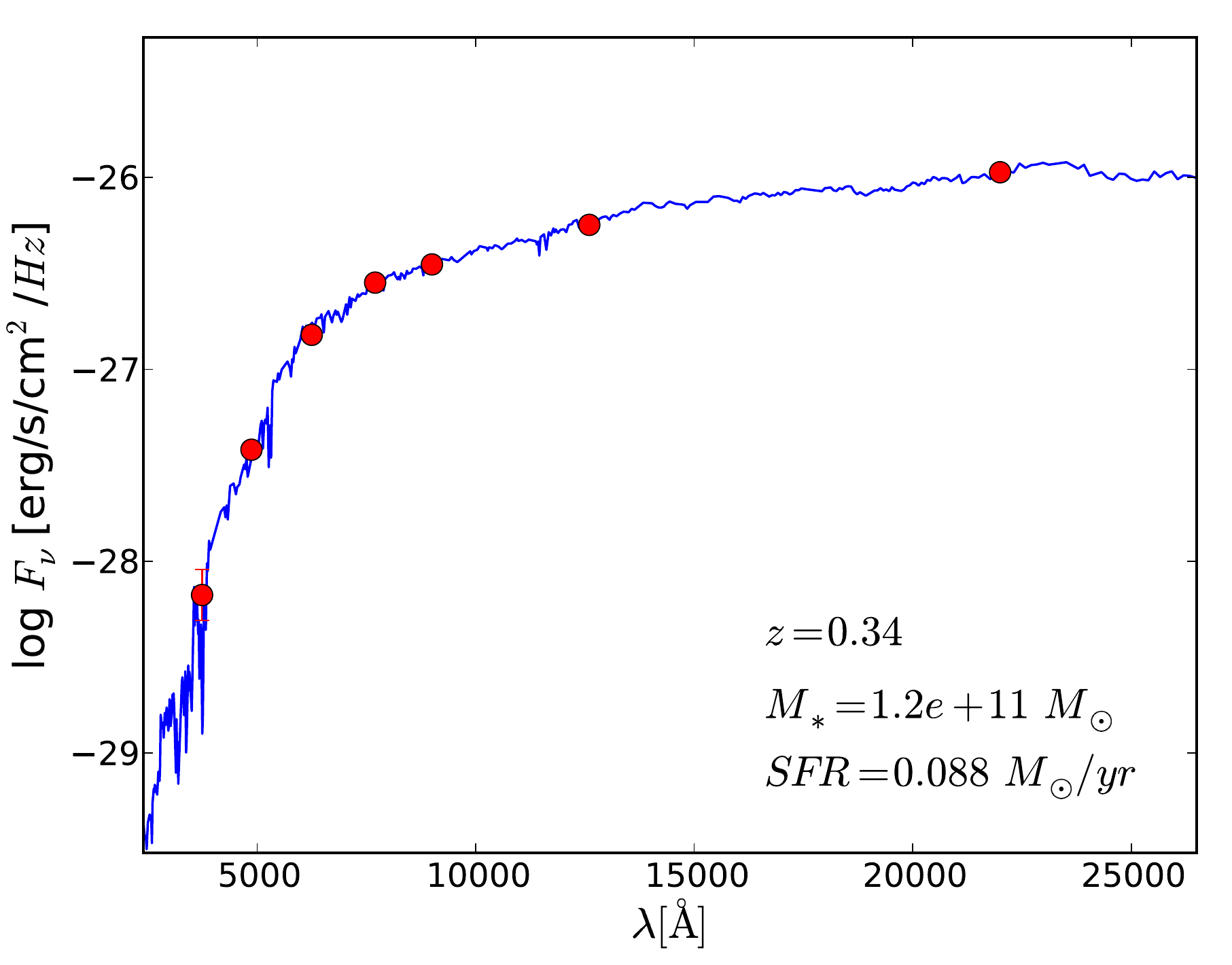}\\		
\includegraphics[width=0.33\textwidth,angle=0]{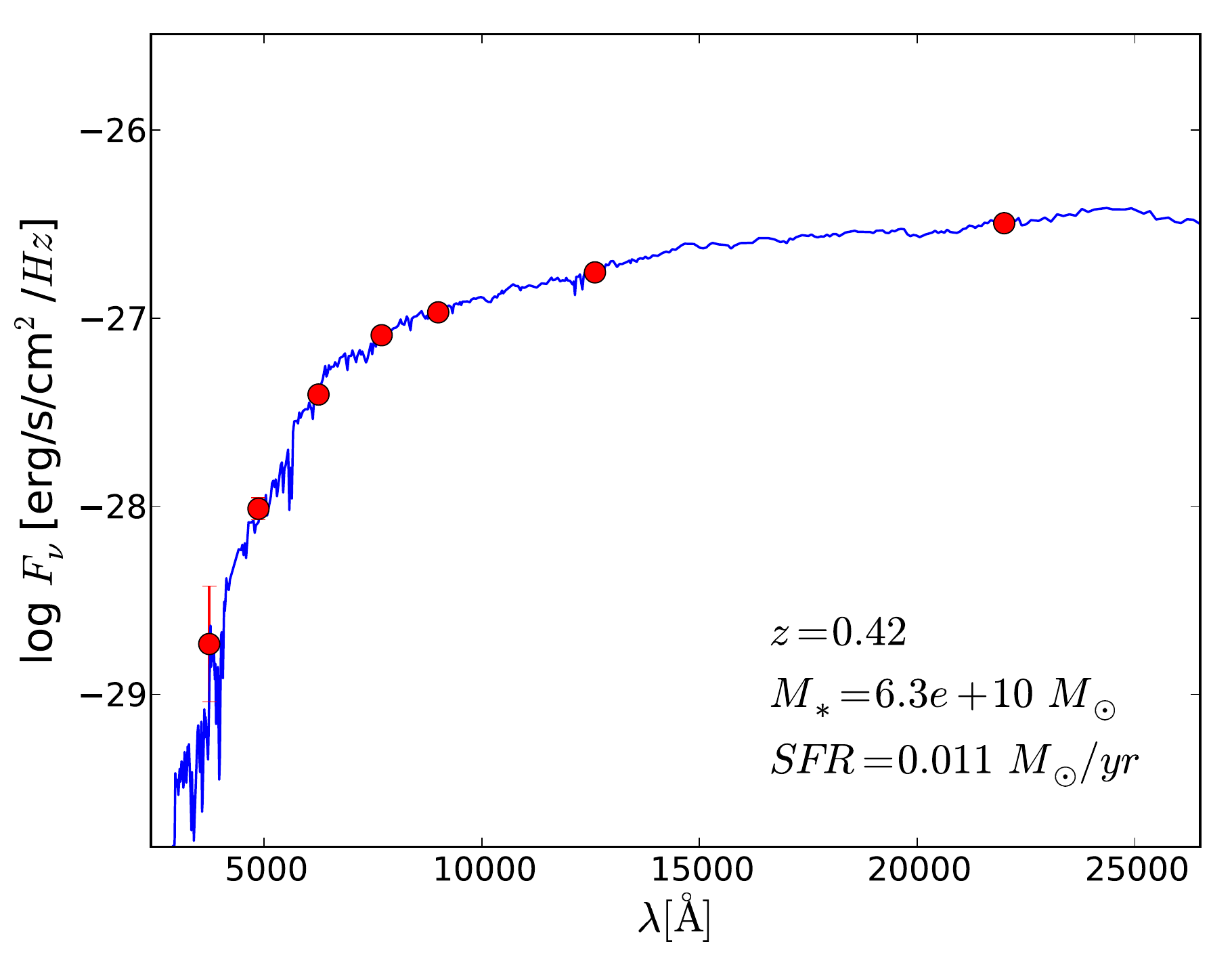}			
\includegraphics[width=0.33\textwidth,angle=0]{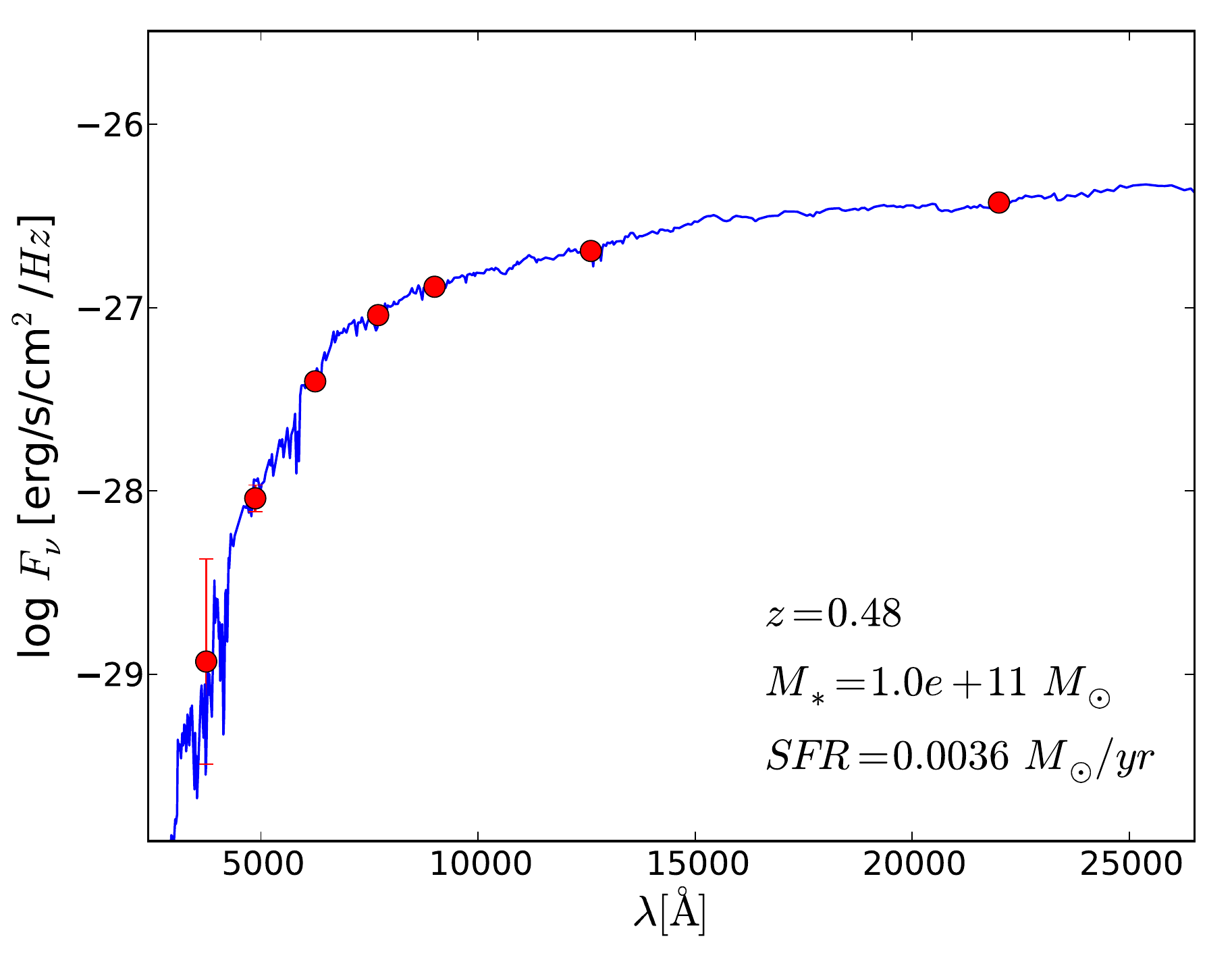}			
\includegraphics[width=0.33\textwidth,angle=0]{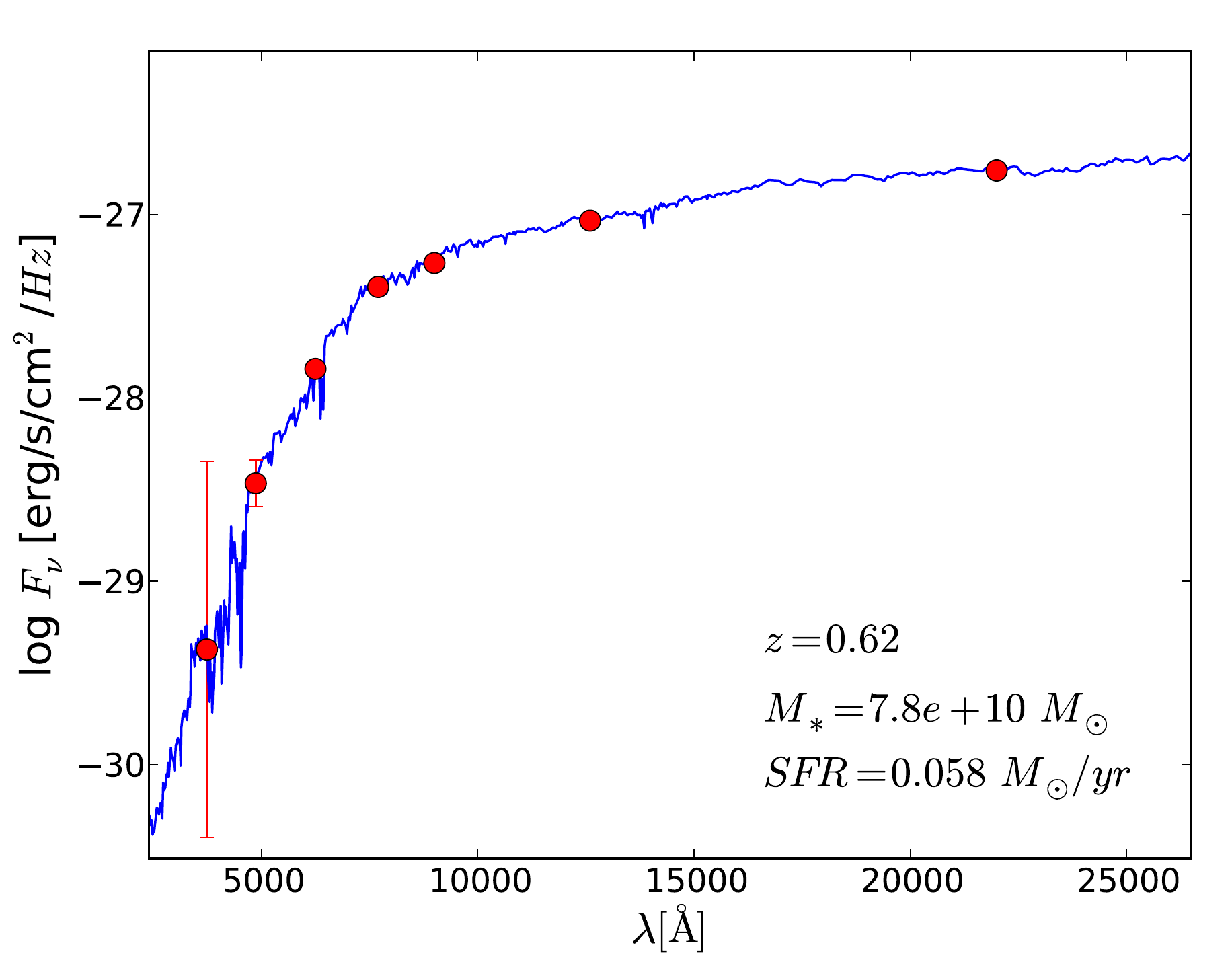}\\		
\caption{Examples of SED fits using our SED fitting code implementing the stellar population synthesis templates from \citet{Bruzual2003} and a delayed star formation history (e.g., \citealt{LeeSK2010}). Examples of radio-AGN sources classified in the OD sub-sample (see Section \ref{sec:SFR}).}
\label{fig:sed1}
\end{center}
\end{figure*}

\begin{figure*}[htbp]
\begin{center}
\includegraphics[width=0.33\textwidth,angle=0]{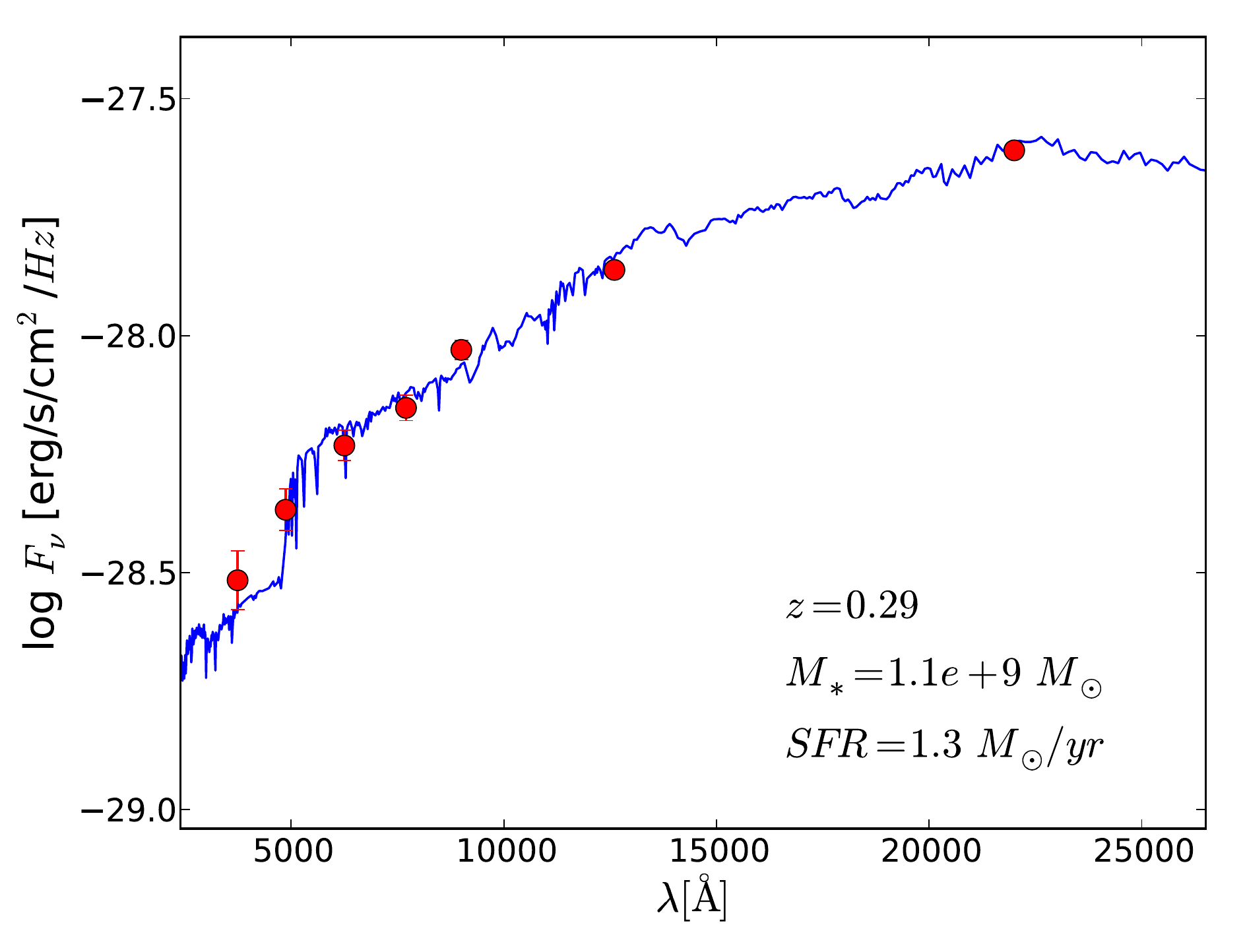}			
\includegraphics[width=0.33\textwidth,angle=0]{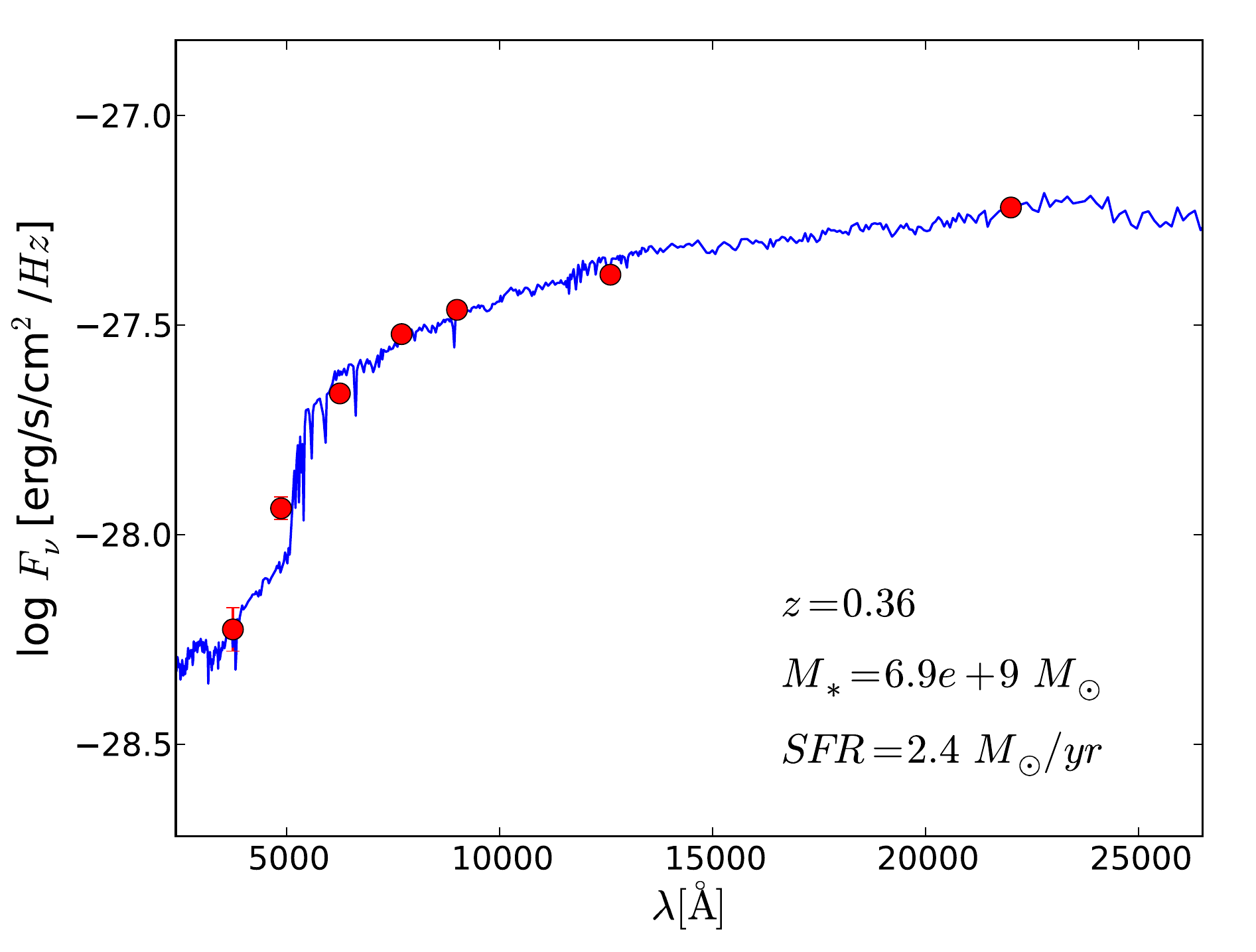}			
\includegraphics[width=0.33\textwidth,angle=0]{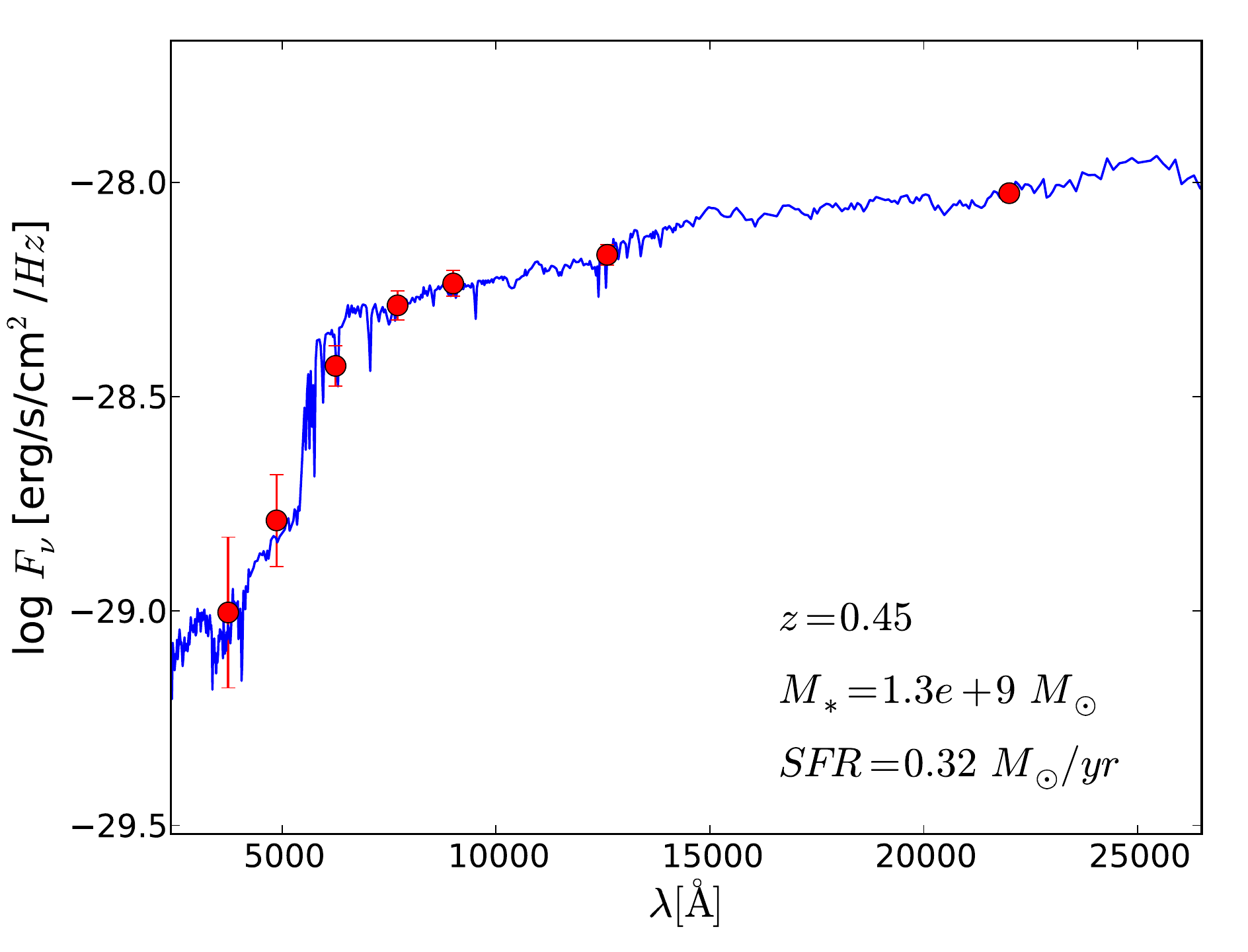}\\		
\includegraphics[width=0.33\textwidth,angle=0]{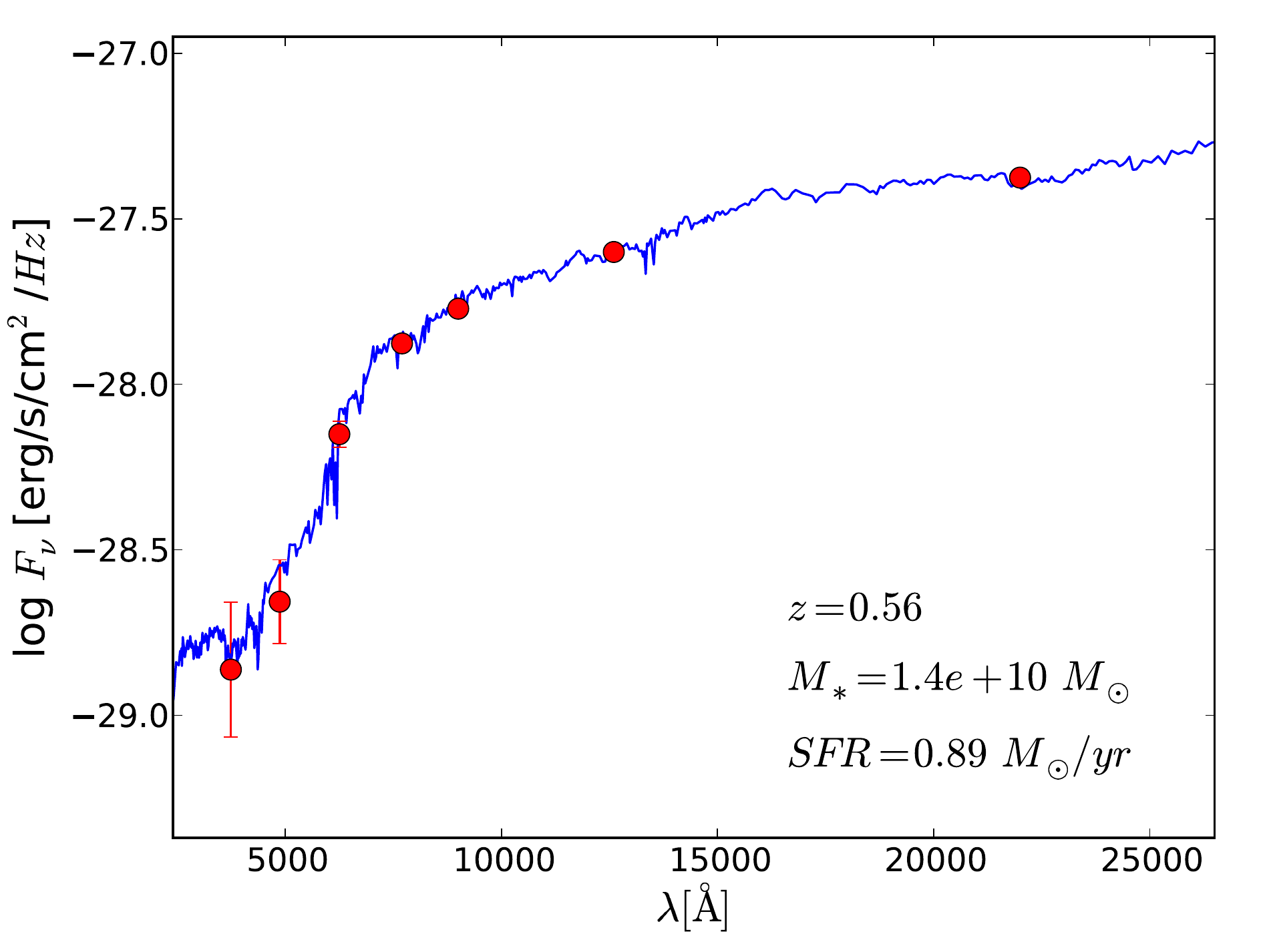}			
\includegraphics[width=0.33\textwidth,angle=0]{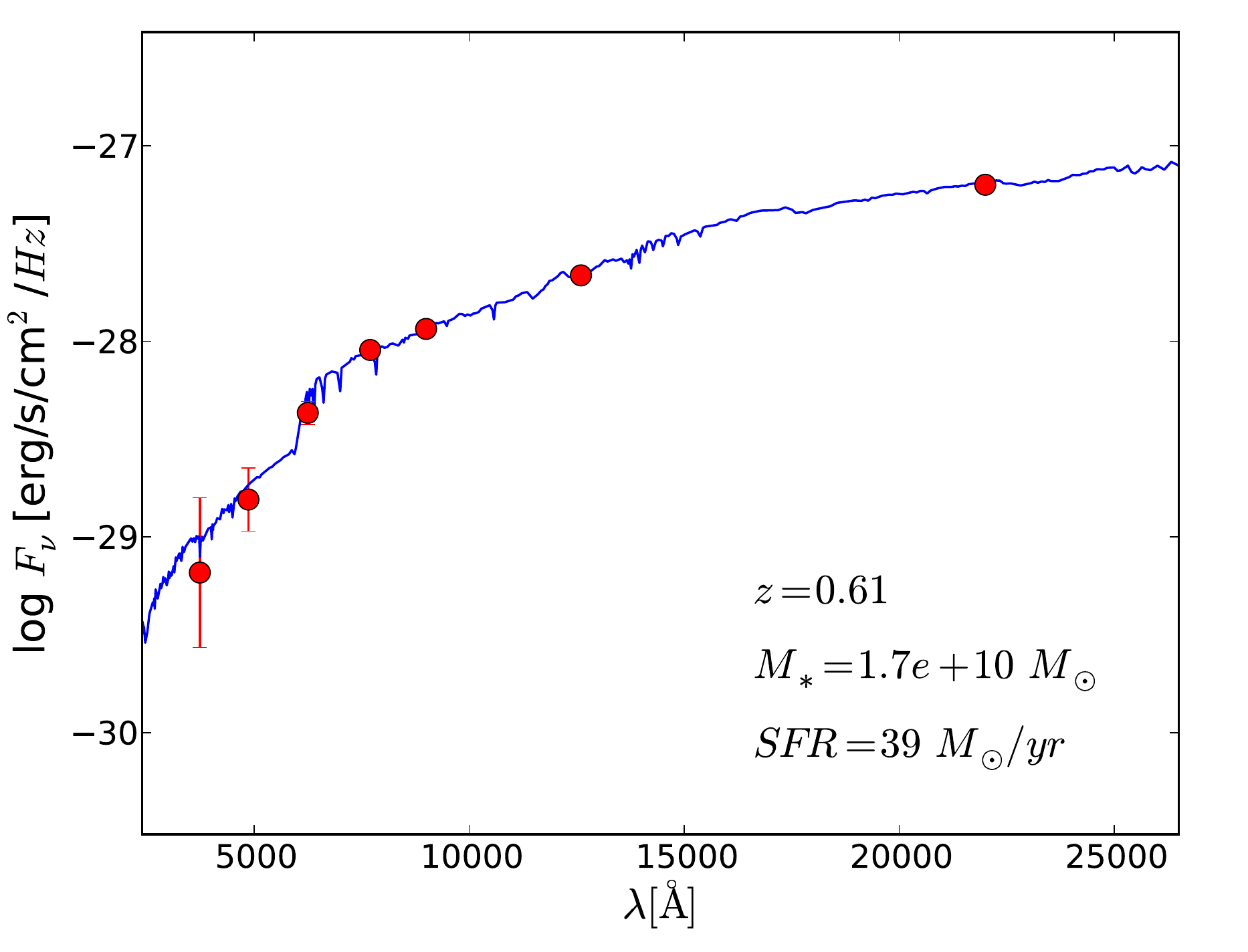}			
\includegraphics[width=0.33\textwidth,angle=0]{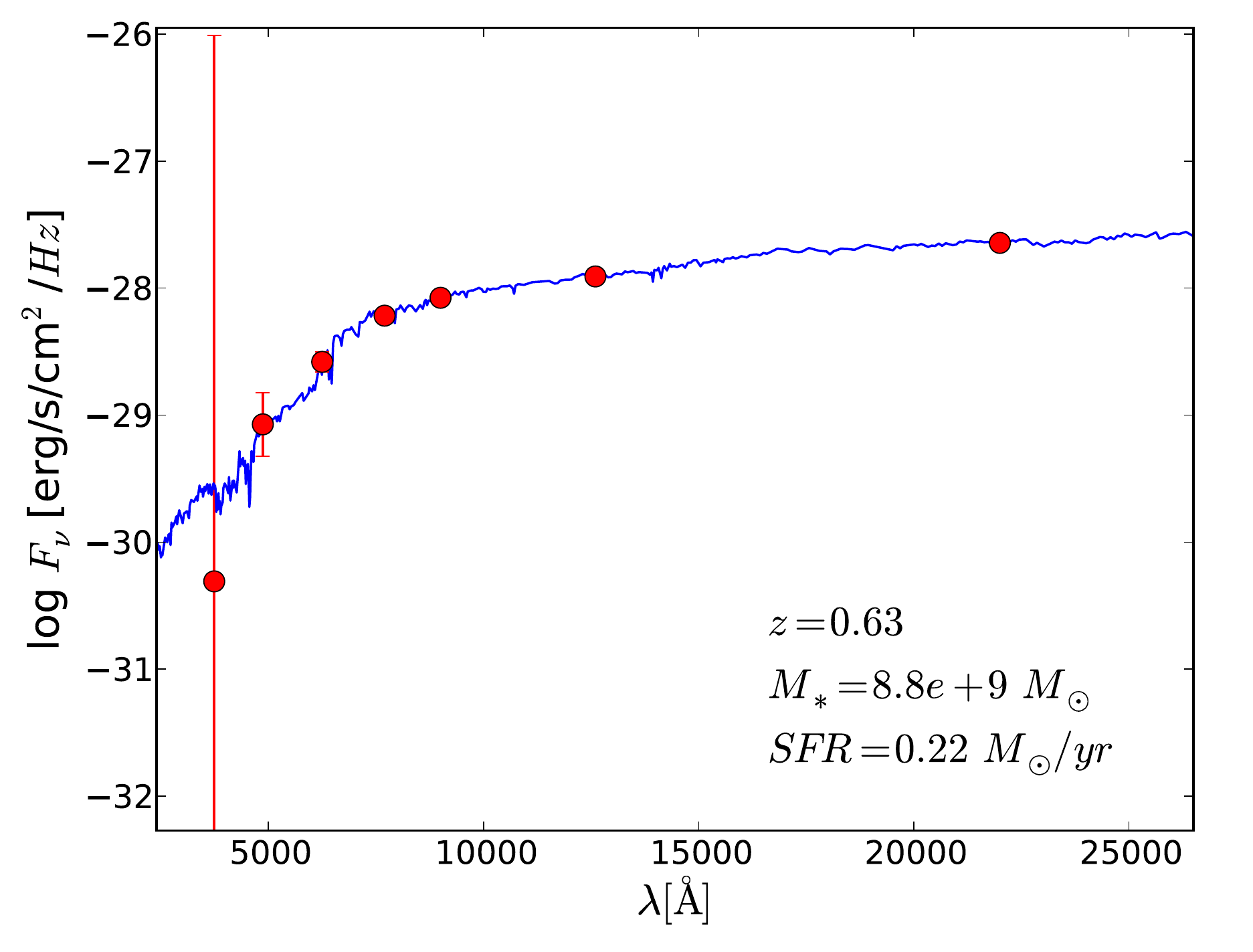}			
\caption{As in Fig. \ref{fig:sed1}. Here we show examples of radio-AGN sources classified in the UD sub-sample (see Section \ref{sec:SFR}).}
\label{fig:sed2}
\end{center}
\end{figure*}

In this Appendix Section we give examples of the best SED fits derived for the radio sources classified as radio-AGN due to their luminosity. In Fig. \ref{fig:sed1} we show examples for radio-AGN in the OD sub-sample, while Fig. \ref{fig:sed2} shows examples for radio-AGN in the UD sub-sample. Redshifts, and derived stellar masses and SFRs are given for each individual source. The SEDs are shown in the observed wavelength.

\bibliographystyle{aa}
\bibliography{bibtex}

\end{document}